\documentclass[aps,
            10 pt,
               pra,
               showpacs,
               amsmath,amssymb,
               nofootinbib,
               superscriptaddress,
               twocolumn,
               longbibliography]{revtex4-2}
\usepackage[ansinew]{inputenc}
\usepackage{bbold}
\usepackage{bm}
\usepackage{amsbsy}
\usepackage{amsthm}
\usepackage{amssymb}
\usepackage{amsfonts}
\usepackage{amsmath, mathtools}
\usepackage{dsfont}
\usepackage{graphicx}
\usepackage{epsfig}
\usepackage{epstopdf}
\usepackage{dsfont}
\usepackage{mathrsfs}
\usepackage{amsmath,amssymb,amsthm,bm,mathtools,amsfonts,mathrsfs,bbm,dsfont}
\usepackage[dvipsnames]{xcolor}
\usepackage[colorlinks]{hyperref}
\makeatletter
\newcommand\org@hypertarget{}
\let\org@hypertarget\hypertarget
\renewcommand\hypertarget[2]{%
  \Hy@raisedlink{\org@hypertarget{#1}{}}#2%
  }
\makeatother
\usepackage[figure,table]{hypcap}
\usepackage{MnSymbol}
\usepackage{enumerate}
\usepackage{float}
\usepackage{comment}
\usepackage{braket}
\usepackage{subfigure}
\usepackage{physics}

\hypersetup{
	bookmarksnumbered,
	pdfstartview={FitH},
	citecolor={darkgreen},
	linkcolor={darkred},
	urlcolor={darkblue},
	pdfpagemode={UseOutlines}}
\definecolor{darkgreen}{RGB}{50,190,50}
\definecolor{darkblue}{RGB}{0,0,190}
\definecolor{darkred}{RGB}{238,0,0}
\usepackage{soul}
\usepackage{orcidlink}
\usepackage{multirow} 
\usepackage[most]{tcolorbox}
\usepackage{hyperref}

\usepackage{mathtools}
\usepackage{xparse}
\makeatletter
\newcommand{\raisemath}[1]{\mathpalette{\raisem@th{#1}}}
\newcommand{\raisem@th}[3]{\raisebox{#1}{$#2#3$}}
\makeatother
\NewDocumentCommand{\newhbar}{O{0pt} O{0pt}}{
  \ensuremath{\mathrlap{\raisemath{#2}{\hspace*{#1}{\mathchar'26\mkern-9mu}}}h}%
}

\def\hbar{\newhbar[0.4pt][-0.35pt]}

\renewcommand{\vec}[1]{\boldsymbol{#1}}

\newcommand{\vA}{\vec{A}}

\newcommand{\ve}{\vec{{\rm e}}}
\renewcommand{\vu}{\vec{{\rm u}}}

\newcommand{\la}{\langle}
\newcommand{\ra}{\rangle}

\newcommand{\cL}{\mathcal{L}}

\newcommand{\Op}[1]{\hat{#1}}

\newcommand{\osigma}{\Op{\sigma}}

\newcommand{\oH}{\Op{H}}

\newcommand{\id}{\ensuremath{\mathbbm 1}}
\renewcommand{\Im}{\ensuremath{{\rm Im}}}
\renewcommand{\Re}{\ensuremath{{\rm Re}}}

\newcommand{\ovsigma}{\Op{\vec{\sigma}}}

\renewcommand{\thesection}{\Roman{section}}
\renewcommand{\thesubsection}{\Roman{section}.\Alph{subsection}}
\renewcommand{\thesubsubsection}{\Roman{section}.\Alph{subsection}.\arabic{subsubsection}}
\makeatletter
\renewcommand{\p@subsection}{}
\renewcommand{\p@subsubsection}{}
\makeatother

\usepackage{tcolorbox}
\definecolor{c1}{HTML}{F26035} 
\colorlet{h1}{c1!30}
\definecolor{c2}{rgb}{0.0, 0.51, 0.5} 
\colorlet{h2}{c2!30}

\begin{document}
\title{
Simulating Microwave-Controlled Spin Imaging with Free-Space Electrons 
}
\author{Santiago Beltr{\'a}n-Romero\,\orcidlink{0009-0000-0310-5551}}
\thanks{santiago.romero@tuwien.ac.at}
\affiliation{Atominstitut, Technische Universit{\"a}t Wien, Stadionallee 2, 1020 Vienna, Austria}
\affiliation{University Service Centre for Transmission Electron Microscopy, TU Wien, Stadionallee 2/E057-02, 1020 Vienna, Austria}

\author{Stefan L{\"o}ffler\,\orcidlink{0000-0003-0080-2495}}
\affiliation{University Service Centre for Transmission Electron Microscopy, TU Wien, Stadionallee 2/E057-02, 1020 Vienna, Austria}

\author{Dennis R{\"a}tzel\,
\orcidlink{0000-0003-3452-6222}}
\thanks{These authors share last authorship
\\
dennis.raetzel@tuwien.ac.at\\
philipp.haslinger@tuwien.ac.at}
\affiliation{Atominstitut, Technische Universit{\"a}t Wien, Stadionallee 2, 1020 Vienna, Austria}
\affiliation{University Service Centre for Transmission Electron Microscopy, TU Wien, Stadionallee 2/E057-02, 1020 Vienna, Austria}
\affiliation{ZARM, University of Bremen, 28359 Bremen, Germany}

\author{Philipp Haslinger\,\orcidlink{0000-0002-2911-4787}}
\thanks{These authors share last authorship
\\
dennis.raetzel@tuwien.ac.at\\
philipp.haslinger@tuwien.ac.at}
\affiliation{Atominstitut, Technische Universit{\"a}t Wien, Stadionallee 2, 1020 Vienna, Austria}
\affiliation{University Service Centre for Transmission Electron Microscopy, TU Wien, Stadionallee 2/E057-02, 1020 Vienna, Austria}

\date{\today}

\begin{abstract}

Coherent spin resonance techniques, such as nuclear and electron spin resonance spectroscopy, have revolutionized non-invasive imaging by providing spectrally resolved information about spin dynamics. Motivated by the recent emergence of electron microscopy methods capable of sensing microwave-excitations~\cite{jaros2025sensingspinsystemstransmission}, we establish a theoretical framework for Spin Resonance Spectroscopy (SRS) in transmission electron microscopy (TEM). This technique combines microwave pump fields with focused electron probe beams to enable state-selective spin imaging at the atomic scale.
Using scattering theory, we model the interaction between free-space electrons and electron spin systems, capturing both elastic and inelastic processes. 
The strongest effect of the spin system on the free electron is a magnetic phase shift. 
Our simulations demonstrate that phase shifts from individual electron spins are detectable in both image mode and diffraction mode. In principle, differential measurements under microwave control allow the extraction of local resonance frequencies that are influenced by the surrounding spin environment. By evaluating the Classical Fisher Information (CFI), we identify imaging conditions that maximize the signal-to-noise ratio (SNR), showing how defocus and beam width affect the measurement sensitivity. These findings establish a foundation for integrating SRS with high-resolution TEM, bridging spin spectroscopy and atomic-scale imaging, and enabling new capabilities in quantum spin research and nanoscale materials characterization.

\end{abstract}

\date{\today}

\maketitle


\section{Introduction}\label{sec:intro}

Recent breakthroughs in Transmission Electron Microscopy (TEM) stem from simultaneous progress on multiple technological fronts. Aberration-corrected lenses now deliver sub-\AA ngstr{\"o}m resolution~\cite{sawada2009, Erni2009, Hawkes2009, ishikawa2022}, while cryogenic sample preparation methods preserve delicate organic structures that would otherwise be inaccessible~\cite{Bai2015}. In parallel, ultrafast probing techniques have extended TEM into the time domain, reaching picosecond to sub-femtosecond temporal resolution~\cite{Lobastov2005, Zewail20064DUE, takubo_2022, mohler2020ultrafast, Morimoto2018, Feist2015, Priebe2017}. The integration of highly coherent pulsed electron sources with picosecond beam blanking~\cite{ZHANG2020112925,reisbick2022stroboscopic} and high-speed direct detectors with sub-nanosecond time-stamping precision~\cite{Llopart_2022} has further transformed the capabilities of modern transmission electron microscopes, creating platforms capable of tracking structural and electronic dynamics with unprecedented precision. This impressive list of TEM techniques may be extended by Spin Resonance Spectroscopy (SRS) if combined with microwave-based spin control as some of us proposed in a recent article~\cite{Haslinger_2024}. 

Conventional microwave (MW) techniques for studying spin dynamics, such as electron spin resonance (ESR) and nuclear magnetic resonance (NMR) spectroscopy, have revolutionized non-invasive imaging across various fields~\cite{Liang2000, Callaghan1993Principles, Bienfait_2015}, including medical diagnostics~\cite{Bullmore2012}, biology~\cite{Borbat2001, sivelli2022nmr}, and chemistry~\cite{Weil1994, Wertz2012}, but also enabled high-precision measurements for fundamental physics~\cite{Ramsey1950}, including searches for dark energy and dark matter~\cite{Safronova2018}.
However, they often fall short in spatial resolution compared to TEM-based methods. While specialized ESR setups like scanning tunneling microscopy and nitrogen-vacancy centers offer high sensitivity~\cite{Simpson2017, Seifert2020}, they are predominantly surface-sensitive. 

In contrast, TEM provides high spatial and temporal resolution to gain insight into bulk samples on the order of several 100~nm, making it uniquely suited for investigating complex quantum states with nanoscale precision.
Beyond our theoretical exploration of the quantum mechanical interaction between microwave-controlled spin states and free electrons~\cite{Haslinger_2024}, stroboscopic MW pump--electron probe techniques have enabled breakthroughs in ultra-fast spin imaging. These include the visualization of magnetic vortex cores~\cite{Moller2020}, beam deflection measurements near MW circuits~\cite{WEELS2022113392}, and the imaging of spin waves in ferromagnetic thin films~\cite{Liu2025CorrelatedSpins}. Additionally, a specially optimized MW sample holder has been developed to harness the strong magnetic environment of a transmission electron microscope, enabling standard ESR spectroscopy of miniaturized samples~\cite{jaros2024electronspinresonancespectroscopy} and facilitating the detection of spin precession contributions to electron beam deflection~\cite{ jaros2025sensingspinsystemstransmission}.

Due to their quantum nature, beam electrons in TEM interact with the sample via multiple coherent pathways. The resulting electron quantum state encodes 
information about the sample, accessible through measurements such as beam deflection angles or phase shifts. Accurate interpretation of these measurements requires detailed simulation of the electron's quantum state evolution. In this work, we present a theoretical framework for simulating SRS in the transmission electron microscope using a pump-probe scheme that combines MW excitation of spins with coherent electron beams as the probe.

Our model describes the interaction between free-space electrons and quantum spin systems in TEM. The incident electron is represented as a Gaussian wavepacket that interacts with the magnetic field generated by the spins of the sample. When the backaction on the spin system is negligible, the state of the electron stays pure and the interaction leads to a position-dependent phase shift, which is commonly known as the Aharonov-Bohm phase, accumulated by the electron wavefunction. This phase manifests in both the momentum (Fourier) space and the real (image) space, forming the physical basis of spin-resolved contrast in TEM, as illustrated in Fig.~\ref{fig:schematic}d. 
In the weak coupling regime considered here, these spin-dependent effects appear primarily to first order in the spin-electron coupling strength and are most effectively revealed through the interference of the scattered and unscattered components of the electron's quantum state, a scheme commonly referred to as holography ~\cite{Lichte_2008, McMorran2011, park2014observation, Boureau_2021, schwartz2019laser}.
In diffraction mode, the interference results in measurable angular deflections and redistribution of intensity in the far-field scattering pattern. In image mode, it manifests as phase contrast or holographic fringes.

To describe the interaction, we employ a scattering theoretical approach grounded in quantum electrodynamics, treating both the electron and spin system as quantum mechanical entities. Therefore, our model allows us to quantify the backaction of the electron on the spin system. Furthermore, it can be used as the basis for an investigation of beam-induced driving effects~\cite{Ratzel2021, Gover2020free, Zhao2021, kolb2025coherentdrivingquantummodulated}, and the loss of coherence of the electron beam due to the interaction with many spins in future studies. 

In practice, the effect of sample spins on the beam electrons is accompanied by the influence of electrostatic potentials and inelastic scattering events within the sample. We treat these as background signals that can be effectively suppressed through a differential imaging approach based on controlled spin manipulation. This forms the basis of SRS in TEM. In SRS, a tunable MW pulse is applied to drive spin transitions, allowing the measurement of local resonance frequencies. These frequencies depend on both an external magnetic field, and the local spin environment, providing spatially resolved insight into spin distributions and interactions. By comparing measurements before and after the MW pulse, one isolates spin-specific magnetic contributions with enhanced sensitivity, enabling indirect imaging of nanoscale spin dynamics.

We further optimize spin detection strategies by evaluating the Classical Fisher Information (CFI) across different experimental configurations. This approach, which quantifies the sensitivity of a measurement to underlying parameters, has recently been applied in TEM for electrostatic phases to identify spatial regions and imaging methods of high informational content~\cite{Koppell2022TransmissionEM, Dwyer2023, Dwyer2024}. By comparing image and diffraction modes, we identify the conditions that maximize sensitivity and signal enhancement. Furthermore, we present an image analysis pipeline that optimizes the Signal-to-Noise Ratio (SNR).

This work provides a theoretical foundation for integrating spin resonance techniques with 
electron microscopy, expanding the capabilities of TEM for atomic-scale spin imaging and manipulation. By optimizing imaging strategies and exploring the metrological bounds of spin detection, we bridge traditional magnetic resonance with ultrafast atomic-scale imaging. These advancements open new possibilities in nanoscale spintronics~\cite{Wang_2018} and materials characterization, including applications in battery research~\cite{Lu2017, Yamamoto2010}.

The article is structured as follows: In Section~\ref{sec:model}, we describe the theoretical model and its relation to the experimental realization within the transmission electron microscope. 
Section~\ref{sec:scattering} presents the final electron state, discussing its scattering amplitudes and accounting for both elastic and inelastic processes. In Section~\ref{sec:hol-imaging}, we present simulations of SRS {in TEM with single-spins in diffraction mode and image mode, including the quantification of SNR for Bohr magneton estimation. We also explore the optimization of data analysis applied to the resulting images.} 
Section~\ref{sec:experimental_feasibility} discusses the experimental feasibility of the spin dynamics detection scheme and the predicted SNR. Finally, in Section~\ref{sec:conclusions}, we provide conclusions and an outlook for future investigations.

\section{Model of MW pumping Spin Probing in TEM}\label{sec:model}

\begin{figure*} \centering \includegraphics[width=\linewidth]{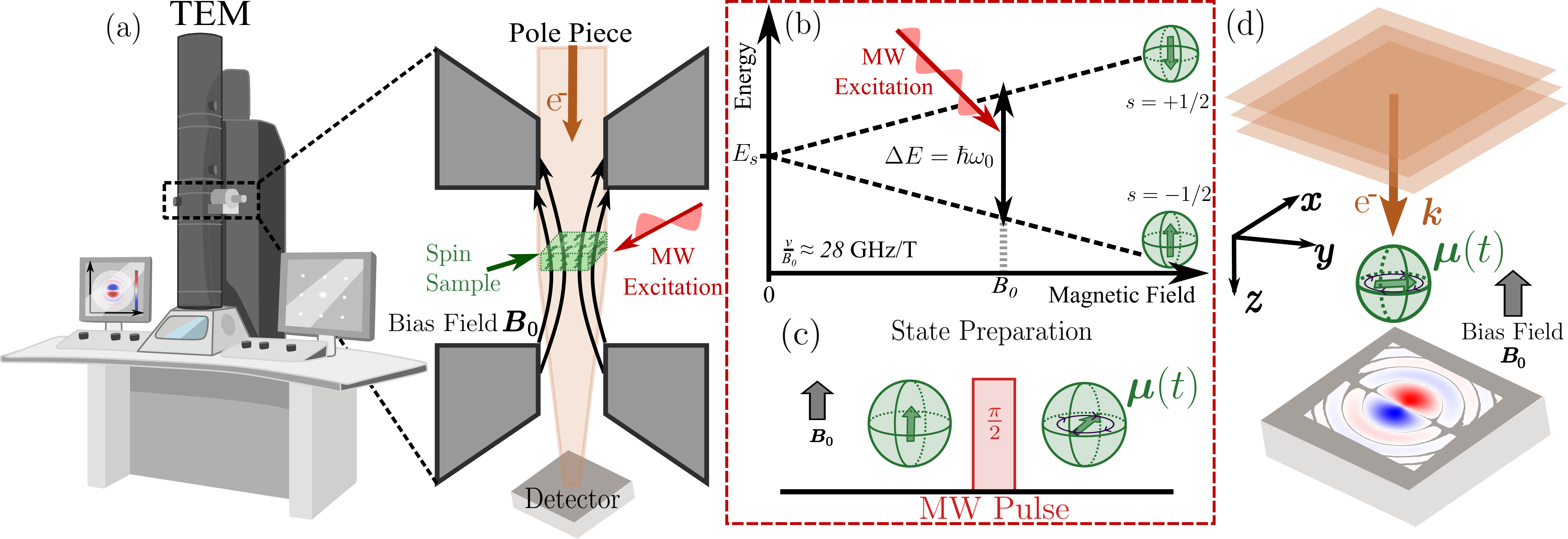} \caption{
(a) Schematic representation of the transmission electron microscope setup for image acquisition, highlighting a zoomed-in view of the sample region positioned between the objective lens pole pieces. A spin sample is placed in a bias magnetic field $\vec{B}_0$, generated by the pole pieces, with the sample subjected to microwave (MW) pulses and probed by an electron beam within the microscope. This configuration enables precise read-out of spin dynamics through the combination of MW excitation and high-resolution electron beam probing.
(b) The Zeeman effect is illustrated, showing the splitting of spin states, where states with $s = \pm1/2$ are energetically separated. Transitions between the spin states are achieved through resonant MW pulses.
(c) The state preparation process begins with the spin's magnetic moment $\vec{\mu} = g_{\rm e} \mu_{\rm B} \langle\ovsigma\rangle/2 \approx -\mu_{\rm B} \langle\ovsigma\rangle$ initially aligned along the bias magnetic field. A resonant $\pi/2$ pulse tilts the spin into the equatorial plane of the Bloch sphere, initiating precession. The precession is visualized by purple arrows encircling the equator.
(d) Graphical representation of the interaction between a precessing spin and an electron plane wave beam characterized by momentum $\hbar\vec{k}$: The electron beam propagates under the influence of the bias field, and its interaction with the spin is captured by the detector. The image displayed here corresponds to image mode, though data acquisition can also be performed in diffraction mode. This dual-observation approach provides a consistent basis for analyzing spin dynamics using complementary imaging techniques.} \label{fig:schematic} 
\end{figure*}

A spin sample inside a transmission electron microscope is subjected to a bias field, $\vec{B}_0$, primarily generated by the microscope's pole pieces, as shown in Figure \ref{fig:schematic}~a. In the vicinity of the sample, this field is aligned with the propagation axis of the beam electrons (the $z$-axis). Non-vanishing components of the electron momentum perpendicular to the $z$-axis then give rise to a Lorentz force which is, however, typically small enough to be neglected in our study\footnote{They are part of the subtracted background, but do not significantly change the results for the spin-induced effects.}.
Spins in the sample experience a Zeeman shift, as illustrated in Figure~\ref{fig:schematic}~b. A MW pump pulse, tuned to the Zeeman splitting, can then be used to control the quantum state of the spin. For instance, a specifically tuned pulse (intensity and duration) called a $\pi/2$ pulse can be used to tilt the spin by 90$^\circ$ to the equator of the Bloch sphere, corresponding to a quantum superposition of the two Zeeman levels. This superposition results in a precession, as shown in Figure~\ref{fig:schematic}~c, which continues until the spin state relaxes back to the lower energetic state corresponding to an anti-alignment of the spin expectation value with the bias field.

By comparing TEM images for the spin aligned perpendicular to the bias field and after relaxation (when anti-aligned with the field), one can isolate spin-induced effects, as illustrated in Figure~\ref{fig:schematic}~d. Other deflections from magnetic fields not sourced by the spin, as well as the mentioned effects of electron velocity components perpendicular to the $z$-axis, electrostatic potentials and electric dipole interactions, appear as constant effects during spin precession and are removed by image subtraction. 

To simulate the probing of electron spins as described above, we model the interaction in the Coulomb gauge using the following Hamiltonian:
\begin{subequations}
    \begin{align}
    \hat{H} &= \hat{H}_0 + \hat{H}_{\rm I},\label{eq:Hmain} \\
    \hat{H}_0 &= \sqrt{(m_{\rm e}c^2)^2 + |\hat{\boldsymbol{p}}|^2 c^2} - \mu \boldsymbol{B}_0 \cdot \hat{\boldsymbol{\sigma}},\label{eq:H0_main} \\
    \hat{H}_{\mathrm{I}} &= \frac{\mu_0 \mu}{4\pi} \left( \hat{\boldsymbol{\sigma}} \cdot \frac{\hat{\boldsymbol{r}}}{\hat{r}^3} \times \hat{\boldsymbol{p}} \right) \frac{e}{m_{\rm e} {\gamma} (\hat{\boldsymbol{p}})}\label{eq:Hint},
    \end{align}
\end{subequations}
where $\mu=-\mu_B$ is the magnetic moment of the electron spin when approximating its g-factor as -2, $\mu_{\rm B}=\frac{e\hbar}{m_{\rm e}}$ is the Bohr magneton, and $\vec{\osigma}=(\osigma_x,\osigma_y,\osigma_z)$ is the vector of Pauli matrices. Note that the case of spin $1/2$ nuclear spins is obtained by replacing $-\mu_B$ with the corresponding nuclear spin magnetic moment.
The free (non-interacting) Hamiltonian $\hat{H}_0$ consists of the relativistic kinetic energy of the probe electron and the Zeeman coupling of the sample spin\footnote{We assume an idealized system of isolated spin, neglecting electrostatic potentials and other non-magnetic interactions as they cancel out in differential measurements.} to the external magnetic field $\vec{B}_0$. The interaction Hamiltonian $\hat{H}_{\rm I}$ describes the relativistically corrected magnetic dipole coupling between the spin-generated electromagnetic field and the momentum of the electron wavepacket. The relativistic correction enters through the Lorentz factor $\gamma(\hat{\boldsymbol{p}}) = \sqrt{1 + (|\hat{\boldsymbol{p}}|/m_{\rm e}c)^2 }$, which effectively modifies the electron's mass and wavelength.  
As shown in Appendix~\ref{ap:relativistic_model}, this Hamiltonian is only approximately Hermitian, since higher-order relativistic terms are neglected in the parallel-illumination regime. 
{This approximation is justified for distances between spin and electron of the order of picometers and larger, considering conventional beam energies of at most MeV. Since electrons and nuclei are delocalized on the picometer scale and electron beams in the microscope are focused at most on this scale, the approximation is valid for all samples we consider.} 
Other possible contributions, such as probe-sample spin-spin coupling, the diamagnetic interaction, and higher-order multipolar terms, are neglected since they are negligible for nanometer-scale probe beams.
In particular, the diamagnetic interaction (from the quadratic vector potential term) is suppressed because it scales quadratically with the Bohr magneton ($\mu_{\rm B}^2$), whereas the dominant interaction considered here scales linearly with $\mu_{\rm B}$.

In the interaction picture, the time-dependent interaction Hamiltonian is given by $\hat{H}_{\mathrm{int}}(t) = \hat{U}_0^{\dagger}(t) \hat{H}_{\rm I} \hat{U}_0(t)$, where the free evolution unitary operator is $\hat{U}_0=\exp\left\{\frac{it}{\hbar}\hat{H}_0\right\}$. At $t=0$, the combined evolution of the free electron and the bound electron spin in the interaction region is described by the scattering operator.
To first order in perturbation theory, this is expressed as $\hat{S}\approx\mathbb{1}+\hat{\tau}^{(1)}$ with $\hat{\tau}^{(1)}= -\frac{i}{\hbar} \int_{-\infty}^{\infty} dt \, \hat{H}_{\rm int}(t)$.
To include the delocalization of the quantum spin system, we average $\hat{\tau}^{(1)}$ over the spatial probability density $|\psi(\vec{r_{\rm s}})|^2$, leading to
\begin{equation}
   \hat{\tau}^{(1)}= -\frac{i}{\hbar} \int_{-\infty}^{\infty} dt \, \int d^3 r_{\rm s} \, 
        |\psi_{\rm s}(\vec{r}_{\rm s})|^2 \hat{H}_{\rm int,r_{\rm s}}(t)\,,
\end{equation}
where $\hat{H}_{\rm int,r_{\rm s}}$ is the interaction Hamiltonian after the transformation $\hat{\vec{r}}\rightarrow \hat{\vec{r}}-\vec{r}_{\rm s}$. Motivated by the imaging of free spins in benchmark ESR systems such as $\alpha,\gamma$-bisdiphenylene-$\beta$-phenylallyl (BDPA) crystals~\cite{BOERO2013133}, we approximate our single-spin system on the fundamental level by the 1s orbital of the Hydrogen atom with Bohr radius $a_0$, which captures the essential spatial extent of the spin wavefunction.

Thus, in first-order perturbation theory, the final state of the system is given by:
\begin{equation}\label{eq:2delectronspin_state}
\begin{split}
    \ket{\psi_{\text{out}}} \approx & \left(\mathbb{1}+\hat{\tau}^{(1)}\right) \ket{\varphi_{\text{in}}}\otimes\ket{s},
\end{split}
\end{equation}
where $\ket{\varphi_{\text{in}}}$ is the incident beam wavefunction and $\ket{s}$ is the state of the spin at $t=0$ after being coherently driven by the MW pulse and evolving for a delay $t_0$.  We assume the delay to be large enough such that the probe electron does not interact with the MW pulse. 
Since the interaction time is on the order of picoseconds, which is much shorter than the spin precession period\footnote{With periods scaling as $\sim(28~\text{GHz/T})^{-1}$ (electrons) and $\sim(42.6~\text{MHz/T})^{-1}$ (nuclei), spin dynamics remain frozen on the interaction timescale for fields ranging from milliTesla to Tesla.}, the wavepacket interacts stroboscopically with a well-defined spin state~(see Appendix~\ref{ap:spectroscopy_imaging}).

The final state can be equivalently expressed in terms of the scattering matrix components $S^{(1)}_{\vec{\xi}, s'; \mathrm{in}, s}$ in a given basis, $ \ket{\vec{\xi}} $, where $ \vec{\xi}= \vec{k}$ for the momentum representation or $ \vec{\xi} = \vec{r}$ for the position representation. 
We find
\begin{equation}\label{eq:2delectronspin_state_2}
\begin{split}
    \ket{\psi_{\text{out}}} \approx & \sum_{s'}\int d^3\xi\, S^{(1)}_{\vec{\xi}, s'; \mathrm{in}, s} \ket{\vec{\xi}}\otimes\ket{s'}\,,
\end{split}
\end{equation}
For a fixed bias field longitudinal to the electron propagation axis, the momentum representation 
takes the form
\begin{subequations}\label{eq:S1_momentum_main}
    \begin{align}
S^{(1)}_{\vec{k}', s'; {\rm in}, s} &= \delta_{s',s} \beta_{1}(\vec{k}) +  \sum_{\varsigma = -1}^{1} \beta_{z,\varsigma}(\vec{k}) \langle s'| \osigma_{z,\varsigma} |s \rangle,  \\
\beta_{1}(\vec{k})&= \varphi_{{\rm in}}(\vec{k}),\\
\beta_{z,\varsigma}(\vec{k}) 
    =& -2^{|\varsigma|/2} \frac{r_{\rm e}}{2\pi} 
        \,{\vec{\ell}}(\vec{k};\varsigma) \cdot \ve_{z,-\varsigma},
    \end{align}
\end{subequations}
where ${r}_e=\frac{e\mu_0\mu_{\rm B}}{2\pi \hbar}\approx 2.82\cdot 10^{-15}~\text{m}$ is the classical electron radius, which is proportional to the Bohr magneton, and the unitary vectors $\ve_{z, \varsigma} = (1-|\varsigma|) \ve_{z} + \frac{|\varsigma|}{\sqrt{2}} (\ve_x + i\varsigma \ve_y)$. The operators $ \hat{\sigma}_{z,\varsigma} = \frac{1}{2} (\hat{\sigma}_x \pm i \varsigma \hat{\sigma}_y) $ correspond to spin-flip transitions with respect to the quantization axis $z$.
Each probability amplitude $\beta_j$ {corresponds to a distinct process:} $\beta_{1}$ corresponds to free propagation, $\beta_{z,0}$ to elastic scattering that preserves the energy of the spin system
but deflects the electron due to the Lorentz force, and $\beta_{z,\pm}$ to inelastic scattering, where energy is exchanged between the systems. 
The vector ${\vec{\ell}}(\vec{k};\varsigma)$, defined in Eq.~\eqref{eq:elldef}, includes the momentum transfer integrated over the momentum space representations of the electron wavefunction and the spin position probability density. 
The variation of the $\beta_j$ coefficients across the detection plane reveals how the spin's magnetic field modulates the electron trajectory, with the resulting image contrast depending sensitively on the spin orientation relative to the beam axis. A full derivation and the general case of an arbitrarily oriented bias field are discussed in Appendix~\ref{ap:general_diffraction}.

For the simulations of TEM spin imaging, we focus exclusively on the degrees of freedom of the beam electron. Accordingly, we obtain the electron's density matrix by tracing out the spin degrees of freedom of the sample, which yields
\begin{equation}
     \rho_{{\rm sc,e}\,\vec{\xi};\vec{\xi}'} = \sum_{s'} S^{(1)}_{\vec{\xi}, s'; {\rm in}, s} S^{(1)~*}_{\vec{\xi}', s'; {\rm in}, s}.
\end{equation}
To simulate the recorded detector images in both diffraction and (defocused) image modes, we define the corresponding probability distributions in accordance with their respective acquisition processes.
Since the duration of a single electron wavepacket(in the range of hundreds of attoseconds) is much shorter than the integration time of the detector, the measured probability distributions reflect a cumulative intensity rather than the temporal evolution of individual wavepackets.

In diffraction mode, the detector measures angles\footnote{Neglecting any energy-dependent aberrations induced by the lensing system, we assume a one-to-one correspondence of angles after the sample and positions in the detector plane in diffraction mode.}. We restrict our consideration to the case that the electron energy is not measured and integrate over multiples of the momentum spread of the initial electron wave packet. As we only consider first order contributions to the probability density and the scattered state amplitudes only appear in a product with the initial state, this integral can be formally extended to an integral over the whole domain of absolute momentum $k$. Consequently, the angular probability distribution is expressed as
\begin{equation}\label{eq:prob_diff_def_main}
    P_{\rm diff}(\vec{\vartheta}) := \mathcal{N}_{\rm diff}\int dk\, k^2 \,\rho_{{\rm sc,e}\,\vec{k}(\phi_k,\vartheta);\vec{k}(\phi_k,\vartheta)}\,,
\end{equation}
where, under the small-angle approximation, $\vec{k}(\phi_k,\vartheta)=k (\vec{\vartheta}+\ve_z)$ with $\vec{\vartheta}=\vartheta(\cos\phi_k,\sin\phi_k)$, and $\mathcal{N}_{\rm diff}$ is a normalization factor ensuring that the angular probability distribution integrates to unity. Note that the angular distribution can be translated into a distribution over transverse momenta by the map $\vec{\vartheta} \rightarrow \vec{k}_{0,\perp} = k_0 \vec{\vartheta}$.

In the (defocused) image mode configuration, the spatial intensity distribution at the detector is obtained by averaging over the time domain. We include the finite optical resolution of the transmission electron microscope due to lens aberrations and detector limitations via a transfer function $M(\vec{k}_{\perp, j})$. In our simulations, this is implemented as a spatial-frequency cutoff corresponding to a real-space resolution of approximately $50~\text{pm}$, by filtering out spatial frequencies above $k_{\rm max} = 2\pi(50~\text{pm})^{-1}$. Combining these ingredients, the probability distribution in the image plane with defocus $z_d$ becomes
\begin{equation}\label{eq:prob_bar_main}
    \begin{aligned}
    P_{\rm  img}(\vec{r}_\perp,z_d) :=&\, \mathcal{N}_{\rm img} \int d^3k_{1}\, d^3k_{2}\, 2\pi\hbar\, \delta\!\left(E(\vec{k}_1)-E(\vec{k}_2)\right) \\
    & e^{i(\vec{k}_{1} - \vec{k}_{2})\cdot \vec{r}}\,
    \rho_{{\rm sc,e}\,\vec{k}_1;\vec{k}_2}\,
    M(\vec{k}_{\perp,1})M(\vec{k}_{\perp,2}),
    \end{aligned}
\end{equation}
where $\vec{r}=(\vec{r}_\perp, z_d)$ with $\vec{r}_\perp$ denoting the position in the image detection plane, $\vec{k}_j = (\vec{k}_{\perp,j}, k_{z,j})$ for $j=\{1,2\}$, and $\mathcal{N}_{\rm img}$ is a normalization constant (see details in Appendix~\ref{ap:general_position_mode}).
{The appearance of the delta function $\delta(E(\vec{k}_1)-E(\vec{k}_2))$ implies that interference is restricted to 
electron wave packet components of equivalent kinetic energy $E(\vec{k}) = \sqrt{(m_{\rm e} c^2)^2 + (\hbar c |\vec{k}|)^2} = \gamma(|\hbar \vec{k}|) m_{\rm e}c^2$. }

Both probability distributions {$P_{\rm diff}$ and $P_{\rm img}$} are evaluated within the paraxial approximation for the incident and scattered electron states, where the longitudinal momentum component dominates over the transverse components across most of the wavepacket. To first order, {the image mode distribution can be expressed as}
\begin{equation}\label{eq:Prob_Pis_def}
    \begin{aligned}
&P_{\rm img}(\vec{r})\approx\tilde\Pi_0(\vec{r})+\Re(\tilde\Pi_{z, +1}(\vec{r})+\tilde\Pi_{z, -1}(\vec{r}))\expval{\osigma_x}\\
&-\Im(\tilde\Pi_{z, +1}(\vec{r})-\tilde\Pi_{z, -1}(\vec{r}))\expval{\osigma_y} +
2\Re(\tilde\Pi_{z, 0}(\vec{r}))\expval{\osigma_z}\,,
    \end{aligned}
\end{equation}
where $\expval{\hat{\sigma}_j}$ denotes the expectation value of the Pauli operator $\hat{\sigma}_j$ for the spin at time $t_0$. 
{Equivalently, $P_{\rm diff}(\vec{\vartheta})$ can be expressed in terms of the functions $\Pi_l(\vartheta)$ for $l\in\{1,(z,\varsigma)\}$ replacing the $\tilde\Pi_l$ in Eq.~\eqref{eq:Prob_Pis_def}.}
The functions ${\Pi}_l(\vec{\vartheta})$ and $\tilde{\Pi}_l(\vec{r})$ are defined in Eqs.~\eqref{eq:Pis_momentum_2level} and~\eqref{eq:pi_pos_def}, respectively.
{Terms of second order in the interaction strength} have been neglected in the Eq.~\eqref{eq:Prob_Pis_def}. 
Although second-order interactions involving angular momentum exchange are theoretically possible, particularly when the spin aligns with the beam axis~\cite{metrologySpinTEM}, we restrict our analysis to the dominant first-order linear coupling, as higher-order corrections are orders of magnitude weaker.
While this approximation could potentially lead to negative probability values at large transverse momenta in the tails of the incident Gaussian wavepackets, such artifacts are negligible in practice. For single-spin imaging in diffraction mode, these deviations are at most $10^{15}$ times smaller than the signal at the center of the zero-deflection peak.
Furthermore, the Gaussian {model for the transverse shape of the electron wave packet }becomes less accurate in the tails, where a polynomial decay is more physically realistic. 

From the expansion in Eq.~\eqref{eq:S1_momentum_main}, it follows that the first-order contributions to the probability distributions arise from the product of the first-order coefficients $\beta_{z,\varsigma}$ with the zeroth-order term $\beta_1$. 
Accordingly, the dominant signatures of the free-electron--spin interaction originate from the interference between the coherent amplitudes of the unscattered (non-deflected) and scattered components of the electron wavefunction. 
{The energy integration in Eq.~\eqref{eq:prob_diff_def_main} as well as the energy-conservation condition in Eq.~\eqref{eq:prob_bar_main} lead to the appearance of the overlap integral of the longitudinal parts of the incident and the scattered electron wave packet in $P_{\rm diff}$ and $P_{\rm img}$, respectively, which determines the visibility of the holographic interference pattern. Aiming for maximal visibility then puts constraints on the longitudinal momentum spread. 
In Sec.~\ref{sec:hol-imaging}, we explicitly derive these constraints.}

\section{Probability Amplitudes of the Free-electron--spin state}\label{sec:scattering}

In this section, we analyze the amplitudes initially in momentum and then in position space.
We model the incident electron beam as a Gaussian wavepacket given by  
$\ket{\varphi_{\rm in}} = \int d^3k \, \varphi_{{\rm in}, \perp} (\vec{k}_\perp) \varphi_{{\rm in}, \|}(k_z) \ket{k_\perp, k_z}$,  
where $\varphi_{\mathrm{in}, \|}(k_z)$ and $\varphi_{\mathrm{in}, \perp}(\vec{k}_\perp)$ are respectively the longitudinal and transversal Gaussian distributions characterized by a mean momentum $k_{z,0} \ve_z$, with $k_{z,0} = 2.5\times 10^{12}~\rm{m}^{-1}$ corresponding to a kinetic energy of $E_{\rm kin} = 200$~keV. 
{The wavepacket $\varphi_{\mathrm{in}, \|}(k_z)$ has the longitudinal standard deviation $\Delta k_z$, while $\varphi_{\mathrm{in}, \perp}(k_\perp)$ has a transversal one $\Delta k_\perp$. The latter determines the beam focusing at the spin position, corresponding to a spatial spread of $\Delta r_\perp = (2\Delta k_\perp)^{-1}\approx\text{FWHM}/2.36$.}

In this work, we restrict our analysis to azimuthally symmetric electron wavepackets, which depend only on the longitudinal momentum component $k_z$ and the magnitude of the transverse momentum $k_\perp$, i.e., $\varphi_{{\rm in}, \perp}(\vec{k}_\perp) = \varphi_{{\rm in}, \perp}(k_\perp)$. 
As a result, the free-propagation amplitude takes the form $\beta_{1}(\vec{k}) = \varphi_{{\rm in}, \perp}({k}_\perp)\,{\varphi}_{{\rm in}, \|}(k_z)$. 
The elastic scattering amplitude $\beta_{z,0}(\vec{k})$, which corresponds to the deflection induced by a spin aligned with the propagation axis, vanishes due to the symmetry considerations. 
Therefore, the dominant contribution to the scattered amplitude, and consequently to the measurable imaging signal, originates from the inelastic components of the interaction.

Specifically, the inelastic scattering amplitudes take the form
$\beta_{z,\varsigma}(\vec{k}) = i\varsigma \sqrt{\tfrac{2}{\pi}}\,r_{\rm e}\, e^{-i\varsigma\phi_k}\, \ell(k_\perp, k_z; \varsigma)$, 
where $\ell(k_\perp, k_z; \varsigma)$ is defined explicitly in Eq.~\eqref{eq:ell_vect_magnitude_general}.
The phase factor $e^{-i\varsigma\phi_k}$ jointly with the form of scattering matrix elements in Eq.~\eqref{eq:S1_momentum_main} reflects angular momentum conservation between the electron and the spin. 
A spin flip from anti-aligned to aligned (or vice versa) with respect to the bias field transfers one unit of angular momentum to the electron, thereby generating a well-defined orbital angular momentum (OAM) in the scattered wave. 
In general, the scattering process produces a coherent superposition of opposite OAM components, effectively resulting in the formation of a Hermite-Gauss beam.
The generation and manipulation of such vortex beams have been experimentally demonstrated~\cite{Uchida2010, Verbeeck2010Production, McMorran2011, Schattschneider_2012Vortex}, 
and their OAM spectra have been successfully characterized using advanced measurement techniques~\cite{Guzzinati2014, Grillo2017, Schattschneider2014Imaging}. 
Finally, the inelastic amplitudes also dominate the electron-induced backaction on the spin system, as discussed in Appendix~\ref{ap:backaction}.
\begin{figure}[h]
    \centering
    \includegraphics[width=1\linewidth]{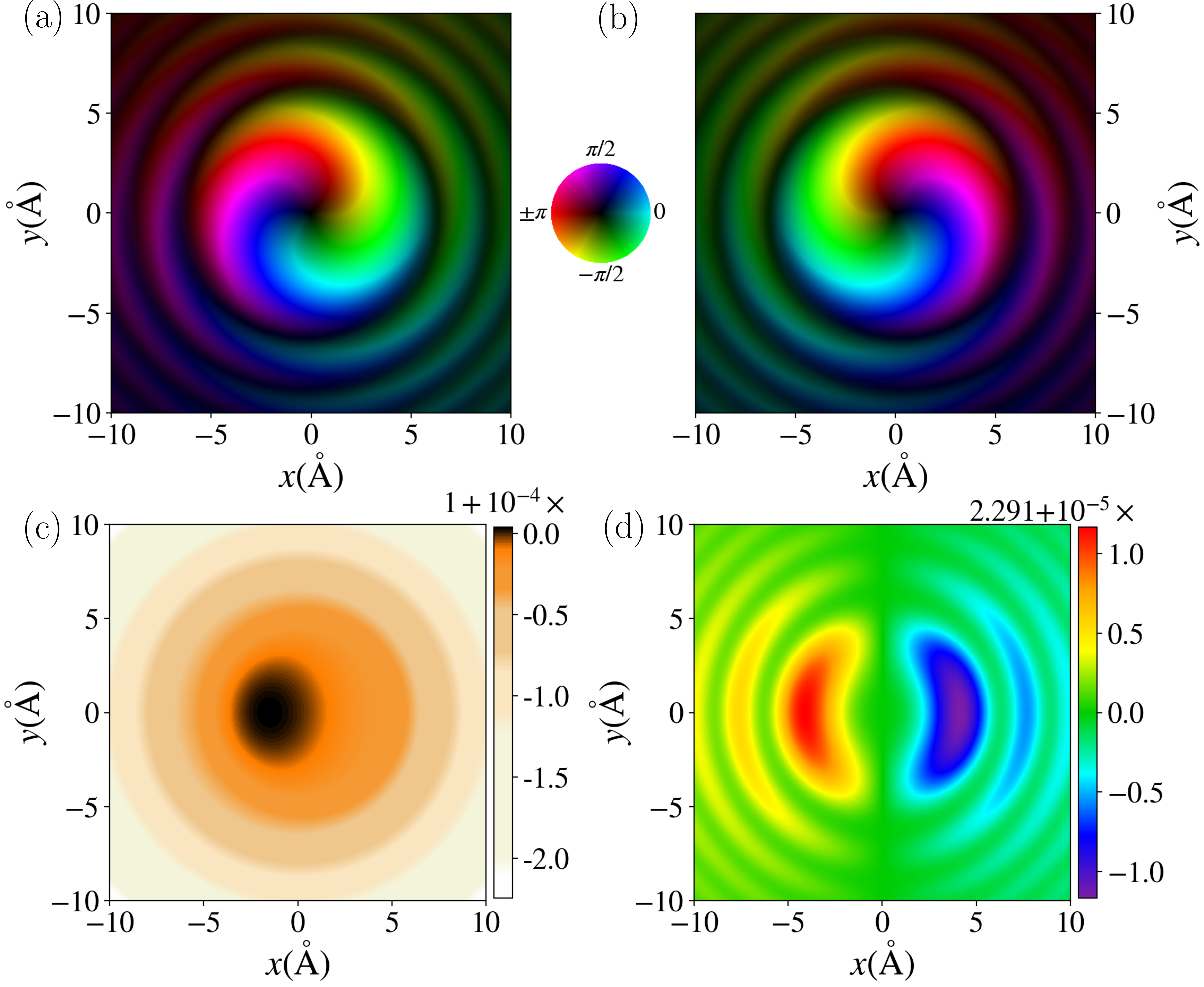}
    \caption{
    Inelastic scattering amplitudes in position representation: (a) $\tilde\beta_{z,+}(\vec{r})$ and (b) $\tilde\beta_{z,-}(\vec{r})$ plotted over the plane transverse to the beam axis for a fixed value of the distance to the sample plane $z = 1.7~\Delta r_\perp=800~\text{\AA}$. The amplitudes correspond to vortex beam components with orbital angular momenta of $-\hbar$ and $+\hbar$ arising from the spin-flip transitions ${\osigma}_{z,+}$ and ${\osigma}_{z,-}$.
    The complex-valued fields are visualized using a standard color-hue convention: the hue represents the local phase, while the brightness scales with the absolute magnitude. These magnitudes are presented in arbitrary units and jointly normalized to a global maximum value of 1.
    When the spin state aligned along $\ve_y$ is conserved, the post-interaction electron state remains pure. Panels (c) and (d) show the amplitude (normalized to unity at the origin in arbitrary units) and phase in radians of the corresponding electron wavefunction represented in the plane transverse to the beam axis at $z$ (see Appendix~\ref{app:coherent_wavefunction}). The simulation assumes an incident Gaussian beam with a FWHM of $0.11~\text{\textmu}$m ($\Delta k_\perp=4.22\times 10^{-6} k_{z,0}$ for $E_{\rm kin} = 200$~keV). The images depict the beam center, where the strongest spin-electron interaction signatures appear. The amplitude shows a smooth parabolic radial profile with an asymmetric modulation due to the magnetic dipole field of the spin, while the phase reflects the Aharonov-Bohm--type shift accumulated along distinct semiclassical paths~\cite{Haslinger_2024}. Interference features in both amplitude and phase arise from the representation of the wavefunction at non-zero $z$.}
    \label{fig:coherent_wavefunction}
\end{figure}

The emergence of vortex beams becomes clearer in the position representation of the scattering matrix elements, $S^{(1)}_{\vec{r}, s'; {\rm in}, s} = \delta_{s',s} \tilde\beta_1(\vec{r}) +  \sum_{\varsigma = -1}^{1} \tilde\beta_{z,\varsigma}(\vec{r}) \langle s'| \osigma_{z,\varsigma} |s \rangle$ with the amplitudes $
\tilde\beta_j(\vec{r}):=\int d^3k\,e^{i\vec{k}\cdot \vec{r}}\beta_j(\vec{k})$
with $j=\{1, (z, \varsigma)\}$. 
Indeed, these amplitudes are eigenfunctions of the orbital angular momentum operator: $\hat{L}_z \tilde\beta_{z,\varsigma}(\vec{r})
    = -\varsigma \hbar\,\tilde\beta_{z,\varsigma}(\vec{r})$. 
This confirms that a change in spin angular momentum is compensated by an opposite change in orbital angular momentum.
As illustrated in Figs.~\ref{fig:coherent_wavefunction}~a and~\ref{fig:coherent_wavefunction}~b, the inelastic scattering amplitudes $\tilde\beta_{z,+}(\vec{r})$ and $\tilde\beta_{z,-}(\vec{r})$, exhibit the characteristic vortex-like structure expected from the transfer of angular momentum accompanying a spin-flip process. These plots have been generated based on the simplified expression given in Appendix~\ref{ap:general_position_mode} Eqs. \eqref{eq:betanvar_pos_paraxial_approx} and \eqref{eq:tildeL_nvar} within the paraxial approximation valid under $\Delta k_\perp < \Delta k_z \ll k_{z,0}$ and under the conditions $k_{z,0} \gg k_a=(a_0)^{-1}$ and $\sqrt{k_a k_{z,0}} \gtrsim \delta k$, with $\delta k= \sqrt{\gamma(\hbar k_{z,0}) m_{\rm e} \omega_0 \hbar}$ representing the momentum transfer associated with the Zeeman energy $\hbar\omega_0= 2 \mu_{\rm B} |\vec{B}_0|$.

From the $\tilde\beta_j(\vec{r})$ amplitudes, we {can construct and visualize} the coherent component of the electron wavefunction {defined as the scattered state of the electron under the condition that the initial state of the spin oriented in the $+y$-direction and does not change in the interaction process.} 
After propagation {to a plane located a distance $z$ from the interaction region,} the spin-induced deflection manifests as measurable spatial displacements in the coherent electron amplitude, as shown in Fig.~\ref{fig:coherent_wavefunction}~c (again based on Eqs. \eqref{eq:betanvar_pos_paraxial_approx} and \eqref{eq:tildeL_nvar}). 
Moreover, the spin interaction imprints an Aharonov-Bohm-type phase shift across the electron wavefront, with opposite signs on either side of the spin (see Fig.~\ref{fig:coherent_wavefunction}~d). 
Importantly, this phase originates from the inelastic scattering coefficients and constitutes the dominant imaging contrast mechanism captured in our simulations.

\section{Holographic Imaging}
\label{sec:hol-imaging}

In Appendices~\ref{ap:general_diffraction} and~\ref{ap:general_position_mode}, we derive the expressions for the diffraction and image modes by applying successive approximations within the paraxial regime and assuming complete overlap of the longitudinal components of the electron wavefunction. These approximations allow us to isolate the effective first-order contributions to the observable probability distributions under experimentally relevant conditions.

The paraxial regime imposes the constraint $\Delta k_\perp,\, \delta k \ll k_{z,0}$, implying that the transverse momentum spread and the characteristic interaction momentum transfer are both much smaller than the mean longitudinal momentum. Physically, this ensures that the scattering angles remain small and that the main propagation direction is preserved in both the scattered and unscattered components of the electron wavefunction. In addition, complete energy overlap between these components imposes the conditions $\Delta k_z > \Delta k_\perp^2/k_{z,0}$ and $\Delta k_z > \delta k^2/k_{z,0}$ for electron probes larger than the scatterer, establishing an upper bound for the coherence length.
If these inequalities are not strictly met, interference visibility is reduced. However, for FWHM~$=1.1$~nm probes, we find the longitudinal overlap remains above $0.997$, rendering the contrast reduction ($<0.3\%$) negligible.

Under these approximations, the probability distribution terms take the effective forms
\begin{subequations}\label{eq:Pis_diffraction_main}
    \begin{align}
{\Pi}_0(\vec{\vartheta}) &\approx\,  k_{z,0}^2\, 
|\varphi_{{\rm in},\perp}(k_{z,0}\vartheta)|^2, \\[4pt]
\begin{split}
{\Pi}_{z,\varsigma}(\vec{\vartheta}) 
&\approx -\frac{2^{|\varsigma|/2}}{\pi}\, r_{\rm e}\, \frac{k_{z,0}^2}{\Delta k_\perp}
\vu^*(\phi_k)\!\cdot\! \ve_{z,-\varsigma}\,
\cL_{\varsigma}(\vartheta;0)\,
f(k^2_{z,0}\vartheta^2),
\end{split}\\[4pt]
\begin{split}
\cL_{\varsigma}(\vartheta;0)
&\approx \int_0^\infty \frac{dq_\perp}{2\Delta k_\perp}\,
f(k_{z,0}^2\vartheta^2 + q_\perp^2)
I_1\!\left(\frac{\vartheta q_\perp k_{z,0}}{2\Delta k_\perp^2}\right)
\mathcal{I}_{\rm es}(q_\perp),\label{eq:Lvarsigma_main}
\end{split}
    \end{align}
\end{subequations}
where we define $\vec{\mathrm{u}}^*(\phi) := (-\sin\phi,\, \cos\phi,\, 0)$, $\vec{q}_\perp=q_\perp(\cos(\phi_q), \sin(\phi_q))$ represents the transverse momentum transfer, $f(k):=\exp\left[-k/(4\Delta k_\perp^2)\right]$, and $\mathcal{I}_{\rm es}(k) = (1 + a_0^2 k^2 / 4)^{-2}$ is the Fourier transform of the spin probability density.

We identify ${\Pi}_0(\vec{\vartheta})$ as the transverse angular distribution of the incident wavefunction, while ${\Pi}_{z,\varsigma}(\vec{\vartheta})$ represents a first-order correction that introduces an azimuthal dependence through the factor $\vu^*(\phi_k) \cdot \ve_{z,-\varsigma}$. This term captures the spin-dependent deflection of the electron beam, consistent with the direction of the magnetic field generated by the localized spin. As expected, for $\varsigma = 0$, this correction vanishes, leading to independence of the total probability distribution on $\expval{\hat{\sigma}_z}$. The function $\cL_{\varsigma}(\vartheta;0)$ describes how the deflection magnitude varies with the scattering angle~$\vartheta$.
When the deflection due to transverse momentum components is negligible, the dependence of the scattering amplitude on the spin-exchange index~$\varsigma$, and thus on the energy exchange between the electron and the spin, effectively disappears. This indicates that the specific energy transfer just affects the dependence on $\phi_k$ in the probability when the interaction due to the longitudinal momentum of the incident beam is dominant.
In Appendix~\ref{ap:general_diffraction} we also discuss the case of an initially defocused beam, characterized by a longitudinal shift~$z_p$, which is relevant for in-line holography~\cite{Lichte_2008}. In this case, the diffraction patterns may exhibit additional features arising from inelastic processes that slightly modify the azimuthal deflection. A detailed quantitative analysis of this effect will be presented in future work.

In an analogous manner, the image-mode probability distribution can be expressed under the same approximations discussed above, with the additional requirement that $k_a \ll k_{z,0}$. Physically, this condition ensures that the maximum possible momentum transfer, which is set by the inverse of the spin smearing length scale, remains much smaller than the mean longitudinal momentum, thereby preserving the validity of the paraxial approximation. As in the diffraction case, the temporal integration of the probability enforces energy overlap between the scattered and unscattered components of the electron state, leading to similar longitudinal coherence conditions (see Appendix~\ref{ap:general_position_mode}).

To exploit the symmetry of the problem, we assume an azimuthally symmetric mask function $M(\vec{k}_\perp) = M(k_\perp)$. Consequently, the effective expressions for the probability distribution terms become
\begin{subequations}\label{eq:Pis_position_main}
\begin{align} 
&\tilde\Pi_0(\vec{r}) \approx \tilde{\mathcal{N}}_{\rm img} \left| \mathcal{B}_1(\vec{r})\right|^2\label{eq:Pi0_pos_main}\\ &\mathcal{B}_1(\vec{r})=2\pi\int {dk_{\perp}\;k_\perp} J_0({{k}_{\perp}{r}_\perp})\, e^{-i\frac{k_{\perp}^2}{2k_{z,0}}z_d}\varphi_{{\rm in},\perp}({k}_{\perp})M({k}_{\perp}). \label{eq:B1_pos_main}\\ 
&\tilde{\Pi}_{z, \varsigma}(\vec{r})=\tilde{\mathcal{N}}_{\rm img} \mathcal{B}^*_1(\vec{r}_\perp) \mathcal{B}_{z,\varsigma}(\vec{r}),\\ 
\begin{split} 
&\mathcal{B}_{z,\varsigma}(\vec{r})=-2i \sqrt{{2}{\pi}}2^{|\varsigma|/2}\;r_{\rm e} \vu^*(\phi_r)\!\cdot\!\ve_{z,-\varsigma}\\ &\quad\int {dk_{\perp}\;k_{\perp}}e^{-i\frac{k_{\perp}^{2}}{2k_{z,0}}z_d}J_1({{k}_{\perp}{r}_\perp})\; \cL_{\varsigma}\left(\tfrac{k_{\perp}}{k_{z,0}}; 0\right)\,M({k}_{\perp}), 
\end{split} 
\end{align} 
\end{subequations}
where $\vec{r}_\perp = r_\perp(\cos\phi_r, \sin\phi_r)$ defines the position in the detection plane. The normalization factor is given by {$\tilde{\mathcal{N}}_{\rm img} = 2\pi\mathcal{N}_{\rm img}/v_0= (2\pi)^{-2}$}
, with $v_0 = \hbar c^2 k_{z,0} / E(k_{z,0})$ representing the mean propagation velocity of the electron wavepacket.
The term $\mathcal{B}_1(\vec{r})$ represents the propagated form of the incident wavefunction to a plane located at a distance $z_d$ from the interaction region. In contrast, $\mathcal{B}_{z,\varsigma}(\vec{r})$ corresponds to the effective scattered component contributing to the image-mode probability distribution. Similar to the diffraction mode, this term exhibits an explicit azimuthal dependence through the factor $\vu^*(\phi_r)\cdot\ve_{z,-\varsigma}$, which suppresses any contribution from spin components parallel to the electron beam axis, resulting in $\mathcal{B}_{z,0}(\vec{r})=0$.
The radial dependence, on the other hand, is governed by the Hankel transform of the product of the momentum-dependent propagation phase, the effective deflection amplitude $\cL_{\varsigma}(k_{\perp}/k_{z,0}; 0)$ (Eq.~\eqref{eq:Lvarsigma_main}), and the mask function $M(k_\perp)$.
Furthermore, in the idealized limit of aberration-free imaging [$M(k_\perp) = 1$], the amplitudes $\mathcal{B}_1(\vec{r})$ and $\mathcal{B}_{z, \varsigma}(\vec{r})$ simplify to the analytical forms given in Eqs.~\eqref{eq:B1pos_defocus_infinity} and~\eqref{eq:Bnpos_defocus_infinity}, respectively.

Building upon the probability distributions derived in both position and angular space, we now analyze the fundamental bound of the measurable magnetic signal. Since the signal amplitude in Eqs.~\eqref{eq:Pis_diffraction_main} and~\eqref{eq:Pis_position_main} arises from the first-order probability terms $\Pi_{z,\varsigma}(\vec{\vartheta})$ and $\tilde{\Pi}_{z,\varsigma}(\vec{r})$, which are each proportional to the classical electron radius $r_{\rm e}$ and therefore to the Bohr magneton $\mu_{\rm B}$, the achievable signal strength directly reflects the sensitivity to the spin magnetic moment.
To quantify the maximum amount of retrievable information from an image, we establish an upper bound on the SNR by connecting it to the precision in estimating the Bohr magneton with (unbiased) estimators:
\begin{equation}\label{eq:SNR_bound}
    \text{SNR} = \frac{\mu_{\rm B}}{\Delta \mu_{\rm B}}
    \leq \mu_{\rm B}\sqrt{N_{\rm e}\,\mathrm{CFI}},
\end{equation}
where $N_{\rm e}$ denotes the number of electrons interacting with the spin,
and $\mathrm{CFI}$ is the classical Fisher information, which quantifies the information content of the recorded intensity distribution (see Eq.~\eqref{eq:cfi_first_order}). This expression follows directly from the Cram\'er-Rao bound for single-parameter estimation~\cite{cramer1946,rao1945}, emphasizing that the attainable SNR is ultimately limited by both the number of detected electrons and the magnetic interaction strength embodied by $\mu_{\rm B}$.

The spin-imaging through the electron intensity distribution described above can be combined with SRS, which operates
by coherently driving a spin system using an oscillatory magnetic field, typically in the MW regime, while sweeping the drive frequency $\omega$ near the system's 
Larmor frequency $\omega_0$. The detuning, $\delta = \omega - \omega_0$, characterizes the effective drive and determines the spin state immediately after excitation, as detailed in Appendix~\ref{ap:spectroscopy_imaging}. In the electron microscopic setup, after excitation, the spin evolves freely before interacting with the probe electron, imprinting distinct spin-dependent signatures in the spatial distribution of the scattered electron.

\begin{figure*}[t]
    \centering
    \includegraphics[width=0.9\linewidth]{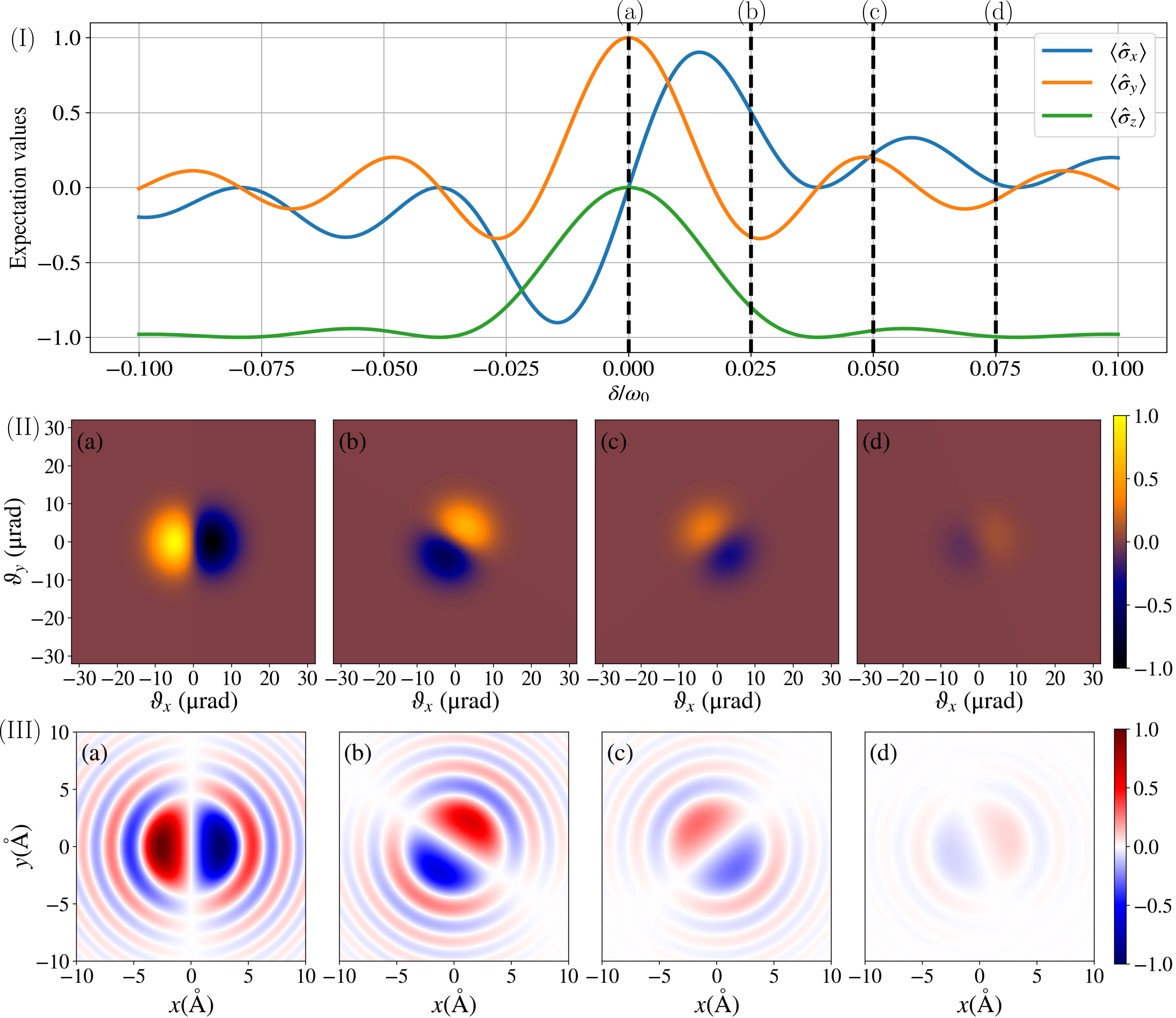}
    \caption{(I) Expectation values of the Pauli operators for a spin initially anti-aligned with the bias magnetic field and driven by a $\pi/2$ MW pulse, shown as a function of detuning $\delta/\omega_0$ for a fixed Rabi frequency $\omega_1 = 0.01\omega_0$. Markers indicate detuning values corresponding to the spin states used for imaging. (II) Differential probability density in angular space with $\vec{\vartheta}=(\vartheta_x, \vartheta_y)$, and (III) in position space at a defocus of $z_d = 1.7~\Delta r_\perp=800~\text{\AA}$, illustrating the spin-dependent interaction with a free electron beam [FWHM~$=0.11~\text{\textmu}\mathrm{m}$ ($\Delta k_\perp=4.22\times 10^{-6} k_{z,0}$)] following the same $\pi/2$ MW pulse. Results are shown for detunings (a) zero detuning, (b)~$\delta = 0.025~\omega_0$, (c)~$\delta = 0.050~\omega_0$, and (d)~$\delta = 0.075~\omega_0$. The color scale represents the normalized intensity variation relative to a far off-resonant excitation ($\delta_0 = 10\omega_0$), corresponding to the spin in its ground state (anti-aligned with the bias field), and is normalized to unity for maximum variation in the on-resonance case.
    }
    \label{fig:spectroscopy_imaging_TEM}
\end{figure*}

By scanning over detuning values, as illustrated in Fig.~\ref{fig:spectroscopy_imaging_TEM}~I, the expectation values of the spin components exhibit a resonance-like dependence on $\delta$, with a maximum transverse component at $\delta = 0$. The resonance width is determined by the ratio between the Rabi frequency $\omega_1 = 2 \mu_{\rm B} B_1/\hbar$ and $\omega_0$. At resonance, the spin is driven to the equator of the Bloch sphere, oriented along the $+y$ axis, where $\expval{\osigma_z} = 0$.
Here, we assume that all decoherence channels on the spin are slower than the driving by the MW pulse (see Section \ref{sec:experimental_feasibility}). 
In addition, we consider the spin orientation immediately after excitation, so that the images correspond to a snapshot of the spin precession as seen by the electron while traversing the spin sample.
As the absolute value of the detuning increases, the rotation of the spin state is decreased and the spin vector $\hat{\boldsymbol{\sigma}}$ becomes increasingly anti-aligned with the $z$-axis, the direction of both the bias magnetic field and the electron beam propagation.

To formalize the differential imaging approach, we calculate the differential probability density by subtracting a reference distribution $P_0$ from the driven distribution $P_{\rm dr}$. The driven term $P_{\rm dr}$ incorporates the spin initialization induced by the MW pulse using the expectation values corresponding to the specific detuning in Figure~\ref{fig:spectroscopy_imaging_TEM}~I. The reference distribution $P_0$ corresponds to a signal acquired after relaxation (or equivalently with no drive pulse or far off resonance) representing the unperturbed state where $\langle \hat{\sigma}_x \rangle = \langle \hat{\sigma}_y \rangle = 0$. 
\subsection{Example in diffraction mode}\label{sec:diffraction}

The angular probability distribution depends on scattering amplitudes associated with inelastic transitions. For spins oriented transversely to the bias magnetic field, these transitions yield non-zero first-order contributions to the angular-space probability density, as described by Eq.~\eqref{eq:Pis_diffraction_main}.
As the beam-transverse component of the spin vector gradually decreases with increasing detuning of the MW pulse, the same is true for the spin-induced features in the differential diffraction mode images as revealed by the frequency sweep in Fig.~\ref{fig:spectroscopy_imaging_TEM}~II. We calculate this differential probability density ($P_{\rm dr}-P_0$) by evaluating the angular probability $P_{\rm diff}$ (Eq.~\eqref{eq:Prob_Pis_def}) using the explicit components from Eq.~\eqref{eq:Pis_diffraction_main} for both the driven and reference states. The analysis is restricted to a detection region of $\vartheta \lesssim 8\Delta k_\perp/k_{z,0}$; this range captures most of the interacting electrons and ensures the probability distributions remain well-defined by excluding regions with negative values\footnote{These are a result of neglecting second order terms in the interaction strength, but are negligible due to their very small absolute value.}.

To rigorously evaluate whether the signal is sufficient to overcome shot noise even at a high dose of $N_{\rm e}= 10^{10}$ interacting electrons (achievable with 1~s exposure at 1.6~nA), we employ the Cram\'er-Rao bound to quantify the maximum achievable sensitivity.
For a single spin on resonance, our analysis of detection size and resolution confirms that the diffraction patterns converge to the diffraction-mode CFI limit established in~\cite{metrologySpinTEM} for a hydrogen 1s orbital spin. This convergence occurs provided the detector possesses sufficient angular resolution to resolve the beam spread ($\sim\Delta k_\perp/k_{z,0}$) and sufficient range to capture large deflection angles ($\sim k_a/k_{z,0}$), which is equivalent to resolving the spin's spatial delocalization.
Quantitatively, for the broad Gaussian beam (FWHM $= 110$~nm; $\Delta k_\perp=4.22\times 10^{-6} k_{z,0}$) shown in Fig.~\ref{fig:spectroscopy_imaging_TEM}~II~a, we find $\mu_{\rm B}^2 \mathrm{CFI} \sim 9.4 \times 10^{-15}$, and an SNR upper bound of $\sim9.7 \times 10^{-3}~\text{s}^{-1/2}\sqrt{t_{\rm acq}}$ at 1.6~nA, or equivalently, a necessary electron dose to achieve unity SNR of about $8\times10^{7}\rm{e}^-/\text{\AA}^2$.  
Employing a localized beam (FWHM~$=1.1$~nm) substantially increases the CFI to $\mu_{\rm B}^2\mathrm{CFI} \sim 8.8\times 10^{-11}$ and the SNR bound to $\sim 0.94~\text{s}^{-1/2}\sqrt{t_{\rm acq}}$ (see Appendix~\ref{ap:optimal_single_spin_diffraction}), and leads to a reduction of the electron dose required for unity SNR by a factor of three. Since the detection range defined by $\vartheta \lesssim 8\Delta k_\perp/k_{z,0}\sim 3\times 10^{-3}$ is almost covering the angular region defined by the spin delocalization scale $k_a/k_{z,0}=(a_0 k_{z,0})^{-1}\sim 10^{-2}$ the CFI is closely approaching the theoretical limit of $1.37\times 10^{-10}$~\cite{metrologySpinTEM} for $\Delta r_\perp=8.92 a_0$ (equivalent to FWHM~$=1.1$~nm). 

With the aim of approaching the theoretical sensitivity limit, we apply a pixelation method that strategically trades angular resolution for enhanced SNR (detailed in Appendix~\ref{ap:optimal_single_spins}). This approach utilizes an optimized masking technique that selectively retains pixels contributing the most significant signal while suppressing those dominated by noise. Specifically, we evaluate the local signal-to-noise ratio for each pixel and retain only those exceeding an optimized threshold. The resulting unmasked pixels are then used in the parameter estimation process, ensuring that only the most informative regions contribute. The unbiased estimator extracts the Bohr magneton $\mu_{\rm B}$ by leveraging the linear relationship between the integrated signal in the unmasked region and $\mu_{\rm B}$. This optimization maximizes information retrieval, achieving an experimental SNR of approximately $1/3$ of the theoretical upper bound (see details in App.~\ref{ap:optimal_single_spin_diffraction}).

\subsection{Example in image mode}\label{sec:image_mode}

Following the analysis of diffraction-mode sensing within the SRS framework, we now consider the complementary case of defocused image-mode detection. For the 
corresponding plots in Figure~\ref{fig:spectroscopy_imaging_TEM}~III, we evaluate the image-mode probability $P_{\rm img}$ (Eq.~\eqref{eq:Prob_Pis_def}) via Eq.~\eqref{eq:Pis_position_main} after applying the mask function $M(k_\perp)=\Theta(k_\perp-k_{\rm max})$. We set the defocus to $z_d = 800~\text{\AA}$ and analyze the probability distribution within a square detection region $X=[-x_{\rm max},x_{\rm max}]^2$.

The interaction between the electron beam and the localized spin imparts momentum transfer and phase shifts as shown in Fig.~\ref{fig:coherent_wavefunction} d. 
Upon free propagation to the defocused plane ($z_d>0$), the resulting phase and amplitude deformations of the scattered wave packet in superposition with the un-deflected wave packet lead to interference fringes.
As in diffraction mode, this signal is most pronounced when the spin lies in the equatorial plane of the Bloch sphere, that is, on resonance and diminishes as detuning increases (Figs.~\ref{fig:spectroscopy_imaging_TEM}~III~b-d).
This implies that, in principle, the resonance condition can be identified through a frequency sweep and image analysis.

The scale of the detection region $x_{\rm max}$ is chosen to cover the portion of the electron wavefunction containing spin-dependent features (see Fig.~\ref{fig:coherent_wavefunction}). A systematic analysis of defocus and detection size (up to $z_d = 10~\Delta r_\perp \approx 4.7\times10^{3}$~\AA{} and $x_{\rm max}=50$~\AA; see Fig.~\ref{fig:CFI_defocus_series}~a) demonstrates that increasing defocus enhances the SNR upper bound. At the same time, free-space propagation broadens the spatial features, and capturing this dispersed information requires a corresponding expansion of the detection region.
For our simulations using an incident Gaussian beam of FWHM $= 110$~nm, we fix the defocus at $z_d = 1.7\Delta r_\perp$ ($800$~\AA). As shown in Fig.~\ref{fig:CFI_defocus_series}~a, the CFI saturates at $x_{\rm max}=10$~\AA, indicating that this window effectively captures the relevant spatial features. 

Under these conditions (on-resonance, Fig.~\ref{fig:spectroscopy_imaging_TEM}~III~a), we obtain $\mu_{\rm B}^2 \, \mathrm{CFI} \sim 4.6 \times 10^{-15}$, corresponding to an SNR upper bound of $\sim 6.8 \times 10^{-3}\,\mathrm{s}^{-1/2}\sqrt{t_{\rm acq}}$ at 1.6~nA. Tightening the beam focus to FWHM $= 1.1$~nm, while keeping other parameters fixed, yields a substantially higher SNR upper bound of $\sim 0.67\,\mathrm{s}^{-1/2}\sqrt{t_{\rm acq}}$. This configuration not only outperforms the broader beam case but also reduces the required electron dose for unity SNR by a factor of three to $5.1\times 10^{7}\, \rm{e}^-/\text{\AA}^2$. Notably, however, even this improved performance is surpassed by that of diffraction mode for equivalent beam parameters.

The impact of significant defocus on the 
fundamental information content, as quantified by the Quantum Fisher Information (QFI), is addressed in a parallel investigation~\cite{metrologySpinTEM}. As established therein, sufficiently increasing both defocus and detection size allows the system to approach the diffraction-mode CFI limit, provided the setup possesses adequate resolution to resolve the smallest spatial scale ($a_0$) while simultaneously covering the beam spread ($\Delta r_\perp$).

In Appendix~\ref{ap:CFI_optimal}, we extend the analysis of SNR limitations to Zernike phase contrast imaging~\cite{ZERNIKE1942686}, which enhances contrast by imparting a $\pi/2$ phase shift to the unscattered amplitude. We find that within a restricted detection region of $x_{\rm max}=10$~\AA, the Zernike configuration yields an SNR upper bound of $\sim1.4\times 10^{-2}\,\mathrm{s}^{-1/2}\sqrt{t_{\rm acq}}$. This value is at least twice that of any defocused imaging mode for the defocus range that we investigated ($z_d = 800$~\AA) and approximately $1.4$ times higher than that of the diffraction mode configuration discussed previously. Nevertheless, as illustrated in Figure~\ref{fig:CFI_defocus_series}~b, this relative advantage diminishes as the defocus increases.  Notably, the Zernike CFI peaks at zero defocus, where it exceeds that of any defocused configuration.
Ideally, with sufficient resolution, the limit of Zernike mode converges to the fundamental diffraction-mode limit for infinite screen size (cf. Figure~\ref{fig:asympthotic_cfi}~a). 
In the narrow beam case (FWHM $= 1.1$~nm) with $x_{\rm max}=10$~\AA, Zernike imaging reaches this limit, resulting in an SNR upper bound of $\sim1.2\,\mathrm{s}^{-1/2}\sqrt{t_{\rm acq}}$ and reducing the electron dose required for unity SNR to $1.7\times 10^{7}~\rm{e}^-/\text{\AA}^2$. 
Despite these theoretical gains, implementing Zernike imaging for spin detection is constrained by the angular proximity of the signal to the zero-deflection peak; isolating the unscattered reference without perturbing the overlapping low-frequency information represents a major technical challenge.

We adapted the analysis framework previously described for diffraction mode to the defocused imaging regime (details in Appendix~\ref{app:position_space_imaging}). A critical distinction lies in the masking strategy: whereas the diffraction mask is typically offset by $\approx 2\Delta k_\perp/k_{z,0}$ to mitigate the dominant shot noise of the zero-deflection peak, the optimal mask for image mode is centered at the origin, where the dominant SNR contribution resides. Consequently, while the mask effectively captures the central feature, it excludes the widely separated interference fringes, which contribute marginally to the total information content. This spatial masking results in an achievable SNR that falls below the theoretical Cram\'er-Rao bound. Specifically, for broad beams, the method captures the majority of available information, coming within a factor of $1.6$ of the limit. Although this discrepancy increases to a factor of $\sim 7$ for narrow beams (FWHM~$=1.1$~nm), where the selectivity of the current masking criterion underperforms due to the extreme signal localization. While advanced segmentation criteria could potentially recover the signal currently lost in peripheral regions, the present framework already yields high-quality results. It demonstrates practical viability using standard analysis tools, providing a robust baseline for implementation without the need for complex post-processing.

\section{Discussion and Experimental feasibility}\label{sec:experimental_feasibility}

This section outlines the practical requirements for the experimental implementation of the previously described spin imaging technique. The electron quantum states and images discussed so far represent snapshots of the spin precession at a specific phase. As detailed in Section~\ref{sec:hol-imaging}, the spin-dependent interaction induces a phase shift that manifests as a slight asymmetry in the electron beam's spatial intensity profile, resulting in a net deflection. However, this deflection averages to zero when integrated over a full Larmor precession cycle. To overcome this, the spin-dependent signal must be isolated by time-tagging electrons based on the arrival time at the sample. Subsequent post-processing, which correlates the electron arrival times with the microwave drive phase, enhances the image contrast under resonant conditions relative to the off-resonance background. These requirements are well matched to ultrafast transmission electron microscopes, which routinely achieve sub-picosecond time resolution. Notably, fast direct electron detectors with temporal resolution on the order of $\sim$260~ps~\cite{Llopart_2022} are sufficient to capture nuclear spin precession dynamics in real time ($\sim 40~\text{MHz}/\text{T}$). For example, even at a bias field of 2~T, the transition frequency $\nu$ remains below 100~MHz. 

The experimental signal is extracted by recording the driven dataset under resonant MW excitation $P_{\rm dr}$ and comparing it to the reference background $P_0$, thereby revealing the dynamic magnetic contributions free from static contrast artifacts.
The difference signal, $P_{\rm dr} - P_{0}$, effectively isolates contributions arising from spin orientation while suppressing background noise from conventional TEM interactions.
This approach enables the detection of subtle spin-dependent effects, even in environments dominated by standard electron--matter interactions.

The detectable signal accumulates linearly if aligned spins are imaged collectively. For example, molecular crystals such as BDPA, a stable paramagnetic organic radical, contain electron spins at a spatial density of $1.5~{\rm spins}/{\rm nm}^3$. Considering a lattice spacing of $a = 8.7~\text{\AA}$, there are $10^2$ spins per column (1-dimensional spin array along the beam axis) for a sample thickness of about 87~nm. Assuming that all spins are fully polarized and neglecting radiation damage, this implies an SNR for defocused imaging on the order of unity after an acquisition time of $30$~ms at a current of $1.6~{\rm nA}$; corresponding to a dose of $7.0 \times 10^5~{\rm e}^-/\text{\AA}^2$ for a Gaussian beam with FWHM = $1.1~\text{nm}$ ($\Delta k_\perp=4.22\times 10^{-4} k_{z,0}$) and using the image analysis method presented in Appendix~\ref{ap:optimal_single_spins}.

However, the use of such extended columns introduces significant practical challenges in the real-space configuration. The longitudinal spread of spins throughout the thickness results in varying local defoci for the top and bottom of the column; while the diffraction geometry is expected to be largely insensitive to this longitudinal variation, such depth-dependent effects can significantly degrade the holographic phase signal. Furthermore, the defocused configuration requires that signals from neighboring spin columns remain spatially isolated to prevent overlap on the detector, which imposes a more stringent limitation on the permissible spin density. These geometric constraints notwithstanding, higher spin densities and larger spin coherence times may be achieved in inorganic ferromagnetic materials, which also offer greater resilience to beam damage than organic radicals.

In practice, the degree of spin polarization of the sample is roughly proportional to the Boltzmann factor $\exp\left(-k_BT/h\nu\right)$, which depends on both temperature $T$ and the transition frequency defined by the external magnetic field as $\frac{\nu}{B_0} \approx 28$ GHz/T for electron spins. Hence, a higher transition frequency $\nu$ is advantageous for achieving greater thermal spin polarization. Since very low temperatures are not yet achievable in transmission electron microscopes, this study investigates electron spin resonance at a frequency of 50~GHz, which lies within the operational range of ultrafast transmission electron microscopes. By reaching dilution refrigerator temperatures below 100 mK~\cite{zu2022development} or employing advanced hyperpolarization techniques~\cite{Le2023, Blanchard2021, Duckett2012}, the stringent requirement for high transition frequencies can be alleviated.

Given that a typical $(100~\text{nm})^3$ BDPA sample hosts $\sim 1.5\times 10^{6}$ spins, yielding $\sim 6\times10^3$ polarized spins even at room temperature (degree of polarization $\sim 0.4\%$ at 50~GHz), we obtain $\text{SNR} > 3$, for an incident beam of FWHM $= 110$~nm, after a signal acquisition time of 6~ms at 1.6~nA corresponding to a dose of $30~\text{e}^-/\text{\AA}^2$, which is well within practical experimental limits. 
In contrast, Ferromagnetic Resonance (FMR) spectroscopy detects the collective precessional dynamics of magnetically ordered moments. For a standard probe such as YIG, the spontaneous magnetization at room temperature corresponds to an effective population of $N_s \approx 1.5 \times 10^7$ polarized electron spins per $(100~\text{nm})^3$ volume~\cite{hansen1974saturation}. This represents a signal enhancement of more than three orders of magnitude compared to the paramagnetic case. Consequently, the acquisition time scales down quadratically, allowing an $\text{SNR} > 3$ to be achieved with ultrafast snapshots (500~ps) at a negligible dose of $\sim 3\times 10^{-6}~\text{e}^-/\text{\AA}^2$.

Turning to nuclear spins, the gyromagnetic ratio ($\nu/B_0 \approx 42.6$~MHz/T) at $B_0 = 2$~T yields a precession frequency of $\sim 85$~MHz, which remains resolvable by state-of-the-art direct electron detectors. 
However, achieving sufficient sensitivity poses a significant challenge; the nuclear magneton is approximately $2000$ times smaller than the electron Bohr magneton, leading to a commensurate attenuation of the scattering signal. 
Consequently, high degrees of polarization are required to compensate for the weak interaction. In order to image nuclear spins, sample temperatures of less than 10~mK might be necessary, leading to a thermal spin polarization of more than $\sim 20\%$. 
Based on the molecular density of vitreous ice~\cite{dubochet1988cryo}, such conditions yield an effective ensemble of $N_s \approx 1.3\times10^7$ polarized spins within a $(100 \text{ nm})^3$ volume.
For a beam with $\text{FWHM} = 110$~nm centered on the sample, an $\text{SNR}>3$ requires an acquisition time of $\sim 0.5$~ms at $1.6$~nA, corresponding to a dose of $2.5~\text{e}^-/\text{\AA}^2$. Crucially, this dose is well within the damage thresholds for standard structural biology, establishing a viable physical baseline for sensing nuclear ensembles in radiation-sensitive materials. This underscores the necessity of technical solutions that enable cryogenic sample holders to reach temperatures in the low-millikelvin range, similar to architectures already realized in scanning probe microscopy~\cite{assig201310mK}.

Apart from material-specific radiation damage, spin-flip processes and coherence loss due to the magnetic interaction with the electron beam are negligible under typical experimental conditions. For a single-particle interaction, where the electron is modeled as a Gaussian wavepacket with a FWHM of $1.1$~nm, the resulting loss in spin purity is of the order of $10^{-10}$, as it scales with the square of the interaction strength (see Appendix~\ref{ap:backaction}). Even under continuous exposure, the cumulative effect remains minimal: simulating $10^{8}$ successive electron detections, which are far exceeding the number typically detected within a single spin relaxation time, results in only a $\sim 1\%$ reduction in purity. Therefore, no beam-induced reset or recovery period is necessary.

Especially investigating the signal from single spins involves measuring small deflections in diffraction mode and subtle variations in intensity near each spin position in image mode. This requires high resolution in both real space and Fourier space~\cite{Ishikawa2023}. 
In this context, STEM techniques are particularly beneficial, as they concentrate electron density at impact parameters where spin-induced deflections become significant ($<100$~pm)~\cite{Haslinger_2024}. 
Provided that a large portion of the electrons in the probe beam interacts coherently at these close distances, a total of $10^{10}$ recorded electrons can be sufficient to reach the SNR necessary for detection~\cite{metrologySpinTEM}.

\section{Conclusions}\label{sec:conclusions}

This work demonstrates the potential of integrating SRS with time-resolved TEM to investigate spin dynamics at the nanoscale. Building on the proposal in~\cite{Haslinger_2024}, we developed a scattering-theoretical framework to model electron-spin interactions, enabling a feasibility study of spin resonance detection in both diffraction and real-space imaging modes. Our analysis confirms the possibility of imaging spin states with minimal backaction, in agreement with theoretical expectations.

From a quantum metrology perspective, diffraction mode and Zernike phase contrast imaging emerge as the most effective among established modalities for SRS, particularly when using narrow Gaussian beams and exploiting the cumulative interaction with multiple spins. The CFI for image mode increases with increasing defocus, approaching the CFI of diffraction mode. While defocused image mode is always suboptimal, the signal also provides information about the position of the spins. 
The example of an image analysis approach for spin sensing presented in this paper leads to SNRs that approximate the classical bounds in both modes. Specifically, the method trades fine spatial or angular detail for maximized sensitivity to the spin-induced contrast.

Although imaging individual spins remains a challenging frontier, detecting spin columns is already within experimental reach. The introduction of next-generation microscope sample holders capable of operating at cryogenic temperatures ($\sim 4$~K)~\cite{Bai2015} will significantly boost spin polarization. 
In the case of nuclear spins, which are more abundant and exhibit extreme spatial localization (on the order of $\sim 10^{-12}$~m), the increased proximity to the electron beam can partially compensate for their intrinsically weaker magnetic interaction.

High SNRs, achievable through tightly focused electron beams, make atomic-scale spin detection increasingly viable and enable the exploration of second-order effects, such as spin noise. Advanced electron imaging techniques, such as DPC and ptychography~\cite{ishikawa2022, Lichte_2008, Edstrom2019, Boureau_2021, Zweck_2016}, offer promising avenues for enhancing sensitivity on the atomic scale~\cite{Koppell2022TransmissionEM, Dwyer2023}. However, their practical implementation remains computationally demanding and is reserved for future studies.

Moreover, the fast spin dynamics open the door to differential measurement strategies analogous to lock-in detection, which allow the selective investigation of spin-dependent signals by suppressing static and off-resonant backgrounds. As TEM continues to advance, pump-probe methodologies, such as microwave spin driving followed by electron beam interrogation, will become essential for probing coherent quantum spin dynamics in the MW frequency regime. Together, these strategies lay the groundwork for a new generation of quantum-enhanced electron microscopy techniques.

\begin{acknowledgments}
    We thank Michael Gaida, Stefan Nimmrichter, Anton\'in Jar{\v o}s,
    Johann Toyfl, Benjamin Czasch, Michael Seifner, and Isobel C. Bicket for helpful discussions. PH thanks the Austrian Science Fund (FWF): P36041, P35953, Y1121 and the FFG project AQUTEM.
\end{acknowledgments}

\section*{Code availability}

The source code used for the simulations and data analysis presented in this study is available at \href{https://doi.org/10.48436/exq5r-6zh78}{this repository}.


\bibliographystyle{apsrev4-1fixed_with_article_titles_full_names_new}
\bibliography{Master_Bib_File.bib}

\hypertarget{sec:appendix}
\appendix

\renewcommand{\thesubsubsection}{\Alph{section}.\Roman{subsection}.\arabic{subsubsection}}
\renewcommand{\thesubsection}{\Alph{section}.\Roman{subsection}}
\renewcommand{\thesection}{\Alph{section}}
\setcounter{equation}{0}
\numberwithin{equation}{section}
\setcounter{figure}{0}
\numberwithin{figure}{section}
\renewcommand{\theequation}{\Alph{section}.\arabic{equation}}
\renewcommand{\thefigure}{\Alph{section}.\arabic{figure}}


\onecolumngrid

\section*{Appendix}

The Appendix provides additional theoretical derivations and technical details supporting the main results. Appendix~\ref{ap:relativistic_model} introduces a relativistic model for the free-electron--spin interaction using a quantum field theoretic approach. 
Appendix~\ref{ap:spectroscopy_imaging} outlines the application of the $\pi/2$ MW pulse to the sample spin.
Appendix~\ref{ap:general_diffraction} derives the electron-spin quantum state and its probability amplitudes in momentum space for both longitudinal and transverse bias fields, including approximations relevant to diffraction mode imaging near the zero-deflection peak. Appendix~\ref{ap:general_position_mode} translates these amplitudes into position space, with a focus on single-spin imaging. Appendix~\ref{ap:backaction} examines the backaction on the spin state and provides a description of the coherent part of the electron beam wavefunction. Finally, Appendix~\ref{ap:CFI_optimal} introduces the CFI framework for quantifying spin information and optimizing data extraction in both momentum and position space.

\section{Relativistic model for the free-electron--spin interaction} \label{ap:relativistic_model}

We develop the following theoretical model in the reference frame of a free electron, modeled as a relativistic free particle. The electron propagates through a sample and interacts with the spin of a bound electron. This bound spin is subject to a static magnetic bias field $\vec{B}_0$, which either has a negligible effect on the free electron, since the electron primarily propagates along the bias field axis, or contributes constant phase shifts that would be removed through image subtraction in practice, as discussed in the main text. Therefore, the effect of the bias field on the free electron is neglected.
The relativistic Hamiltonian describing the free electron interacting with the bound electron's spin, using the minimal coupling scheme, is given by  
\begin{subequations}\label{eq:full_hamiltonian}
\begin{align}
\oH=& \oH_{\rm e}+\oH_{Z},\\
\begin{split}
    \text{where}~\oH_{\rm e} =&\sqrt{(m_{\rm e}c^2)^2+(\vec{\hat{p}}+e\vec{\hat{A}})^2c^2},\\
=&\sqrt{(m_{\rm e}c^2)^2+|\vec{\hat{p}}|^2c^2+e(\vec{\hat{p}}\cdot\vec{\hat{A}}+\vec{\hat{A}}\cdot \vec{\hat{p}})c^2+e^2|\vec{\hat{A}}|^2c^2},
\end{split}
\end{align}
\end{subequations}
and the Zeeman interaction is 
\begin{equation}
\oH_Z = -\mu \vec{B}_0 \cdot \vec{\osigma},    
\end{equation}
where $\mu=-\mu_B$ is the magnetic moment of the electron spin when approximating the electron's spin g-factor as -2 and $\mu_{\rm B}$ is the Bohr magneton.
In Coulomb gauge, the vector potential generated by a spin is given by  
\begin{equation}\label{eq:vecpot}
    \vec{\hat{A}} = \frac{\mu_0 \mu}{4\pi} \frac{\vec{\hat{\sigma}} \times \hat{\vec{r}}}{\hat{r}^3},
\end{equation}
where $\vec{\hat{\sigma}} = (\hat{\sigma}_x, \hat{\sigma}_y, \hat{\sigma}_z)$ is the Pauli matrix vector.
To analyze the interaction term from minimal coupling, which appears in the third term under the square root in Eq.~\eqref{eq:full_hamiltonian}, we compute the anti-commutator between $\hat{\vec{p}}$ and $\vec{\hat{A}}$. Using the identity
\begin{equation}
    \begin{aligned}
        \hat{\vec{p}} \cdot \vec{\hat{A}} + \vec{\hat{A}} \cdot \hat{\vec{p}} &= 2\vec{\hat{A}} \cdot \hat{\vec{p}} + i\hbar \nabla \cdot \vec{\hat{A}} \\
        &= 2\vec{\hat{A}} \cdot \hat{\vec{p}} - i\hbar \frac{\mu_0 \mu}{4\pi} \vec{\hat{\sigma}} \cdot \left( \vec{\nabla} \times \frac{\hat{\vec{r}}}{\hat{r}^3} \right) \\
        &= 2\vec{\hat{A}} \cdot \hat{\vec{p}},
    \end{aligned}
\end{equation}
we conclude that the interaction term can be expressed simply as $\vec{\hat{A}} \cdot \hat{\vec{p}}$.

Comparing the paramagnetic interaction term $\frac{2e}{m_{\rm e}}\vec{\hat{A}} \cdot \hat{\vec{p}}$ with the magnetic vector potential self-interaction or diamagnetic interaction $\frac{e^2}{m_{\rm e}}|\vec{\hat{A}}|^2$, we obtain, respectively,
\begin{subequations}
    \begin{align}
        \frac{2e}{m_{\rm e}}\vec{\hat{A}} \cdot \hat{\vec{p}} &= \frac{\hbar^2}{m_{\rm e}}\left(\frac{r_{\rm e}}{\hat{r}^3}\right)\frac{\mu}{\mu_{\rm B}}\frac{\vec{\hat{r}} \times \vec{\hat{p}}}{\hbar} \cdot \vec{\osigma}, \\
        \frac{e^2}{m_{\rm e}}|\vec{\hat{A}}|^2 &= \frac{3\hbar^2}{4m_{\rm e}}\left(\frac{r_{\rm e}^2}{\hat{r}^4}\right)\left(\frac{\mu}{\mu_{\rm B}}\right)^2\left(1 - \left(\frac{\vec{\osigma} \cdot \vec{\hat{r}}}{\hat{r}}\right)^2\right),
    \end{align}
\end{subequations}
where $r_{\rm e} = \frac{e \mu_0 \mu_{\rm B}}{2\pi \hbar} \approx 2.82\cdot 10^{-15}\,\text{m}$ is the classical electron radius. Since the paramagnetic term scales linearly with $r_{\rm e}/\hat{r}$ and the diamagnetic term scales quadratically with $r_{\rm e}/\hat{r}$, the latter is suppressed by at least five orders of magnitude for electron beams with transverse extension on the nanometer scale. Therefore, under conditions of parallel illumination, where electrons do not closely approach the sample spin, the diamagnetic contribution $\frac{e}{m_{\rm e}}|\vec{\hat{A}}|^2$ can be safely neglected in the Hamiltonian.

We then expand the Hamiltonian in a power series of the energy terms with respect to the rest energy of the electron $mc^2$, obtaining
\begin{equation}\label{eq:he_expansion}
    \begin{aligned}
        \hat{H}_\text{e} \approx& \sqrt{(m_{\rm e}c^2)^2 + |\vec{\hat{p}}|^2c^2 + 2e\vec{\hat{A}} \cdot \vec{\hat{p}} c^2} \\
        =& m_{\rm e}c^2 \sum_{j=0}^\infty \begin{pmatrix}
            1/2\\
            j
        \end{pmatrix} \left( \frac{|\hat{\vec{p}}|^2}{(m_{\rm e}c)^2} + \frac{2e}{m^2_{\rm e}c^2} \hat{\vec{A}} \cdot \hat{\vec{p}} \right)^j \\
        =& \underbrace{m_{\rm e}c^2\sum_{j=0}^\infty
            \begin{pmatrix}
                1/2\\
                j
            \end{pmatrix} \left( \frac{|\hat{\vec{p}}|}{m_{\rm e}c} \right)^{2j}}_{\mathcal{O}(m_{\rm e}c^2)}
            + \underbrace{ \sum_{j=1}^\infty \begin{pmatrix}
                -1/2\\
                j-1
            \end{pmatrix} \frac{e}{m_{\rm e}} \hat{\vec{A}} \cdot \hat{\vec{p}} \left( \frac{|\hat{\vec{p}}|}{m_{\rm e}c} \right)^{2(j-1)} + \Omega_j(\hat{\vec{A}}, \hat{\vec{p}}) }_{\mathcal{O}(\frac{e}{m_{\rm e}}\hat{\vec{A}}\cdot \hat{\vec{p}})} 
            + \mathcal{O} \left( \left( \frac{e}{m_{\rm e}} \hat{\vec{A}} \cdot \hat{\vec{p}} \right)^2 \right).
    \end{aligned}
\end{equation}
In the third line, we reorganize the sum as factors with respect to the interaction Hamiltonian $\frac{e}{m_{\rm e}}\hat{\vec{A}}\cdot \hat{\vec{p}}$. Then we can apply a similar argument as we used for neglecting the $|\vec{\hat{A}}|^2$ term, that is, we neglect powers of the interaction term $\frac{e}{m_{\rm e}}\hat{\vec{A}}\cdot \hat{\vec{p}}$ larger than one, because the coupling strength is proportional to $r_{\rm e}/\hat{r}$.Additionally, the kinetic terms are reorganized to the right-hand side of each term in the sum. The commutation terms arising from this ordering are included in the operator $\Omega_j(\hat{\vec{A}}, \hat{\vec{p}})$.  
We observe that the first sum in the third line corresponds to the Taylor series of the free electron relativistic energy, while the second sum represents the Taylor expansion of the inverse of the Lorentz factor ${\gamma}(\hat{\vec{p}})=  \sqrt{1 + |\vec{\hat{p}}|^2/(m_{\rm e}c)^2)}$ times $\frac{e}{m_{\rm e}} \hat{\vec{A}} \cdot \hat{\vec{p}}$. This naturally leads to the identification of $\oH_{\text{e},0}$ and $\oH_{\text{I}}$ as 
\begin{equation}
\hat{H}_\text{e} = \underbrace{m_{\rm e}\gamma(\hat{\vec{p}})c^2}_{:=\oH_{\text{e},0}}+\underbrace{\hat{\vec{A}}\cdot \hat{\vec{p}} \frac{e}{m_{\rm e} \gamma(\hat{\vec{p}})}}_{:=\oH_{\text{I}}}+\underbrace{\sum_{j=1}^\infty \Omega_j(\hat{\vec{A}}, \hat{\vec{p}})}_{:=\Omega(\hat{\vec{A}}, \hat{\vec{p}})}+\mathcal{O}\left(\left(\frac{e}{m_{\rm e}}\hat{\vec{A}}\cdot \hat{\vec{p}}\right)^2\right).
\end{equation}
The terms $\Omega_j(\hat{\vec{A}}, \hat{\vec{p}})$ account for commutator contributions from operator ordering, where for instance
\begin{equation}
    \begin{aligned}
\Omega_1(\hat{\vec{A}}, \hat{\vec{p}})=&0\,,\\
\Omega_2(\hat{\vec{A}}, \hat{\vec{p}})=&-\frac{e}{4m_{\rm e}^3c^2}[\hat{\vec{p}}^2, \hat{\vec{A}}\cdot \hat{\vec{p}}]=-\frac{e}{4m_{\rm e}^3c^2}\left[-2i\hbar \vec{\nabla}(\hat{\vec{A}}\cdot \hat{\vec{p}})\cdot \hat{\vec{p}}-\hbar^2\nabla^2(\hat{\vec{A}}\cdot \hat{\vec{p}})\right]\,,\\
\Omega_3(
{\hat{\vec{A}}}, \hat{\vec{p}})=&\frac{e}{8m_{\rm e}^5c^4}\left[3[\hat{\vec{p}}^2, \hat{\vec{A}}\cdot \hat{\vec{p}}] \hat{\vec{p}}^2-2i\hbar \vec{\nabla}( [\hat{\vec{p}}^2, \hat{\vec{A}}\cdot \hat{\vec{p}}])\cdot \hat{\vec{p}}-\hbar^2\nabla^2( [\hat{\vec{p}}^2, \hat{\vec{A}}\cdot \hat{\vec{p}}])\right]\,.
    \end{aligned}
\end{equation}
These operators in $\Omega_2(\hat{\vec{A}}, \hat{\vec{p}})$ scale as  
\begin{equation}
\frac{1}{m_{\rm e}^2 c^2}i\hbar \nabla\!\left(\frac{e}{m_{\rm e}}\vec{A}\cdot \vec{p}\right)\cdot \vec{p}
\ \sim\ \frac{\lambda_{\rm c}^2}{\lambda_{\rm e}r}\,
\left|\frac{e}{m_{\rm e}}\vec{A}\cdot \vec{p}\right|
\end{equation}
and  
\begin{equation}
\frac{1}{m^2 c^2}\hbar^2 \nabla^2\!\left(\frac{e}{m}\vec{A}\cdot \vec{p}\right)
\ \sim\ \frac{\lambda_{\rm c}^2}{r^2}\,
\left|\frac{e}{m}\vec{A}\cdot \vec{p}\right|,
\end{equation}
where the reduced Compton wavelength is $\lambda_{\rm c}=\hbar/(m_{\rm e} c)\approx 0.386~\mathrm{pm}$ and the electron wavelength for a $200~\mathrm{keV}$ beam is $\lambda_{\rm e}=2.51~\mathrm{pm}$. 
For electron wave functions with an extension of the order of nanometers, we have $r\gg \lambda_{\rm c}$ and the former term is always larger than the latter as $\lambda_{\rm c}^2/\lambda_{\rm e}r\approx 0.15 \lambda_{\rm c}/r\gg \lambda^2_{\rm c}/r^2$.

Therefore, the largest correction from the commuting terms arises from the contributions proportional to 
$\vec{\nabla}\!\left(\tfrac{e}{m_{\rm e}}\hat{\vec{A}}\cdot \hat{\vec{p}}\right)$, 
since they scale with $\hat{r}^{-3}$, while the other terms scale at least with $\hat{r}^{-4}$ leading to their suppression at distances much larger than $\lambda_c$ and $\lambda_e$. 
By considering only the interaction term proportional to $\vec{\nabla}\!\left(\tfrac{e}{m_{\rm e}}\hat{\vec{A}}\cdot \hat{\vec{p}}\right)$, we find
\begin{equation}
    \Omega_2(\hat{A}, \hat{p})\sim-\frac{3i\lambda_{\rm c}^2 r_{\rm e}}{4 m_{\rm e}}\frac{\mu}{\mu_{\rm B}}\left(\frac{(\vec{\osigma}\times\hat{\vec{r}})\cdot \hat{\vec{p}}}{\hat{r}^5}\hat{\vec{r}}\cdot \hat{\vec{p}}\right)+\mathcal{O}\left(\frac{\lambda_{\rm c}^2}{r^2}\,
\left|\frac{e}{m_{\rm e}}\vec{A}\cdot \vec{p}\right|\right),
\end{equation}
since
\begin{equation}
    \vec{\nabla}\left(\frac{e}{m_{\rm e}}\hat{\vec{A}}\cdot \hat{\vec{p}}\right)=\frac{\hbar r_{\rm e}}{2 m_{\rm e}}\frac{\mu}{\mu_{\rm B}}\left[\frac{\vec{\osigma}\times \hat{\vec{p}}}{\hat{r}^3}-3\frac{(\vec{\osigma}\times\hat{\vec{r}})\cdot \hat{\vec{p}}}{\hat{r}^5}\hat{\vec{r}}\right].
\end{equation}
However, for every $j>1$, there exists an interaction term that 
contains a contribution proportional to $\vec{\nabla}\!\left(\tfrac{e}{m_{\rm e}}\hat{\vec{A}}\cdot \hat{\vec{p}}\right)$ which dominates at distances much larger than $\lambda_c$ and $\lambda_e$:
\begin{equation}
     \begin{aligned}
 \Omega_j(\hat{A}, \hat{p})\sim&\frac{e}{m_{\rm e}^3 c^2}\begin{pmatrix}
         1/2\\
         j
     \end{pmatrix}j(j-1)[\hat{\vec{p}}^2, \hat{\vec{A}}\cdot \hat{\vec{p}}]\left(\frac{{|\hat{\vec{p}}|}}{m_{\rm e}c}\right)^{2(j-2)}\\
     \sim& 3i\begin{pmatrix}
         1/2\\
         j
     \end{pmatrix}j(j-1)\frac{\lambda^2_{\rm c} r_{\rm e}}{m_{\rm e}}\frac{\mu}{\mu_{\rm B}} \left(\frac{(\vec{\osigma}\times\hat{\vec{r}})\cdot \hat{\vec{p}}}{\hat{r}^5}\hat{\vec{r}}\cdot \hat{\vec{p}}\right)\left(\frac{{|\hat{\vec{p}}|}}{m_{\rm e}c}\right)^{2(j-2)},
     \end{aligned}
\end{equation}
because these terms originate from the commutator $[\hat{\vec{p}}^2, \hat{\vec{A}}\cdot \hat{\vec{p}}]$.  
Reordering the kinetic terms to the right-hand side (RHS) introduces a multiplicative factor $j(j-1)/2$, corresponding to the number of commutations.
For each power $j$, there are $j-1$ terms proportional to $\vA \cdot \vec{\hat{p}}$. 
To push the kinetic term accompanying each $\vA \cdot \vec{\hat{p}}$ to the RHS, one requires 
a sequence of $j-1, j-2, \dots, 0$ commutations, respectively. 
Since our interest lies in the commutators $[|\vec{\hat{p}}|^2, \vA \cdot \vec{\hat{p}}]$, 
the total number of them for a given $j$ is precisely the triangular number 
$\tfrac{j(j-1)}{2}$.

In the second line, we retain only the term $\vec{\nabla}( \hat{\vec{A}}\cdot \hat{\vec{p}})$, which we identify with the $\hat{r}^{-3}$ radial scaling of the interaction.  
Furthermore, the sum over $j$ acts only on the momentum variable. Hence,
\begin{equation}
    \begin{aligned}
\Omega(\hat{\vec{A}}, \hat{\vec{p}})=\sum_{j=2}^\infty \Omega_j( \hat{\vec{A}}, \hat{\vec{p}})\sim& 3i \frac{\lambda^2_{\rm c} r_{\rm e}}{m}\frac{\mu}{\mu_{\rm B}} \left(\frac{(\vec{\osigma}\times\hat{\vec{r}})\cdot \hat{\vec{p}}}{\hat{r}^5}\hat{\vec{r}}\cdot \hat{\vec{p}}\right)\sum_{j=2}^\infty\begin{pmatrix}
         1/2\\
         j
     \end{pmatrix}j(j-1)\left(\frac{|\hat{\vec{p}}|}{m_{\rm e}c}\right)^{2(j-2)}\\
     =&-\frac{3}{4}i \frac{\lambda^2_{\rm c} r_{\rm e}}{m_{\rm e}}\frac{\mu}{\mu_{\rm B}} \left(\frac{(\vec{\osigma}\times\hat{\vec{r}})\cdot \hat{\vec{p}}}{\hat{r}^5}\hat{\vec{r}}\cdot \hat{\vec{p}}\right)\frac{1}{\gamma^3(\hat{\vec{p}})},
    \end{aligned}
\end{equation}
where we have used the identity
\begin{equation}
    \sum_{j=2}^\infty\begin{pmatrix}
         1/2\\
         j
     \end{pmatrix}j(j-1)t^{j-2}=\frac{d^2}{dt^2}\sum_{j=0}^\infty\begin{pmatrix}
         1/2\\
         j
     \end{pmatrix}t^j=\frac{d^2}{dt^2}\sqrt{1+t}=-\frac{1}{4}(1+t)^{-3/2}.
\end{equation}
In terms of scaling, we find
\[
|\hat{H}_{\rm I}|\sim \frac{\hbar^2}{m_{\rm e} \gamma}\frac{\mu}{\mu_{\rm B}}\left(\frac{r_{\rm e}}{r^2 \lambda_{\rm e}}\right),
\qquad
\left|\Omega(\hat{\vec{A}}, \hat{\vec{p}})\right|\sim\frac{\hbar^2}{m_{\rm e} \gamma^3}\frac{\mu}{\mu_{\rm B}}\left(\frac{\lambda_{\rm c}^2 r_e}{r^3 \lambda^2_{\rm e}}\right),
\]
where $\gamma$ is the scalar Lorentz factor of the incident beam.  
The commuting correction is negligible, provided 
\[
r\gg \frac{\lambda_{\rm c}^2}{\gamma^2\lambda_{\rm e}}\approx 0.030~\text{pm},
\] 
a condition that is unlikely to be violated
at the nanometer scale.  
For instance, a Gaussian beam wavefunction centered at the origin, where the spin is located, with a width of $50~\text{nm}$ (corresponding to a FWHM of $110~\mathrm{nm}$ or $\Delta k_\perp=4.22\times 10^{-6} k_{z,0}$) has only $10^{-6}$  
of the total number of electrons localized at distances $r<100~\text{pm}$.  
At such distances, the commuting correction remains four orders of magnitude smaller than $\hat{H}_{\rm I}$.

The approximate Hamiltonian can be written as
\begin{subequations}
\begin{align}
    \oH &\approx \oH_0 + \oH_{\text{I}}, \\
    \oH_0 &= \oH_{\text{e},0} + \oH_Z ,
\end{align}
\end{subequations}
where contributions from $\Omega(\hat{\vec{A}}, \hat{\vec{p}})$ and $\vec{\hat{A}}^2$ have been neglected, as their magnitude is much smaller than that of $\oH_{\text{I}}$ in the far-field regime.  
This effective Hamiltonian describes the interaction between the free electron and the localized sample spin. The Lorentz factor $\gamma(\hat{\vec{p}})$ enters the electron's kinetic term, providing relativistic corrections to its effective mass and energy. These corrections also modify the electron's wavelength, in agreement with standard treatments in relativistic electron scattering theory~\cite{1961_fujiwara}.  
Finally, we note that $\hat{H}_{\text{I}}$ is only approximately Hermitian. The difference from its symmetric form arises from commuting terms in $\Omega(\hat{\vec{A}}, \hat{\vec{p}})$, which are negligible for probe beams larger than a few picometers. Under this approximation, the Hamiltonian remains suitable for describing the effective interaction between the two quantum systems.

\section{Microwave driving and spin initialization}\label{ap:spectroscopy_imaging}

The proposed technique for SRS imaging of a spin-based sample inside a transmission electron microscope relies on MW driving fields used to initialize the spin states. We consider an experimental setup where the bias magnetic field is applied by the microscope objective pole pieces and aligned parallel to the electron beam propagation axis. 
A transverse MW $\pi/2$ pulse, oriented along the $y$-axis, is applied to the spin system. On resonance, this pulse coherently rotates the spin onto the equatorial plane of the Bloch sphere, after which the spin undergoes free precession. Off resonance, the spin is only partially driven, resulting in a tilt away from the bias field axis. The $\pi/2$ pulse is assumed to be sufficiently short to ensure coherent evolution, so that loss of spin purity during this process is negligible.
In general, the initial spin state is a pure state, lying on the surface of the Bloch sphere, with expectation values given by $\expval{\vec{\osigma}(\frac{\pi}{2\omega_1})} = \left(s_x(\frac{\pi}{2\omega_1}), s_y(\frac{\pi}{2\omega_1}), s_z(\frac{\pi}{2\omega_1})\right)$, where $\omega_1=2\mu_{\rm B}B_1$ is the Rabi frequency. The time evolution of the expectation values of the Pauli matrices, based on the solution without relaxation in~\cite{Johnston2020_BlochEq}, is given by:
\begin{equation}
    s_x(t) = -s_1(t)\cos(\omega_0t) - s_2(t)\sin(\omega_0t), \qquad 
    s_y(t) = s_2(t)\cos(\omega_0t) - s_1(t)\sin(\omega_0t),
\end{equation}
where $s_j(t)$ represents the spin components in the rotating frame. These evolve according to:
\begin{subequations}
\begin{align}
\begin{split}
    s_1(t) =& \frac{\omega_1}{\omega_1^2 + \delta^2} \left( \omega_1 s_x(0) + \delta s_z(0) \right) \\
    &\qquad \qquad \qquad+ \frac{\delta}{\omega_1^2 + \delta^2} \left[ \left( \delta s_x(0) - \omega_1 s_z(0) \right) \cos\left( \sqrt{\omega_1^2 + \delta^2} \, t \right) 
    + \sqrt{\omega_1^2 + \delta^2} \, s_y(0) \sin\left( \sqrt{\omega_1^2 + \delta^2} \, t \right) \right],
\end{split}\\
    s_2(t) =&\, s_y(0) \cos\left( \sqrt{\omega_1^2 + \delta^2} \, t \right) 
    + \frac{\omega_1 s_z(0) - \delta s_x(0)}{\sqrt{\omega_1^2 + \delta^2}} 
    \sin\left( \sqrt{\omega_1^2 + \delta^2} \, t \right), \\
\begin{split}
    s_z(t) =& \frac{\delta}{\omega_1^2 + \delta^2} \left( \omega_1 s_x(0) + \delta s_z(0) \right) 
    \\
    &\qquad \qquad \qquad- \frac{\omega_1}{\omega_1^2 + \delta^2} \left[ \left( \delta s_x(0) - \omega_1 s_z(0) \right) \cos\left( \sqrt{\omega_1^2 + \delta^2} \, t \right) 
    + \sqrt{\omega_1^2 + \delta^2} \, s_y(0) \sin\left( \sqrt{\omega_1^2 + \delta^2} \, t \right) \right],
\end{split}    
\end{align}
\end{subequations}
where the detuning is defined as $\delta = \omega - \omega_0$, with $\omega$ the applied oscillating MW frequency, and $\omega_0$ the natural Larmor frequency determined by the Zeeman energy splitting. Additionally, the time $t=0$ refers to the spin state before the MW driving, where the spin is anti-aligned to the bias field, i.e. $\expval{\vec{\osigma}(0)}=(s_x(0), s_y(0), s_z(0))=(0,0,-1)$.

From this point onward, we define the time $t = 0$ as the moment immediately following the MW driving pulse. The spin-resolved images discussed in this work will correspond either to this initial time or to later times that are integer multiples of the spin precession period, $\text{T}_{\rm prec} = \frac{2\pi}{\omega_0}$, during which relaxation effects are negligible. Under these conditions, the spin remains in a pure state, and the corresponding expectation values of the spin vector are shown in Fig.~\ref{fig:spectroscopy_imaging_TEM}I in the main text for various detunings.
As illustrated by the curve for $\expval{\hat{\sigma}_z}$, only under resonant driving does the spin lie entirely in the equatorial plane, becoming aligned along the positive $y$-axis. As the detuning increases, the spin tilts progressively out of the plane and eventually becomes anti-aligned with the bias field. 
In this work, we investigate spin dynamics and imaging performance for four representative detuning values, as shown in Table~\ref{tab:cases_SRS_TEM}.

\begin{table}[h]
    \centering
    \begin{tabular}{c|c|c|c|c}\hline\hline
    Label & Detuning $\delta/\omega_0$ &  $\expval{\osigma_x}$ & $\expval{\osigma_y}$ & $\expval{\osigma_z}$ \\ \hline
    (a)  &  0.000 (on resonance)    &  0.00 & 1.00 & 0.00\\
    (b)  &  0.025     &  0.51 & -0.33 & 0.80\\
    (c)  &  0.050    &  0.22 & 0.19 & 0.96\\
    (d)  &  0.075     &  0.03 & -0.08 & 0.99\\\hline\hline
    \end{tabular}
    \caption{Selected detunings of study for SRS in TEM and the respective Pauli matrices expected values of the initialized spin pure states used in Fig.~\ref{fig:spectroscopy_imaging_TEM} in the main text.}
    \label{tab:cases_SRS_TEM}
\end{table}

\section{Probability amplitudes of the free electron-sample spin state in momentum representation}\label{ap:general_diffraction}

We model the incident beam electron as a pure state in the momentum eigenbasis given by:
\begin{equation}
\ket{\varphi_{\rm in}} = \int d^3k \, \varphi_{{\rm in}} (\vec{k}) \ket{\vec{k}},
\end{equation}
where the wavepacket is assumed to be azimuthally symmetric, which means that it only depends on the longitudinal momentum component $k_z$ and the transverse momentum component $k_\perp$, i.e. $\varphi_{{\rm in}} (\vec{k})=\bar\varphi_{{\rm in}} (k_\perp, k_z)$.
The sample spin is initialized in a pure state $\ket{s}$, with its spatial degrees of freedom delocalized according to a probability density $|\psi_{\rm s}(\vec{r}_{\rm s})|^2$ centered around the origin. Consequently, since the interaction depends on the relative position between the free beam electron $\vec{r}_{\rm e}$ and the bound electron (sample spin) $\vec{r}_{\rm s}$, it must be averaged over the spatial delocalization of both quantum systems. In this context, we assume that the spatial distribution of the bound electron is not significantly changed due to the interaction with the free electron.

The interaction Hamiltonian in the interaction picture corresponds to:
\begin{equation}
    \hat{H}_{\rm int}(t):=\hat{U}_0(t)\hat{H}_{\rm I} \hat{U}^\dagger_0(t),
\end{equation}
where the free evolution unitary is $\hat{U}_0(t)=\exp{\frac{i}{\hbar}\hat{H}_0 t}$. Since the evolution of the free Hamiltonian of the spin and the one of the free electron commute ($\hat{U}_0(t)=\exp{\frac{i}{\hbar}\hat{H}_{\rm e, 0} t}\exp{\frac{i}{\hbar}\hat{H}_Z t}$), we can apply first the free evolution of the spin, and then the evolution with the free-electron kinetic energy. In effect,
\begin{equation}
    \hat{H}_{\rm II}(t):=e^{-i\mu\vec{B_0}\cdot \vec{\osigma}t/\hbar}\hat{H}_{\rm I}e^{i\mu\vec{B_0}\cdot \vec{\osigma}t/\hbar}=\vec{\hat{A}}(t)\cdot \vec{\hat{p}}\frac{e}{m_{\rm e}\gamma(\vec{\hat{p}})}\,.
\end{equation}
From now on, we replace $\vec{r} \to \vec{r}_{\rm e}-\vec{r}_{\rm s}$ in Eq.~\eqref{eq:vecpot} so that the electromagnetic potential also depends explicitly on the spin position,  
\begin{equation}\label{eq:vecpot2}
    \vec{\hat{A}}(t) = \frac{\mu_0 \mu}{4\pi} 
    \frac{\vec{\hat{\sigma}}(t) \times (\vec{\hat{r}}-\vec{r}_{\rm s})}{|\vec{\hat{r}}-\vec{r}_{\rm s}|^3}.
\end{equation}
 Then, the interaction Hamiltonian is
\begin{equation}
    \begin{aligned}
\hat{H}_{\mathrm{int}, r_{\rm s}}(t)=&\exp{\frac{i}{\hbar}\hat{H}_{{\rm e},0}t}\hat{H}_{\rm II}\exp{-\frac{i}{\hbar}\hat{H}_{{\rm e},0}t}\,.
\end{aligned}
\end{equation}
The temporal dependence on the Pauli matrices corresponds to:
\begin{equation}
    \begin{split}
     \hat{\vec{\sigma}}(t)=& e^{i \frac{\omega_0 t}{2}\left(\ve_{n, 0} \cdot \hat{\vec{\sigma}}\right)} ~ \hat{\vec{\sigma}} ~  e^{-i \frac{\omega_0 t}{2}\left(\ve_{n, 0} \cdot \hat{\vec{\sigma}}\right)}\\
     =&
  \hat{\vec{\sigma}}~\cos (\omega_0 t) + \ve_{n, 0} \times \hat{\vec{\sigma}} ~ \sin(\omega_0 t) + \ve_{n, 0} ~ (\ve_{n, 0} \cdot \hat{\vec{\sigma}}) ~ (1 - \cos(\omega_0 t))\,,
    \end{split}
\end{equation}
for general orientation vectors $\ve_{n, 0}$ of the magnetic field, where $\omega_0 = 2\mu_{\rm B}|\vec{B}_0|/\hbar$ is the Larmor frequency (for the electron spin g-factor set to $-2$). 
Without loss of generality, we can choose any unitary orientation vector $\ve_{m, 0}$ perpendicular to $\ve_{n, 0}$, both of which define a Cartesian coordinate system with $\ve_{l, 0}=\ve_{n, 0}\times\ve_{m, 0}$. Then, it is possible to rewrite
\begin{equation}
    \hat{\vec{\sigma}}(t)={\sqrt{2}}\left(e^{i\omega_0 t}\osigma_{n, +}\ve_{n, -}+e^{-i\omega_0 t}\osigma_{n, -}\ve_{n, +}\right)+\osigma_{n,0} \ve_{n, 0}\,,
\end{equation}
where $\ve_{n, \pm}=\frac{1}{\sqrt{2}}(\ve_{m, 0}\pm i\ve_{l, 0})$ and $\osigma_{n,0}= \ve_{n, 0}\cdot \hat{\vec{\sigma}} $. Additionally, the Pauli matrices $\osigma_{n, \pm}=\frac{1}{2}(\osigma_{m,0}\pm i \osigma_{l,0})$ define the creation and annihilation operators with respect to the basis established by $\osigma_{n,0}$ and $\osigma_{m,0}= \ve_{m, 0}\cdot \hat{\vec{\sigma}} $ and $\osigma_{l,0}= \ve_{l, 0}\cdot \hat{\vec{\sigma}} $.

The two-particle quantum state after the interaction is given by
\begin{subequations}\label{eq:psi_final_pure}
    \begin{align}
        \ket{{\psi}_{\rm out}} 
        =  \hat{S} \ket{\varphi_{\rm in}} \otimes \ket{s},
    \end{align}
\end{subequations}
where the scattering operator is formally defined as 
\begin{equation}
    \hat{S} = \mathcal{T} \exp \left( -\frac{i}{\hbar} \int d^3 r_{\rm s} \, 
        |\psi_{\rm s}(\vec{r}_{\rm s})|^2\int_{-\infty}^{\infty} dt\, \hat{H}_{\rm int, r_{\rm s}}(t) \right),
\end{equation}
with $\mathcal{T}$ denoting time ordering and the interaction is averaged over the spin's spatial probability density $|\psi_{\rm s}(\vec{r}_{\rm s})|^2$ to account for its delocalization. Since we are interested in the leading-order contributions to the interaction between the free electron and the sample spin, we restrict ourselves to first-order perturbation theory. In this approximation, the scattering operator becomes
\begin{equation}\label{eq:scattoperator_first_order}
    \hat{S} \approx \mathbb{1}+\hat{\tau}^{(1)}:=\mathbb{1} -\int d^3 r_{\rm s} \, 
        |\psi_{\rm s}(\vec{r}_{\rm s})|^2\left( \frac{i}{\hbar} \int_{-\infty}^{\infty} dt \, \hat{H}_{\rm int, r_{\rm s}}(t)\right).
\end{equation}
Next, the matrix elements of the first-order scattering operator due to the interaction with a spatially delocalized spin are given by
\begin{equation}\label{eq:S_ks_ins}
    \begin{aligned}
        S^{(1)}_{\vec{k}', s'; {\rm in}, s} 
        &:= \int d^3 r_{\rm s} \, |\psi_{\rm s}(\vec{r}_{\rm s})|^2 \,
        \bra{\vec{k}', s'} 
        \left[ \mathbb{1} - \frac{i}{\hbar} \int_{-\infty}^{\infty} dt \, \hat{H}_{\rm int, r_{\rm s}}(t) \right]
        \ket{\varphi_{\rm in}} \otimes \ket{s} \\
        &= \int d^3k \, \bar\varphi_{{\rm in}} (k_\perp, k_z) \, 
        S^{(1)}_{\vec{k}', s'; \vec{k}, s},
    \end{aligned}
\end{equation}
where $S^{(1)}_{\vec{k}', s'; \vec{k}, s}$ are the S-matrix elements that correspond to an incident plane wavefunction of the electron. Thus, the S-matrix elements of the incident Gaussian wave packet are obtained by averaging over the momentum distribution of the free electron wavefunction.

The $S^{(1)}_{\vec{k}', s'; \vec{k}, s}$ elements are given by
\begin{equation}\label{eq:scatt_matrix_momentum_origin_v2}
    \begin{aligned}
    S^{(1)}_{\vec{k}',s';\vec{k},s} :=&\int d^3 r_{\rm s} \, |\psi_{\rm s}(\vec{r}_{\rm s})|^2 \,
        \bra{\vec{k}', s'} 
        \left[ \mathbb{1} - \frac{i}{\hbar} \int_{-\infty}^{\infty} dt \, \hat{H}_{\rm int, r_{\rm s}}(t) \right]
        \ket{\vec{k}} \otimes \ket{s}\\
\approx&  \la s'|s\ra \delta(\vec{k}'-\vec{k}) - \la s'|i\frac{e \mu_0 \mu}{4\pi m_{\rm e} \gamma(|\hbar \vec{k}|)}\frac{1}{(2\pi)^3}\int d^3{r_{\rm s}} \int d^3{{r}_{\rm e}} |\psi(\vec{r}_s)|^2\\
    &\qquad\qquad\qquad\qquad\qquad\qquad \int_{-\infty}^\infty dt\, e^{-i\vec{k'}\cdot \vec{r}_{\rm e}}  \hat{\vec{\sigma}}(t) \cdot \left( \exp{\frac{it}{\hbar}E(\vec{k}')}  \frac{\vec{r}}{r^3} \exp{-\frac{it}{\hbar}E(\vec{k})} \times \vec{k} \right)e^{i\vec{k}\cdot \vec{r}_{\rm e}} |s\ra \, \\
    \approx& \la s'|s\ra \delta(\vec{k}'-\vec{k}) - \la s'|i\frac{e \mu_0 \mu}{4\pi m_{\rm e} \gamma(|\hbar \vec{k}|)}\frac{1}{(2\pi)^2} \sum_{\varsigma_={-1}}^1 2^{|\varsigma|/2}\osigma_{n,\varsigma}\delta\left(\frac{1}{\hbar}(E(\vec{k'})-E(\vec{k}))+\varsigma\omega_0\right)\\
    &\qquad\qquad\qquad\qquad\qquad\qquad\qquad\qquad\qquad\qquad\qquad\qquad\ve_{n, -\varsigma}\cdot\left(  \int d^3{r} \, e^{-i\vec{q}\cdot \vec{r}}    \frac{\vec{r}}{r^3}  \times \vec{k} \right) \mathcal{I}(\vec{q})|s\ra \,, 
    \end{aligned}
\end{equation}
where $E(\vec{k}) = \sqrt{(mc^2)^2 + \hbar^2 |\vec{k}|^2 c^2}= m_{\rm e}\gamma(|\hbar \vec{k}|)c^2$ is the relativistic energy, $\vec{r}=\vec{r}_{\rm e}-\vec{r}_{\rm s}$, and we define the momentum kick $\vec{q}:=\vec{k'}-\vec{k}$, and $\varsigma = \{-1, 0, 1\}$ corresponds to $\hbar \omega_0$, $0$, and $-\hbar \omega_0$ energy transfer from the sample spin to the beam electron, respectively. 
In the second line, we write the momentum wave functions in position representation, with the factor $(2\pi)^{-3}$ arising from the normalization convention of the Fourier transform between position and momentum space. In the third line, the temporal integration yields 
a delta function corresponding to energy conservation.
In addition, we apply the transformation of coordinates $\vec{r}=\vec{r}_{\rm e}-\vec{r}_{\rm s}$ to split up the integrals, where the new integral over $\vec{r}$ is applied to the interaction, and the integral over $\vec{r}_s$ is a Fourier transform of the  probability density of the spin position leading to 
\begin{equation}\label{eq:FT_spindens}
    \mathcal{I}(\vec{q}):=\int d^3r_{\rm s} |\psi_{\rm s}(\vec{r}_{\rm s})|^2e   ^{-i\vec{q}\cdot \vec{r}_{\rm s}}.
\end{equation}
In particular, we model the delocalization of our electron-spin as a hydrogen-like atom. Then we obtain the following  probability density of the spin position and its Fourier transform
\begin{equation}
    \begin{aligned}
|\psi_{\rm s}(\vec{r}_{\rm s})|^2=&\frac{1}{\pi a_o^3} e^{-2r_{\rm s}/a_0},\\
\mathcal{I}_{\rm es}(q)=&4\pi \int_0^\infty r_s  dr_s |\psi({r}_{\rm s})|^2\frac{\sin(q r_{\rm s})}{q}
= \frac{4}{a_0^3q}\int_0^\infty r_{\rm s} dr_{\rm s} e^{-2r_{\rm s}/a_0} \sin(q r_{\rm s})
= \frac{16}{(4+a_0^2q^2)^2},
    \end{aligned}
\end{equation}
where we take advantage of the isotropy of this density, and we write only the dependence on the magnitude of the wavevector $\vec{q}$.

From the argument of the delta functions in Eq.~\eqref{eq:scatt_matrix_momentum_origin_v2}, 
we solve explicitly 
for the longitudinal momentum component $k_z$ of the initial wave function corresponding to the final momentum $\vec{k}'$:
\begin{equation}\label{eq_kzsol_general}
    \begin{aligned}
        \quad k^{\, {\rm sol}}_{z, \varsigma} &:= \frac{1}{\hbar c} \sqrt{(E(\vec{k}') +  \varsigma \, \hbar\omega_0)^2 -\hbar^2k^{2}_\perp c^2-(mc^2)^2}, 
    \end{aligned}
\end{equation}
where $k_\perp$ is the magnitude of the transversal component of the wavevector $\vec{k}$. 
Meanwhile, the integral with respect to the relative position, 
 on which the interaction depends, 
corresponds to:
\begin{equation}\label{eq:integral_interaction}
    \begin{aligned}
\int d^3{r} \, e^{-i\vec{q}\cdot \vec{r}}    \frac{\vec{r}}{r^3}=&-\int d^3{r} \, e^{-i\vec{q}\cdot \vec{r}}  \vec{\nabla}\left(\frac{1}{r}\right)
=\int d^3{r} \, \vec{\nabla}(e^{-i\vec{q}\cdot \vec{r}})  \frac{1}{r}= -i\vec{q}\int d^3r (e^{-i\vec{q}\cdot \vec{r}})  \frac{1}{r}
=-4\pi i \frac{\vec{q}}{q^2}.
    \end{aligned}
\end{equation}

By inserting Eqs.~\eqref{eq:FT_spindens} and~\eqref{eq:integral_interaction}, along with the energy conservation condition, the scattering matrix can be expressed in $\vec{k}$-space as the product of three distinct functions:
\begin{subequations}\label{eq:smatrix_momentum_planewave}
    \begin{align}
        S^{(1)}_{\vec{k}',s';\vec{k},s} \approx{}& \langle s' | s \rangle \, \delta(\vec{k}' - \vec{k}) 
        + \langle s' | \frac{r_{\rm e}}{2\pi} \vec{\Sigma}(\vec{k}', \vec{k}) \cdot \vec{\mathcal{M}}(\vec{k}', \vec{k}) \, \mathcal{I}_{\rm es}(|\vec{k}' - \vec{k}|) | s \rangle\,, \\
        \vec{\Sigma}(\vec{k}', \vec{k}) :=& \sum_{\varsigma=-1}^1 2^{|\varsigma|/2}\osigma_{n,\varsigma} \ve_{n,-\varsigma} \frac{\delta(k_z - k_{z,\varsigma}^{\,{\rm sol}})}{ k_{z,\varsigma}^{{\rm sol}}}\,,
         \\
        \vec{\mathcal{M}}(\vec{k}', \vec{k}) :=& \frac{\vec{k}' \times \vec{k}}{|\vec{k}' - \vec{k}|^2}\,,
    \end{align}
\end{subequations}
where we used $\mu=-\mu_B$ for the electron spin and $r_{\rm e} = \frac{e \mu_0 \mu_{\rm B}}{2\pi \hbar} \approx 2.82\cdot 10^{-15}\,\text{m}$ is the classical electron radius.
Here, $\vec{\Sigma}(\vec{k}', \vec{k})$ encodes the energy conservation constraints,
and $\vec{\mathcal{M}}(\vec{k}', \vec{k})$ corresponds to the interaction in momentum representation.
Then, S-matrix elements in Eq.~\eqref{eq:S_ks_ins} are
\begin{equation}\label{eq:scatt_momentum_azi_smearing}
    \begin{aligned}
S^{(1)}_{\vec{k}', s'; {\rm in}, s} =&
        \delta_{s', s}\, \varphi_{{\rm in}} (\vec{k}') - \frac{r_{\rm e}}{2\pi} \sum_{\varsigma = -1}^{1} 2^{|\varsigma|/2} \, \tilde{\vec{\ell}}(\vec{k}'; \varsigma) 
        \cdot \ve_{n,-\varsigma}
        \langle s' | \osigma_{n,\varsigma}    | s \rangle \\
        =& \delta_{s',s} \beta_{1}(\vec{k}') +  \sum_{\varsigma = -1}^{1} \beta_{n,\varsigma}(\vec{k}') \langle s'| \osigma_{n,\varsigma} |s \rangle 
    \end{aligned}
\end{equation}
where the $\beta$-amplitudes are defined as
\begin{subequations}\label{eq:beta_mom_smear_n}
    \begin{align}
        \beta_{1}(\vec{k})&= \varphi_{{\rm in}} (\vec{k})\,\\
        \beta_{n,\varsigma}(\vec{k}) 
        &=-2^{|\varsigma|/2}\frac{r_{\rm e}}{2\pi}  {\vec{\ell}}(\vec{k};\varsigma) \cdot \ve_{n,-\varsigma} \,.
    \end{align}
\end{subequations}
We reorganize the vectorial products to define the following vector
\begin{equation}\label{eq:elldef}
    \begin{aligned}
    {\vec{\ell}}(\vec{k}'; \varsigma)=& \left.\int d^3k \, \varphi_{{\rm in}} (\vec{k}) \, \frac{ \delta(k_z - k_{z,\varsigma}^{\,{\rm sol}})}{k_{z,\varsigma}^{\, {\rm sol}}}\frac{\vec{k}'\times(\vec{k}'-\vec{k})}{|\vec{k}'-\vec{k}|^2}\mathcal{I}_{\rm es}(|\vec{k}' - \vec{k}|)\right|_{ k_z = k_{z,\varsigma}^{{\rm sol}}}\\
        =& \int d^2k_\perp \, \left.
        \frac{1}{k_{z,\varsigma}^{ \, {\rm sol}}}\varphi_{{\rm in}} (\vec{k})
        \frac{\vec{k}'\times(\vec{k}'-\vec{k})}{|\vec{k}'-\vec{k}|^2}
        \mathcal{I}_{\rm es}(|\vec{k}' - \vec{k}|) 
        \right|_{k_z = k_{z,\varsigma}^{{\rm sol}}} \\
        =&  
        \int d^2q_\perp \, \left.
        \frac{1}{k_{z,\varsigma}^{ \, {\rm sol}}}\varphi_{{\rm in}}((\vec{k}'_\perp-\vec{q}_\perp,k_{z,\varsigma}^{{\rm sol}})) 
        \frac{\vec{k}'\times\vec{q}}{|\vec{q}|^2}
        \mathcal{I}_{\rm es}(|\vec{q}|) 
        \right|_{q_z = q_{z,\varsigma}^{{\rm sol}}}\,,
    \end{aligned}
\end{equation}
where $\vec{q}_\perp = \vec{k}'_\perp - \vec{k}_\perp$, and the longitudinal component $q_{z,\varsigma}^{{\rm sol}} = k'_{z} - k_{z,\varsigma}^{\rm sol}$.
In this vector, the interaction in momentum representation is integrated
together with the incident wavefunction and the Fourier transform of the probability density of the spin position
under the energy-conservation condition. We can handle the azimuthal integral in Eq.~\eqref{eq:elldef} by exploiting the azimuthal symmetry of the initial wave function in the transverse plane;
defining $\bar\varphi_{{\rm in}} (k_\perp, k_z):= \varphi_{{\rm in}} (\vec{k})$. 
To this end, we introduce polar coordinates $\vec{q}_\perp = q_\perp (\cos\phi_q, \sin\phi_q, 0)$ and $\vec{k}'_\perp = k'_\perp (\cos\phi'_k, \sin\phi'_k, 0)$ and define $\vec{{\rm u}}(\phi):=(\cos\phi, \sin\phi, 0)$ and $\vec{{\rm u}}^*(\phi):=(-\sin\phi, \cos\phi, 0)$.
The azimuthal dependence of the integrand in Eq.~\eqref{eq:elldef} arises both through $\vec{q}_\perp$ itself and through the scalar product $\vec{k}'_\perp \cdot \vec{q}_\perp=k'_\perp q_\perp \cos(\phi_q-\phi'_k)$, which enters the magnitude of the transverse momentum ${k}_\perp=|\vec{k}'_\perp-\vec{q}_\perp|=(k^{'~2}_\perp+q_\perp^2-2\vec{k}'_\perp\cdot\vec{q}_\perp)^{1/2}$. The latter dependence appears within the incident wavefunction $\bar\varphi_{{\rm in}}(|\vec{k}'_\perp-\vec{q}_\perp|,k_{z,\varsigma}^{{\rm sol}})$ and in the longitudinal components $k_{z,\varsigma}^{\mathrm{sol}}$ and $q_{z,\varsigma}^{\mathrm{sol}}$ which are therefore all even functions of $\phi_q-\phi'_k$. 
 Taking into account that that $\vec{k}'_\perp \times \ve_z = -k'_\perp\vec{{\rm u}}^*(\phi'_k)$ and $\vec{k}'_\perp\times  \vec{q}_\perp = k'_\perp q_\perp (0,0,\sin(\phi_q-\phi'_k))$ and using that integrals of anti-symmetric functions over a symmetric interval vanish, we find
\begin{equation}\label{eq:ell_vect_split}
\begin{aligned}
 {\vec{\ell}}(\vec{k}';\varsigma)=&
         \int_0^{\infty} dq_\perp \, q_\perp \int_0^{2\pi} d\phi_q\, \left.
        \frac{1}{k_{z,\varsigma}^{ {\rm sol}}}\bar\varphi_{{\rm in}}(k_\perp,k_{z,\varsigma}^{{\rm sol}}) 
        \frac{q_\perp \vec{k}'_\perp\times  \vec{{\rm u}}(\phi_q)  + q_z\vec{k}'_\perp\times  \ve_z}{|\vec{q}|^2}
        \mathcal{I}_{\rm es}(|\vec{q}|) 
        \right|_{q_z = q_{z,\varsigma}^{{\rm sol}}}\\
    &+ k_z'\int_0^{\infty} dq_\perp \, q_\perp \int_0^{2\pi} d\phi_q\,  \left.\frac{1}{k_{z,\varsigma}^{ {\rm sol}}}
        \bar\varphi_{{\rm in}}(k_\perp,k_{z,\varsigma}^{{\rm sol}})   \frac{q_\perp \vec{{\rm u}}^{*}(\phi_q)}{|\vec{q}|^2}
        \mathcal{I}_{\rm es}(|\vec{q}|) 
        \right|_{q_z = q_{z,\varsigma}^{{\rm sol}}} \\
    =& 2\sqrt{2\pi}\ell(k'_\perp,k'_z;\varsigma)\vec{{\rm u}}^*(\phi'_k)\,,
   \end{aligned}
\end{equation}
where
\begin{subequations}\label{eq:ell_vect_magnitude_general}
    \begin{align}
\ell(k'_\perp,k'_z;\varsigma) =& \ell_\perp(k'_\perp,k'_z;\varsigma)+\ell_\|(k'_\perp,k'_z;\varsigma),\\[4pt]
\ell_\perp(k'_\perp,k'_z;\varsigma)=& \int_0^\infty dq_\perp\, \frac{ q_\perp^2}{2\sqrt{2\pi}}\int_0^{2\pi}\!d\phi_q\, \cos(\phi_q-\phi'_k)\,\bar\varphi_{{\rm in}}(k_\perp,k_{z,\varsigma}^{{\rm sol}}) \,\left.
         \frac{k'_z}{k_{z,\varsigma}^{ {\rm sol}}}\frac{1}{|\vec{q}|^2}
        \mathcal{I}_{\rm es}(|\vec{q}|) 
        \right|_{q_z = q_{z,\varsigma}^{{\rm sol}}},\\[6pt]
\ell_\|(k'_\perp,k'_z; \varsigma)=&- \int_0^\infty dq_\perp\, \frac{ q_\perp k'_\perp}{2\sqrt{2\pi}} 
        \int_0^{2\pi}\!d\phi_q\, \bar\varphi_{{\rm in}}(k_\perp,k_{z,\varsigma}^{{\rm sol}}) \left.  \frac{q_z}{k_{z,\varsigma}^{{\rm sol}}} \frac{1}{|\vec{q}|^2}
        \mathcal{I}_{\rm es}(|\vec{q}|) 
        \right|_{q_z = q_{z,\varsigma}^{{\rm sol}}}\,,
    \end{align}
\end{subequations}
where $\ell_\perp$ and $\ell_\|$ arise from the first and second integral in Eq.~\eqref{eq:ell_vect_split}, respectively. They contain the contributions of the transverse and longitudinal components of the magnetic field generated by the spin, respectively, applied to an azimuthally symmetric incident electron beam wavefunction.

The above expressions retain the full structure of the electron wavefunction, providing a complete description of the interaction without invoking approximations besides the first order expansion of the interaction, that is,
\begin{equation}\label{eq:scatt_state}
 \varphi_{\rm sc}(\vec{k},s') = S^{(1)}_{\vec{k}, s'; {\rm in}, s} \,.
\end{equation}
The corresponding density matrix is
\begin{equation}\label{eq:scatt_state_density}
 \rho_{{\rm sc}\,\vec{k}',s';\vec{k}'',s''} = S^{(1)}_{\vec{k}', s'; {\rm in}, s} S^{(1)~*}_{\vec{k}'', s''; {\rm in}, s}\,,
\end{equation}
and the reduced density matrix of the electron
\begin{equation}\label{eq:scatt_state_density_electron}
\begin{aligned}
 \rho_{{\rm sc,e}\,\vec{k};\vec{k}'} &= \sum_{s'} S^{(1)}_{\vec{k}, s'; {\rm in}, s} S^{(1)~*}_{\vec{k}', s'; {\rm in}, s} \\
 &=  \,\beta_{1}(\vec{k})\beta_{1}(\vec{k}')^* +  \left(\beta_{n,0}(\vec{k}) \beta_{1}(\vec{k}')^* + \beta_{n,0}(\vec{k}')^* \beta_{1}(\vec{k})\right) \langle \hat{\sigma}_{n,0} \rangle\\
 &\quad +  \frac{1}{2}\left(\beta_{1}(\vec{k}')^*(\beta_{n,+}(\vec{k}) + \beta_{n,-}(\vec{k})) + \beta_{1}(\vec{k})(\beta_{n,+}(\vec{k}')^* + \beta_{n,-}(\vec{k}')^*) \right) \langle \hat{\sigma}_{m,0} \rangle \\
 & \quad + \frac{i}{2} \left(\beta_{1}(\vec{k}')^*(\beta_{n,+}(\vec{k}) - \beta_{n,-}(\vec{k})) - \beta_{1}(\vec{k})(\beta_{n,+}(\vec{k}')^* - \beta_{n,-}(\vec{k}')^*)\right) \langle \hat{\sigma}_{l,0} \rangle  \,,
\end{aligned}
\end{equation}
where the second line is given to first order in the interaction strength, and we have used the decomposition $\osigma_{n, \pm}=\frac{1}{2}(\osigma_{m,0}\pm i \osigma_{l,0})$.
To extract probabilities over momentum space, we only need the diagonal elements of the reduced density matrix, which can be represented as
\begin{equation}\label{eq:rho_sc_e_momentum}
    \begin{aligned}
    \rho_{{\rm sc,e}\, \vec{k};\vec{k}} &:= \sum_{s'} S^{(1)}_{\vec{k}, s'; {\rm in}, s} S^{(1)~*}_{\vec{k}', s'; {\rm in}, s} \\
    &=  \,|\beta_1|^2  + \Re{\left( \beta_1^*(\beta_{n,+} + \beta_{n,-}) \right)} \langle \hat{\sigma}_{m,0} \rangle  - \Im{\left( \beta_1^*(\beta_{n,+} - \beta_{n,-}) \right)} \langle \hat{\sigma}_{l,0} \rangle+ 2 \Re\left({\beta_{n,0}\beta_1^*} \right)\langle \hat{\sigma}_{n,0} \rangle\,,
    \end{aligned}
\end{equation}
in which we dropped the arguments of the $\beta$-amplitudes in the second line to simplify the presentation.
To predict images in diffraction mode, we are interested in the probability distribution of angles averaging over the electron energy
\begin{equation}
    P(\phi_k,\vartheta)= \sin\vartheta \int dk\, k^2 \,\rho_{{\rm sc,e}\,\vec{k}(\phi_k,\vartheta);\vec{k}(\phi_k,\vartheta)},
\end{equation}
where $\vec{k}(\phi_k,\vartheta)=k(\cos\phi_k\sin\vartheta,\sin\phi_k\sin\vartheta,\cos\vartheta)$ and $\mathcal{N}_{\rm diff}$ renormalizes the angular probability distribution. Then, the volume element becomes $\sin\vartheta \,d\vartheta d\phi_k$ and the probability distribution in the transversal plane is
\begin{equation}\label{eq:prob_diff_def}
    P_{\rm diff}(\vartheta(\cos\phi_k,\sin\phi_k)):= \mathcal{N}_{\rm diff}\int dk\, k^2 \,\rho_{{\rm sc,e}\,\vec{k}(\phi_k,\vartheta);\vec{k}(\phi_k,\vartheta)}\,.
\end{equation}

\subsection{Approximations in paraxial regime}

The incident electron beam is modeled as a Gaussian wavepacket in momentum space, ensuring a normalizable final state and a well-defined probability density over the transverse momenta. To introduce suitable approximations in the paraxial regime for the incident electron wavepacket, we decompose it into transverse and longitudinal components as $\varphi_{\rm in}(\vec{k}) = \varphi_{{\rm in},\perp}(\vec{k}_\perp)\,\varphi_{{\rm in},\|}(k_z)$ with
\begin{subequations}\label{eq:incident_Gaussian_wavefunction}\allowdisplaybreaks
    \begin{align}
\varphi_{{\rm in},\perp}(\vec{k}_\perp) &= 
    \frac{1}{\sqrt{2\pi}\,\Delta k_\perp}\,
    \exp\!\left[-\frac{|\vec{k}_\perp|^2}{4\Delta k_\perp^2}\right],\\[4pt]
\varphi_{{\rm in},\|}(k_z) &= 
    \frac{1}{(2\pi \Delta k_z^2)^{1/4}}\,
    \exp\!\left[-\frac{(k_z - k_{z,0})^2}{4\Delta k_z^2}\right]
    e^{i k_z z_p},
    \end{align}
\end{subequations}
where the transverse and longitudinal components have different variances, $\Delta k_\perp$ and $\Delta k_z$, both much smaller than the central longitudinal momentum, i.e., $\Delta k_z, \Delta k_\perp \ll k_{z,0}$. The parameter $z_p$ represents a defocus distance of the electron probe with respect of the sample plane.

Under the paraxial approximation, the longitudinal momentum distribution can be regarded as sharply peaked around $k_{z,0}$. In this work, we consider electron energies on the order of $200~\mathrm{keV}$, corresponding to $k_{z,0} \sim 10^{12}\,\mathrm{m}^{-1}$. Assuming a wavefunction energy spread of approximately $2~\mathrm{eV}$, we obtain $\Delta k_z \sim 10^{6}\,\mathrm{m}^{-1} \ll k_{z,0}$. 
For small angles $\vartheta \ll 1$ and given the narrow energy spread of the incident wavefunction, the $k$-integral in Eq.~\eqref{eq:prob_diff_def} can be rewritten as a longitudinal momentum $k_z$ integral as
\begin{equation}\label{eq:prob_diff_def_approx}
P_{\rm diff}(\vartheta(\cos\phi_k,\sin\phi_k))\approx \mathcal{N}_{\rm diff}\int_{-\infty}^\infty dk_z\, k_{z}^2 \,\rho_{{\rm sc,e}\,\vec{k}(\phi_k,\vartheta);\vec{k}(\phi_k,\vartheta)}\,.   
\end{equation}
Here we have used the approximation $\vec{k}(\phi_k,\vartheta)\approx k_z(\vartheta\vu(\phi_k)+\ve_z)$.
As a result, the angular probability density can be approximately represented as an angular probability distribution over the two-dimensional plane spanned by $\vartheta(\cos\phi_k,\sin\phi_k)$, where $\vartheta$ acts as a dimensionless radial coordinate and the volume element becomes $\vartheta\, d\vartheta\, d\phi_k$. Using the reduced density matrix elements in Eq.~\eqref{eq:rho_sc_e_momentum}, we define the following components of the probability distribution:
\begin{subequations}\label{eq:Pis_momentum_2level}
    \begin{align}
{\Pi}_0(\vartheta(\cos\phi_k,\sin\phi_k)) :=& \,\mathcal{N}_{\rm diff}\int_{-\infty}^\infty dk_z\,k_z^2|\beta_1(k_z(\vartheta\vu(\phi_k)+\ve_z))|^2\,, \\
\begin{split}
    {\Pi}_{n, \varsigma}(\vartheta(\cos\phi_k,\sin\phi_k)) :=&\, \mathcal{N}_{\rm diff}\int_{-\infty}^\infty dk_z\,k_z^2\beta_{n,\varsigma}(k_z(\vartheta\vu(\phi_k)+\ve_z))\beta_1(k_z(\vartheta\vu(\phi_k)+\ve_z))^*,
\end{split}
    \end{align}
\end{subequations}
which correspond to the non-scattered and first-order scattered contributions to the angular probability, respectively. Then, the probability distribution can be written as
\begin{equation}
    \begin{aligned}
&P_{\rm diff} \approx \Pi_0+\Re(\Pi_{n, +1}+\Pi_{n, -1})\expval{\osigma_{m,0}}-\Im(\Pi_{n, +1}-\Pi_{n, -1})\expval{\osigma_{l,0}}+2\Re(\Pi_{n, 0})\expval{\osigma_{n,0}}\,,
    \end{aligned}
\end{equation}
where we have dropped the explicit dependence on $\vartheta(\cos\phi_k,\sin\phi_k)$ for brevity.

The non-scattered term, using the incident wavefunction defined in Eq.~\eqref{eq:incident_Gaussian_wavefunction}, is given explicitly by
\begin{equation}\label{eq:Pi0_momentum_2level}
    \begin{aligned}
{\Pi}_0(\vartheta(\cos\phi_k,\sin\phi_k)) \approx& \,\mathcal{N}_{\rm diff}\int_{-\infty}^\infty d\kappa_z\,\frac{(k_{z,0}+\kappa_z)^2}{2\pi \Delta k_\perp^2\sqrt{2\pi \Delta k_z^2}}\exp\left[-\frac{(k_{z,0}+\kappa_z)^2\vartheta^2}{2\Delta k_\perp^2}\right]\exp\left[-\frac{\kappa_z^2}{2\Delta k_z^2}\right]\\
= & \mathcal{N}_{\rm diff}\frac{ \Delta k_\perp  \left(k_{z,0}^2 \Delta k_\perp^2 + \vartheta ^2 \Delta k_z^4+\Delta k_\perp^2 \Delta k_z^2\right)}{2 \pi \left(\vartheta ^2 \Delta k_z^2+\Delta k_\perp^2\right)^{5/2}}  \exp\left[-\frac{ k_{z,0}^2 \vartheta^2}{2 \left(\vartheta^2 \Delta k_z^2+\Delta k_\perp^2\right)}\right]  \\
\approx& \mathcal{N}_{\rm diff}\frac{k_{z,0}^2}{2\pi \Delta k_\perp^2}\exp\left[-\frac{k_{z,0}^2\vartheta^2}{2\Delta k_\perp^2}\right]= \mathcal{N}_{\rm diff}{k_{z,0}^2}|\varphi_{\rm in, \perp}(k_{z,0}\vartheta)|^2,
    \end{aligned}
\end{equation}
where we have introduced $\kappa_z = k_z - k_{z,0}$ and we have used that $k_{z,0}\gg \Delta k_\perp, \Delta k_z$, and $\Delta k_\perp\gg \vartheta \Delta k_z \sim \Delta k_\perp \Delta k_z/k_{z,0} $ to arrive at the third line.
Furthermore, the paraxial and narrow longitudinal momentum distribution lead to the approximated expression for the incident electron energy given by
\begin{equation}\label{eq:energy_paraxial}
    \begin{aligned}
        E(\vec{k}')
        &= \sqrt{E(k_{z,0})^2 
            + \hbar^2 (k_z^2 - k_{z,0}^2)c^2 
            + \hbar^2 k_\perp^2 c^2} \approx E_0 
            + \frac{\hbar^2}{\gamma_0m_{\rm e}}
            \left(\kappa_z k_{z,0} + \frac{1}{2 \gamma_0^2}\kappa_z^2 + \frac{1}{2}k_\perp^{'~2}\right)\,,
    \end{aligned}
\end{equation}
where $E_0=E(k_{z,0})$ and $\gamma_0=\gamma(\hbar k_{z,0})$ refer to the mean energy and Lorentz factor respectively. 
Following the same procedure as in Eq.~\eqref{eq:energy_paraxial}, we expand $k^{\mathrm{sol}}_{z,\varsigma}$ around a small longitudinal deviation $\kappa_z$, which is defined in this case as $\kappa_z=k'_z-k_{z,0}$, obtaining
\begin{equation}\label{eq:kzsol_paraxial}
    \begin{aligned}
        k^{\mathrm{sol}}_{z,\varsigma}
        &= \sqrt{(k_{z,0}+\kappa_z)^2
            + 2\varsigma\frac{\omega_0}{\hbar c^2}E(\vec{k'})
            + \varsigma^2\frac{\omega_0^2}{c^2}
            - (k_\perp^2 - k_\perp^{'2})} \\[4pt]
        &\approx \sqrt{k_{z,0}^2 + 2k_{z,0}\kappa_z + \kappa_z^2 
            + 2\varsigma\frac{\omega_0}{\hbar c^2}\!\left(
                E_0 
            + \frac{\hbar^2}{\gamma_0m_{\rm e}}
            \left(\kappa_z k_{z,0} + \frac{1}{2 \gamma_0^2}\kappa_z^2 + \frac{1}{2}k_\perp^{'~2}\right)
            \right)
            + \varsigma^2\frac{\omega_0^2}{c^2}
            - (k_\perp^2 - k_\perp^{'2})} \\[4pt]
        &\approx k_{z,0} + \kappa_z 
            + \frac{k_\perp^{'2}-k_\perp^2}{2k_{z,0}}
            + \varsigma\,\frac{\delta k^2}{k_{z,0}},
    \end{aligned}
\end{equation}
where we define 
$\delta k = \sqrt{\gamma_0\, m_{\rm e} \omega_0 / \hbar}$,
which characterizes the momentum transfer associated with the energy exchange between the free electron and the sample.
In the last step, we retain only the leading-order terms based on a scale analysis of the square-root expansion.  
Each contribution inside the square root in the second line is compared to the dominant term $k_{z,0}^2$.  
We find that the relative magnitudes are:
$\kappa_z/k_{z,0}\sim10^{-6}$;
$\kappa_z^2/k_{z,0}^2\sim10^{-12}$,
$E_0\omega_0/(\hbar c^2 k_{z,0}^2)\sim10^{-9}$ for 
$\omega_0/c = 2\pi\times  28\,\mathrm{GHz/T} \times  1.8\,\mathrm{T}/c \sim 10^{3}\,\mathrm{m}^{-1}$; 
$\hbar\omega_0 \kappa_z/(E_0 k_{z,0})\sim10^{-15}$; 
$\hbar\omega_0 \kappa_z^2/(E_0 k_{z,0}^2)\sim10^{-21}$, $\omega_0^2/(c^2k_{z,0}^2)\sim10^{-18}$; $\hbar\omega_0 k_\perp^{'~2}/(E_0 k^2_{z,0})\sim10^{-13}$, 
{assuming small angle deflections where $k'_\perp \lesssim  10^{10}\,\mathrm{m}^{-1}$}; and $|k_\perp^2 - k_\perp^{'2}|/k_{z,0}^2 \gtrsim 10^{-10}$.  
Therefore, only the terms that are at least of the order of $10^{-12}$ are kept in the final expression.

{Since the non-scattered term dominates ($\Pi_0 \gg \Pi_{n, \varsigma}$), we determine $\mathcal{N}_{\rm diff}$ by requiring $\Pi_0$ to be normalized to unity. Integrating the simplified Gaussian form in Eq.~\eqref{eq:Pi0_momentum_2level} over the detection angles yields 
\begin{equation}
    \int \vartheta d\vartheta d\phi_k \, \Pi_0(\vartheta(\cos\phi_k,\sin\phi_k)) \approx \mathcal{N}_{\rm diff} \int d^2k_\perp |\varphi_{\rm in, \perp}(k_\perp)|^2 = \mathcal{N}_{\rm diff}\,.
\end{equation}
Consequently, we set $\mathcal{N}_{\rm diff} = 1$.}

By employing the expansion in energy and longitudinal wavenumber, the first-order interference terms in the angular probability distribution can be written as
\begin{equation}\label{eq:Pi_nvarsig_momentum_2level_pre}\allowdisplaybreaks
    \begin{aligned}
{\Pi}_{n, \varsigma}(\vartheta(\cos\phi_k,\sin\phi_k)) =& \,-2^{|\varsigma|/2}\sqrt{\frac{2}{\pi}}r_{\rm e}\mathcal{N}_{\rm diff}\vu^*(\phi_k)\cdot \ve_{n, -\varsigma}\int_{-\infty}^\infty dk_z\,k_z^2\ell(k_z\vartheta, k_z;\varsigma)\varphi_{\rm in, \perp}(k_z\vartheta)^*\varphi_{\rm in, \|}(k_z)^*\\
=& \,-2^{|\varsigma|/2}\sqrt{\frac{2}{\pi}}r_{\rm e}\mathcal{N}_{\rm diff}\vu^*(\phi_k)\cdot \ve_{n, -\varsigma}\int_{-\infty}^\infty dk_z\,k_z^2\int_0^\infty dq_\perp\, \frac{ q_\perp}{2\sqrt{2\pi}}\int_0^{2\pi}\!d\phi_q\,\left.
        \frac{1}{|\vec{q}|^2}
        \mathcal{I}_{\rm es}(|\vec{q}|) 
        \right|_{q_z = q_{z,\varsigma}^{{\rm sol}}}\\
&\qquad \varphi_{{\rm in, \perp}}(|{k}_z\vartheta\vu(\phi_k)-\vec{q}_\perp|)\varphi_{{\rm in, \|}}(k_{z,\varsigma}^{{\rm sol}}) \,\varphi_{\rm in, \perp}({k}_{z}\vartheta)^*\varphi_{\rm in, \|}(k_z)^* \frac{k_z}{k_{z, \varsigma}^{\rm sol}}\left(q_\perp\cos(\phi_q-\phi_k)-\vartheta q_{z,\varsigma}^{{\rm sol}}\right)\\
\approx& \,-2^{|\varsigma|/2}\sqrt{\frac{2}{\pi}}r_{\rm e}\mathcal{N}_{\rm diff}\vu^*(\phi_k)\cdot \ve_{n, -\varsigma}\int_0^\infty dq_\perp\, \frac{ q_\perp}{2\sqrt{2\pi}}\int_0^{2\pi}\!d\phi_q\,\left.
        \frac{1}{|\vec{q}|^2}
        \mathcal{I}_{\rm es}(|\vec{q}|) 
        \right|_{q_z = q_{z,\varsigma}^{{\rm sol, 0}}}\\
&\qquad \qquad \qquad \varphi_{{\rm in, \perp}}(|{k}_{z,0}\vartheta\vu(\phi_k)-\vec{q}_\perp|)\varphi_{\rm in, \perp}({k}_{z,0}\vartheta)^*k^2_{z,0} \frac{k_{z,0}}{k_{z, \varsigma}^{\rm sol,0}}\left(q_\perp\cos(\phi_q-\phi_k)-\vartheta q_{z,\varsigma}^{{\rm sol,0}}\right)\\
&\qquad \qquad \qquad \underbrace{\left.\int_{-\infty}^\infty dk_z\,\varphi_{{\rm in, \|}}(k_{z,\varsigma}^{{\rm sol}}) \,\varphi_{\rm in, \|}(k_z)^*\right|_{ k'_\perp=k_{z,0}\vartheta}}_{:=D_{\varsigma}(\vartheta, q_\perp, \phi_q-\phi_k)}\,.
    \end{aligned}
\end{equation}
In the third line, the dominant contributions in the $k_z$-integral arise from the narrow Gaussian distributions in the longitudinal momentum. Therefore, all other factors in the integrand are evaluated at  $k_z = k_{z,0}$, as the corresponding corrections are proportional to $\Delta k_z/k_{z,0} \sim 10^{-6}$. This results in the approximated longitudinal wavenumbers:
\begin{equation}\label{eq:kz_qz_sol_approx}
    k_{z,\varsigma}^{{\rm sol},0}
    = k_{z,0} - q_{z,\varsigma}^{{\rm sol},0},
    \qquad
    q_{z,\varsigma}^{{\rm sol},0}
    = -\varsigma\,\frac{\delta k^2}{k_{z,0}}+\frac{q_\perp^2}{2k_{z,0}}
    -\vartheta q_\perp \cos(\phi_q-\phi_k),
\end{equation}
where the momentum transfer $q_{z,\varsigma}^{{\rm sol},0}$ is composed of a term induced by the energy change of the electron and two terms representing the change of longitudinal momentum corresponding to the deflection of the electron.

As expected, the last term in Eq.~\eqref{eq:Pi_nvarsig_momentum_2level_pre} represents the dependence on the overlap between the longitudinal components of the scattered and unscattered wavefunctions, which is given by
\begin{equation}\label{eq:Dvarsigma_overlap}
    \begin{aligned}
D_{\varsigma}(k_\perp, q_\perp, \phi_q-\phi_k)=&\int_{-\infty}^\infty \frac{dk_z}{\sqrt{2\pi}\Delta k_z}\,
    \exp\!\left[-\frac{(k_{z,\varsigma}^{\rm sol,0}-k_{z,0})^2}{4\Delta k_z^2}\right]
    e^{-i q_{z,\varsigma}^{\rm sol, 0} z_p}\exp\!\left[-\frac{(k_z-k_{z,0})^2}{4\Delta k_z^2}\right]\\
    =&\,e^{-i q_{z,\varsigma}^{\rm sol, 0} z_p}\exp\left[-\frac{(q^{\rm sol,0}_{z,\varsigma})^2}{8\Delta k_z^2}\right].
    \end{aligned}
\end{equation}
where we use initially $k_{z,\varsigma}^{\rm sol}-k_{z,0}\approx k_{z,\varsigma}^{\rm sol,0}-k_{z,0}$.
The magnitude of this overlap depends directly on the longitudinal momentum transfer relative to the width of the initial momentum distribution.
Replacing the overlap and reorganizing the terms, we find
\begin{subequations}\label{eq:Pi_nvarsig_momentum_2level}
    \begin{align}
{\Pi}_{n, \varsigma}(\vartheta(\cos\phi_k,\sin\phi_k)) 
\approx& \,-2^{|\varsigma|/2}\sqrt{\frac{2}{\pi}}r_{\rm e}k_{z,0}^2\mathcal{N}_{\rm diff}\vu^*(\phi_k)\cdot \ve_{n, -\varsigma}\cL_{\varsigma}(\vartheta; z_p)\varphi_{\rm in, \perp}({k}_{z,0}\vartheta)^*\,,\label{eq:Pi_nvarsigma_deflection}\\
\begin{split}
    \cL_{\varsigma}(\vartheta; z_p):=&\int_0^\infty dq_\perp\, \frac{ q_\perp}{2\sqrt{2\pi}}\int_0^{2\pi}\!d \phi_q\,\frac{k_{z,0}}{k_{z, \varsigma}^{\rm sol,0}}\left(q_\perp\cos(\phi_q-\phi_k)-\vartheta q_{z,\varsigma}^{{\rm sol,0}}\right)\varphi_{{\rm in, \perp}}(|{k}_{z,0}\vartheta\vu(\phi_k)-\vec{q}_\perp|)\\
&\qquad\qquad\qquad\qquad\qquad\qquad\qquad\left.e^{-i q_{z,\varsigma}^{\rm sol, 0} z_p}\exp\left[-\frac{(q^{\rm sol,0}_{z,\varsigma})^2}{8\Delta k_z^2}\right]
        \frac{\mathcal{I}_{\rm es}(|\vec{q}|)}{|\vec{q}|^2} 
        \right|_{q_z = q_{z,\varsigma}^{{\rm sol, 0}}}\,.
\end{split}\label{eq:L_nvarsigma_deflection}
    \end{align}
\end{subequations}

Since the interference terms ${\Pi}_{n,\varsigma}$ in Eq.~\eqref{eq:Pi_nvarsigma_deflection} contain the factor $\varphi_{\rm in,\perp}(k_{z,0}\vartheta)$, we expect that the significant part of the probability distribution is concentrated in a region of width $\sim\Delta k_\perp/k_{z,0}$ around $\vartheta=0$. Then, the transverse initial electron wave function in $\cL_{\varsigma}(\vartheta; z_p)$ implies that the relevant region of the $q_\perp$ integration domain is of width $\sim\Delta k_\perp$ and centered at $q_\perp=0$. A  scale analysis of $q_{z,\varsigma}^{\rm sol,0}$ in Eq.~\eqref{eq:kz_qz_sol_approx} shows that $\delta k\sim 10^8~\text{m}^{-1}$ for spin transition frequencies up to a few hundred~GHz. Therefore, we conclude that  $q_\perp\sim \min(\Delta k_\perp, k_a)$ with $k_a$ the scale implied by the Fourier transform of the spin distribution $\mathcal{I}_{\rm es}(|\vec{q}|)$ which may be smaller than $\Delta k_\perp$ and put stronger bounds on $q_\perp$. For example, $k_a$ is of the order of the inverse Bohr radius if the bound electron is an unpaired electron in an s-orbital of an atom. Hence, from Eq.~\eqref{eq:kz_qz_sol_approx}, we conclude that $q_{z,\varsigma}^{\rm sol,0}\sim \max(\min(\Delta k_\perp, k_a)^2/k_{z,0},\min(\Delta k_\perp, k_a)\Delta k_\perp/k_{z,0},\delta k_\perp^2/k_{z,0})$, depending on which contribution to the longitudinal momentum transfer dominates. The longitudinal overlap term approaches unity whenever 
$\Delta k_z>\min(\Delta k_\perp, k_a)^2/(4 k_{z,0})$ and
$\Delta k_z>\delta k_\perp^2/(2\sqrt2 k_{z,0})$. 
This condition is satisfied in our case, where $\Delta k_z\sim 10^{6}~\text{m}^{-1}$ and $\Delta k_\perp\lesssim 10^{8}~\text{m}^{-1}<k_a$ (see Table~\ref{tab:simulation_constants}).  
Consequently, we can write
\begin{equation}\label{eq:L_nvarsigma_deflection_3level}
    \begin{aligned}
\cL_{\varsigma}(\vartheta; z_p)\approx&\int_0^\infty dq_\perp\, \frac{ q_\perp}{2\sqrt{2\pi}}\int_0^{2\pi}\!d\phi_q\,\frac{k_{z,0}}{k_{z, \varsigma}^{\rm sol,0}}\left(q_\perp\cos(\phi_q-\phi_k)-\vartheta q_{z,\varsigma}^{{\rm sol,0}}\right)\varphi_{{\rm in, \perp}}(|{k}_{z,0}\vartheta\vu(\phi_k)-\vec{q}_\perp|)e^{-i q_{z,\varsigma}^{\rm sol, 0} z_p}\left.\frac{\mathcal{I}_{\rm es}(|\vec{q}|) }{|\vec{q}|^2}        
        \right|_{q_z = q_{z,\varsigma}^{{\rm sol, 0}}}\,.
    \end{aligned}
\end{equation}
If the above condition is not fulfilled, the interference term of the incident and scattered wave function may be suppressed, and second-order terms in the scattering matrix become relevant.  

\begin{table}[h!]
    \centering
    \begin{tabular}{c|c|c}
    \hline \hline
    Parameter & Symbol & Value / Range \\
    \hline
    Electron beam kinetic energy & $E_{\rm kin} = \bar E_{\rm kin} + \Delta E$ & $200~\mathrm{keV} \pm 2~\mathrm{eV}$ \\
    Zeeman splitting & $\hbar \omega_0$ & $0.208~\mathrm{meV}$ \\
    Mean longitudinal wavenumber & $k_{z,0}$ & $2.51 \times 10^{12}~\mathrm{m}^{-1}$ \\
    Longitudinal wavenumber spread & $\Delta k_z$ & $1.06 \times 10^{7}~\mathrm{m}^{-1}$ \\
    Transverse wavenumber spread & $\Delta k_\perp$ & $(1.06 \times 10^{7}$--$1.06 \times 10^{9})~\mathrm{m}^{-1}$ \\
    Bohr radius & $a_0$ & $5.29 \times 10^{-11}~\mathrm{m}$ \\
    Spin localization wavenumber & $k_a = 1/a_0$ & $1.89 \times 10^{10}~\mathrm{m}^{-1}$ \\
    Energy transfer wavenumber & $\delta k$ & $6.17 \times 10^{7}~\mathrm{m}^{-1}$ \\
    Resolution limit wavenumber & $k_{\mathrm{max}}$ & $2\pi/(50~\mathrm{pm}) = 1.26 \times 10^{11}~\mathrm{m}^{-1}$ \\
    Defocus distance and $z$ position & $z_d, z$ & $\leq 1.0 \times 10^{-7}~\mathrm{m}$ \\
    \hline \hline
    \end{tabular}
    \caption{Physical parameters and constants used in the simulations.}
    \label{tab:simulation_constants}
\end{table}

Using Eq.~\eqref{eq:L_nvarsigma_deflection_3level}, the above scale analysis implies that $k_{z,0}/k_{z, \varsigma}^{\rm sol,0}\sim 1$ up to a correction of the order of ${q_{z,\varsigma}^{\rm sol,0}}/{k_{z,0}}\sim \min(\Delta k_\perp,k_a)^2/k_{z,0}^2\sim 10^{-8}$. Moreover, for the parameters considered in this article, the transverse momentum transfer satisfies $q_\perp\sim \Delta k_\perp \gg \vartheta q_{z,\varsigma}^{\rm sol,0} \sim \Delta k_\perp q_{z,\varsigma}^{\rm sol,0}/k_{z,0}\sim 10^{-8} \Delta k_\perp$. 
Hence, the term $\vartheta\, q_{z,\varsigma}^{\rm sol,0}$ is negligibly small and can be safely neglected within the paraxial regime.
This indicates that the dominant contribution arises from the transverse component $\ell_\perp$ in Eq.~\eqref{eq:ell_vect_magnitude_general}, which represents the momentum transfer perpendicular to the electron-beam axis. 
Finally, using the relation $q_\perp\sim \Delta k_\perp \gg q^{\rm sol,0}_{z,\varsigma}\sim \Delta k_\perp^2/k_{z,0}$, we can approximate $|\vec{q}|\approx q_\perp$ and rewrite Eq.~\eqref{eq:L_nvarsigma_deflection_3level} as
\begin{equation}\label{eq:L_nvarsigma_deflection_4level}
    \begin{aligned}
\cL_{\varsigma}(\vartheta; z_p)\approx&\int_0^\infty  \frac{dq_\perp}{4\pi \Delta k_\perp}\int_0^{2\pi}\!d\phi\,\cos(\phi)\exp\left[-\frac{{k}^2_{z,0}\vartheta^2+q_\perp^2-2{k}_{z,0}\vartheta q_\perp\cos(\phi)}{4\Delta k_\perp^2}\right]e^{-i \left(-\varsigma\,\frac{\delta k^2}{k_{z,0}}+\frac{q_\perp^2}{2k_{z,0}}
    -\vartheta q_\perp \cos(\phi)\right)z_p}
        \mathcal{I}_{\rm es}(q_\perp)\\
        =&\int_0^\infty \frac{dq_\perp}{4\pi \Delta k_\perp}\int_0^{2\pi}\!d\phi\,\cos(\phi)\exp\left[-\frac{{k}^2_{z,0}\vartheta^2+q_\perp^2}{4\Delta k_\perp^2}\right]e^{i\left(\varsigma \frac{\delta k^2}{k_{z,0}}-\frac{q_\perp^2}{2k_{z,0}}\right)z_p}
        \exp\left[\frac{\vartheta q_\perp k_{z,0}}{2\Delta k_\perp^2}\zeta(z_p)\cos\phi\right]\mathcal{I}_{\rm es}(q_\perp)\\
        =&\int_0^\infty \, \frac{ dq_\perp}{2\Delta k_\perp}\exp\left[-\frac{{k}^2_{z,0}\vartheta^2+q_\perp^2}{4\Delta k_\perp^2}\right]e^{i\left(\varsigma \frac{\delta k^2}{k_{z,0}}-\frac{q_\perp^2}{2k_{z,0}}\right)z_p}
        I_1\left[\frac{\vartheta q_\perp k_{z,0}}{2\Delta k_\perp^2}\zeta(z_p)\right]\mathcal{I}_{\rm es}(q_\perp),
    \end{aligned}
\end{equation}
where we have introduced the integration variable $\phi=\phi_q-\phi_k$ and the complex parameter 
\begin{equation}
    \zeta(z_p)=1+\frac{2i\Delta k_\perp^2 z_p}{k_{z,0}}.
\end{equation}
In the second line, we rearranged the terms to employ the standard identity (valid for complex arguments \(k'_\perp q_\perp \zeta(z_p)\)):
\begin{equation}
    \int_0^{2\pi}\!d\phi\,\cos\phi\,e^{\frac{k'_\perp q_\perp}{2\Delta k_\perp^2}\zeta(z_p)\cos\phi}
=2\pi I_1\!\left(\frac{k'_\perp q_\perp}{2\Delta k_\perp^2}\zeta(z_p)\right),
\end{equation}
where \(I_1\) denotes the modified Bessel function of the first kind.  
In Eq.~\eqref{eq:L_nvarsigma_deflection_4level}, we observe that for defocused incident beam imaging conditions ($z_p\neq0$), the back-propagated phase must be integrated over the transverse momentum transfer. The resulting defocus effect depends not only on the propagation distance $z_p$ but also on the energy exchange between the interacting particles, which can be comparable in magnitude to the deflection contribution.  

For the scope of this work, we focus on the diffraction-imaging regime where the beam is longitudinally focused at the sample plane ($z_p=0$). Under this condition, $\cL_{\varsigma}(\vartheta;0)$ becomes independent of $\varsigma$, and $\tilde{\Pi}_{n,\varsigma}$ is real and $\varsigma$-independent except for $\ve_{n,-\varsigma}$ and a factor $2^{|\varsigma|/2}$. The corresponding real and imaginary parts act directly on these vector components according to
\begin{equation}
    \Re[\ve_{n,-1}+\ve_{n,+1}]=\sqrt{2}\,\ve_{m,0},\quad
\Im[\ve_{n,-1}-\ve_{n,+1}]=-\sqrt{2}\,\ve_{l,0},\quad
\Re[\ve_{n,0}]=\ve_{n,0}.
\end{equation}
Hence, the angular probability takes the form
\begin{equation}\label{eq:barP_diffractionzp0}
    P_{\rm diff}(\vartheta(\cos\phi_k, \sin\phi_k))
    \approx \mathcal{N}_{\rm diff}k_{z,0}^2\!\left(
    |\varphi_{{\rm in},\perp}(k_{z,0}\vartheta)|^2
    - 2 r_{\rm e} \sqrt{\frac{2}{\pi}}\,
    \varphi_{\rm in,\perp}(k_{z,0}\vartheta)\, \cL_{0}(\vartheta; 0)
    \vu^*(\phi'_k) \cdot  \langle \vec{\osigma}\rangle\right),
\end{equation}
where $\langle \vec{\osigma}\rangle=\sum_{j \in \{m, l, n\}}\ve_{j,0}\,
    \langle \hat{\sigma}_{j,0} \rangle$ and we have taken into account that $\varphi_{\rm in,\perp}$ is real.

For numerical evaluation, since the first-order interaction effects are concentrated near the zero-deflection peak, we compute diffraction-mode images within this region, restricting $\vartheta$ to 
$\vartheta \leq 8\,\Delta k_\perp/k_{z,0}$. 
The integration over transverse momentum transfer is limited to 
$q_\perp\in[\max(0,\,\vartheta k_{z,0} - 10\Delta k_\perp),\,\vartheta k_{z,0} + 10\Delta k_\perp]$, 
consistent with the observation that the integrand in $\cL_0$ is negligibly small if $q_\perp$ is not in a region of width of the order of $\Delta k_\perp$ around $\vartheta k_{z,0}$.

An analytical result can be obtained by taking into account that $\Delta k_\perp\ll (2a_0)^{-1}$, for the simulations performed for this work and approximating $\mathcal{I}_{\rm es}(q_\perp) \approx \mathcal{I}_{\rm es}(0) = 1$ in \eqref{eq:L_nvarsigma_deflection_4level} because its variation across the relevant integration domain is negligible.
Thus, in the large probe size regime, Eq.~\eqref{eq:L_nvarsigma_deflection_4level} simplifies to
\begin{equation}\label{eq:L_nvarsigma_deflection_5level}
    \begin{aligned}
\cL_{\varsigma}(\vartheta; z_p)\approx&\exp\left[-\frac{{k}^2_{z,0}\vartheta^2}{4\Delta k_\perp^2}\right]e^{i\varsigma \frac{\delta k^2}{k_{z,0}}z_p}\int_0^\infty \, \frac{ dq_\perp}{2\Delta k_\perp}\exp\left[-\frac{q_\perp^2\zeta(z_p)}{4\Delta k_\perp^2}\right] I_1\left[\frac{\vartheta q_\perp k_{z,0}}{2\Delta k_\perp^2}\zeta(z_p)\right]\\
=&\exp\left[-\frac{{k}^2_{z,0}\vartheta^2}{4\Delta k_\perp^2}\right]e^{i\varsigma \frac{\delta k^2}{k_{z,0}}z_p}\frac{\Delta k_\perp}{\vartheta k_{z,0}\zeta(z_p)}\left(\exp\left[\frac{\vartheta^2k_{z,0}^2\zeta(z_p)}{4\Delta k^2_\perp}\right]-1\right),
    \end{aligned}
\end{equation}
which highlights that the transverse response smoothly increases with $\vartheta$ and approaches a behavior of $1/\vartheta$ for large deflections, providing both physical insight and a convenient analytical form for numerical simulations.

\section{Probability amplitudes of the free electron-sample spin state in position representation}\label{ap:general_position_mode}

In the previous sections of the appendix, analytical integral expressions for the probability amplitudes in momentum representation were given based on scattering theory. From these expressions, it is possible to derive the probability amplitudes of the final state in position representation. This representation in the three-dimensional position space corresponds to:
\begin{equation}
\begin{aligned}
     \rho_{{\rm sc,e}\,\vec{r};\vec{r}'} =& \int d^3k_1 \,d^3k_2\, e^{i(\vec{k}_1\cdot \vec{r} - \vec{k}_2\cdot \vec{r}')} \rho_{{\rm sc,e}\,\vec{k}_1;\vec{k}_2}= \sum_{s'} \underbrace{\int d^3k_1 \, e^{i\vec{k}_1\cdot \vec{r}}S^{(1)}_{\vec{k}_1, s'; {\rm in}, s}}_{S^{(1)}_{\vec{r}, s'; {\rm in}, s}} \underbrace{\int d^3k_2\, e^{-i \vec{k}_2\cdot \vec{r}}S^{(1)~*}_{\vec{k}_2, s'; {\rm in}, s}}_{S^{(1)~*}_{\vec{r}, s'; {\rm in}, s}}\,,
\end{aligned}
\end{equation}
which is simply given by the Fourier transformation of the S-matrix elements from momentum to real space. 
The first-order scattering matrix elements in position can be expressed in a similar way as in Eq.~\eqref{eq:scatt_momentum_azi_smearing} as follows
\begin{subequations}\label{eq:scatt_posiiton_azi_smearing}
    \begin{align}
S^{(1)}_{\vec{r}, s'; {\rm in}, s} =& \delta_{s',s} \tilde\beta_1(\vec{r}) +  \sum_{\varsigma = -1}^{1} \tilde\beta_{n,\varsigma}(\vec{r}) \langle s'| \osigma_{n,\varsigma} |s \rangle\,,\\
\tilde\beta_j(\vec{r}):=&\int d^3k\,e^{i\vec{k}\cdot \vec{r}}\beta_j(\vec{k})\,,\label{eq:tildebeta_pos_def}
    \end{align}
\end{subequations}
where the $\beta_j$ coefficients are explicitly given in Eq.~\eqref{eq:beta_mom_smear_n} with $j=\{1, (n, \varsigma)\}$.

{We exploit the azimuthal symmetry of the scattering process and employ the narrow longitudinal wavefunction approximation ($\Delta k_z \ll k_{z,0}$) with small deflection angles, consistent with Eq.~\eqref{eq:Pi0_momentum_2level}. Under these conditions, the non-scattered amplitude becomes}
\begin{equation}\label{eq:beta1_pos_paraxial_approx}
    \begin{aligned}
        \tilde\beta_1((\vec{r}_\perp, z_d))
        &\approx \int d^2 k_\perp 
        \int_{-\infty}^{\infty} d\kappa_z \,
        e^{i (k_{z,0}+\kappa_z) z_d}
        e^{i\vec{k}_\perp\cdot\vec{r}_\perp}
        \varphi_{{\rm in},\|}(\kappa_z + k_{z,0})
        \varphi_{{\rm in},\perp}(k'_\perp) \\
        &= 2\pi e^{i k_{z,0}(z_d + z_p) }
        \tilde{\varphi}_{{\rm in},\|}\!\left(z_d + z_p\right) \int_0^\infty d k_\perp \, k_\perp J_0(k_\perp r_\perp) 
        \varphi_{{\rm in},\perp}(k_\perp)\\
&=2\sqrt{2\pi}\Delta k_\perp e^{i k_{z,0}(z_d + z_p) }
        \tilde{\varphi}_{{\rm in},\|}(z_d+z_p)\!\exp\left[-r_\perp^2\Delta k_\perp^2\right],
    \end{aligned}
\end{equation}
where {$\vec{r}=(\vec{r}_\perp, z_d)$ denotes the position vector, with $z_d$ representing the defocus distance from the sample plane. The longitudinal envelope function} $\tilde{\varphi}_{{\rm in},\|}(z)$ is defined as
\begin{equation}\label{eq:varphitilde_z}
    \tilde{\varphi}_{{\rm in},\|}(z)
= \left|\int dk_z \, \varphi_{{\rm in},\|}(k_z)\,
e^{i k_z z}\right|
= (8\pi \Delta k_z^2)^{1/4}
\exp\!\left[-{z^2 \Delta k_z^2}\right]\,.
\end{equation}
In the second step, we write $\vec{r}_\perp=r_\perp(\cos\phi_r,\sin\phi_r)$ and solve the azimuthal integral using
\begin{equation}
\int_0^{2\pi} d\phi_k \exp{i k_\perp r_\perp \cos(\phi_k-\phi_r)}= 2 \pi J_0(k_\perp r_\perp),
\end{equation}
and for the final step we  used the standard integral
\begin{equation}
\int_0^\infty\, dk\, k J_0(k r) \,e^{-\frac{1}{4\Delta k_\perp^2}\zeta^2 k^2}=\frac{2\Delta k_\perp^2}{\zeta^2}e^{-r^2\Delta k_\perp^2/\zeta^2}\quad \text{for }\Re (\zeta)>0.
\end{equation}

Following the same procedure as in Eq.~\eqref{eq:Pi_nvarsig_momentum_2level_pre} {defining azimuthal coordinates for $\vec{k}_\perp$ by $\vec{k}_\perp = k_\perp (\cos\phi_k, \sin\phi_k, 0)$, as well as $\phi=\phi_q-\phi_k$ and noting that $q_{z,\varsigma}^{{\rm sol,0}}$ is a function of $\phi$}, the scattered probability amplitudes in Eq.~\eqref{eq:tildebeta_pos_def} can be expressed as
\begin{equation}\label{eq:betanvar_pos_paraxial_approx}\allowdisplaybreaks
    \begin{aligned}
        \tilde\beta_{n,\varsigma}((\vec{r}_\perp, z_d))
        \approx& -2^{|\varsigma|/2}\sqrt{\frac{2}{\pi}}{r_{\rm e}}\int d^2 k_\perp 
        \int_{-\infty}^{\infty} dk_z \,
        e^{i k_z z_d}
        e^{i\vec{k}_\perp\cdot\vec{r}_\perp}
        \ell(k_\perp,k_z;\varsigma)\vec{{\rm u}}^*({\phi_k})\cdot \ve_{n, -\varsigma}\\
        \approx& -2^{|\varsigma|/2}\sqrt{\frac{2}{\pi}}{r_{\rm e}}\int d^2 k_\perp e^{i\vec{k}_\perp\cdot\vec{r}_\perp} \int_0^\infty dq_\perp\, \frac{ q_\perp}{2\sqrt{2\pi}}\int_0^{2\pi}\!d\phi\,\left.
        \frac{1}{|\vec{q}|^2}
        \mathcal{I}_{\rm es}(|\vec{q}|) 
        \right|_{q_z = q_{z,\varsigma}^{{\rm sol,0}}}{ \frac{k_{z,0}}{k_{z, \varsigma}^{\rm sol,0}}}\left(q_\perp\cos\phi-{\frac{k_\perp}{k_{z,0}}} q_{z,\varsigma}^{{\rm sol,0}}\right)\\
&\qquad \qquad \qquad \qquad \qquad \vec{{\rm u}}^*({\phi_k})\cdot \ve_{n, -\varsigma}\varphi_{{\rm in, \perp}}({ \sqrt{k_\perp^2+q_\perp^2 - 2k_\perp q_\perp \cos\phi}}
)\int_{-\infty}^{\infty} d\kappa_z \,
        e^{i (k_{z,0}+\kappa_z) z_d}\varphi_{{\rm in, \|}}(k_{z,\varsigma}^{{\rm sol}})\\
        =&-2\pi i 2^{|\varsigma|/2}\sqrt{\frac{2}{\pi}}{r_{\rm e}}e^{ik_{z,0}(z_d+z_p)}\tilde\varphi_{{\rm in, \|}}(z_d+z_p)\int_0^\infty d k_\perp \, k_\perp J_1({k}_\perp{r}_\perp)\vu^*(\phi_r) \cdot \ve_{n, -\varsigma} \int_0^\infty dq_\perp\, \frac{ q_\perp}{2\sqrt{2\pi}} \\
&\qquad \int_0^{2\pi}\!d\phi\,\left.{ \frac{k_{z,0}}{k_{z, \varsigma}^{\rm sol,0}}}\left(q_\perp\cos\phi-{\frac{k_\perp}{k_{z,0}}} q_{z,\varsigma}^{{\rm sol,0}}\right)\varphi_{{\rm in, \perp}}({ \sqrt{k_\perp^2+q_\perp^2 - 2k_\perp q_\perp \cos\phi}})
        e^{i q_{z, \varsigma}^{\rm sol,0}  z_d}
        \frac{1}{|\vec{q}|^2}
        \mathcal{I}_{\rm es}(|\vec{q}|) 
        \right|_{q_z = q_{z,\varsigma}^{{\rm sol,0}}}\\
=&-2\pi i 2^{|\varsigma|/2}\sqrt{\frac{2}{\pi}}{r_{\rm e}}e^{ik_{z,0}(z_d+z_p)}\tilde\varphi_{{\rm in, \|}}(z_d+z_p)\vu^*(\phi_r) \cdot \ve_{n, -\varsigma}\underbrace{\int_0^\infty d k_\perp \, k_\perp J_1({k}_\perp{r}_\perp) \cL_{\varsigma}(k_\perp/k_{z,0};-z_d)}_{:=\tilde \cL_{\varsigma}(r_\perp;-z_d)}.
    \end{aligned}
\end{equation}
{In the second line, we expand $k_{z, \varsigma}^{\rm sol}$ as in Eq.~\eqref{eq:kzsol_paraxial} and evaluate all terms at $\kappa_z=0$, except for the longitudinal Gaussian in momentum. This treatment, consistent with the approximations in Eq.~\eqref{eq:Pi_nvarsig_momentum_2level_pre}, relies on the narrow longitudinal wavepacket assumption and neglects corrections of order $10^{-6}$.}
As a result, we obtain a dominant propagation phase over a distance $z_d+z_p$, determined by the mean longitudinal wavenumber $k_{z,0}$. An additional phase $ e^{i q_{z, \varsigma}^{\rm sol,0}  z_d}$ integrated over the wavefunction arises from the longitudinal momentum transfer associated with the electron-spin interaction. 
Since the $\phi_k$ dependence appears only in the vector term, the corresponding integral is evaluated as
\begin{equation}
\int_0^{2\pi} d\phi_k \exp{i k_\perp r_\perp \cos(\phi_k-\phi_r)}\vu^*(\phi_k)
=2\pi i J_1(k_\perp r_\perp)\,\vu^*(\phi_r).
\end{equation}
The amplitude $\cL_{\varsigma}(k_\perp/k_{z,0};-z_d)$ is introduced in the last step by noting that it coincides with the expression in Eq.~\eqref{eq:L_nvarsigma_deflection_3level} when the condition for complete longitudinal overlap between the scattered and unscattered wavefunction components is satisfied. In the present case, this condition is not necessary, as the overlap naturally appears when computing probability contributions.

To illustrate the emergence of orbital angular momentum (OAM) in the scattered electron wavefunction, we consider the paradigmatic case of a longitudinally aligned bias magnetic field. In this configuration, the inelastic scattering amplitudes take the approximate form
$\tilde\beta_{z,\varsigma}((\vec{r}_\perp, z_d)) \approx 
b(r_\perp, z_d)\,\vu^*(\phi_r) \cdot \ve_{z,-\varsigma},$
where $\ve_{z, \varsigma} = (1-|\varsigma|) \ve_{z} + \frac{|\varsigma|}{\sqrt{2}} (\ve_x + i\varsigma \ve_y)$ and the azimuthal dependence can be explicitly expressed as
\begin{equation}
    \vu^*(\phi_r) \cdot \ve_{z,-\varsigma}
    = -\frac{i|\varsigma|}{2\sqrt{2}}
    \left[(\varsigma - 1)e^{i\phi_r} + (\varsigma + 1)e^{-i\phi_r}\right].
\end{equation}
Acting with the angular momentum operator $\hat{L}_z = -i\hbar\,\partial/\partial\phi_r$ on the scattered amplitude yields
\begin{equation}\label{eq:demo_OAM_amplitudes}
    \begin{split}
    \hat{L}_z \tilde\beta_{z,\varsigma}((\vec{r}_\perp, z_d))
    &=
    -i\hbar\frac{\partial}{\partial\phi_r}
    \left[b(r_\perp, z_d)\,\vu^*(\phi_r) \cdot \ve_{z,-\varsigma}\right]= \hbar \, b(r_\perp, z_d) \left(-\frac{i|\varsigma|}{2\sqrt{2}}\right) \left[(\varsigma - 1)e^{i\phi_r} - (\varsigma + 1)e^{-i\phi_r}\right]
    \\
    &= -\varsigma \hbar\,\tilde\beta_{z,\varsigma}((\vec{r}_\perp, z_d)),
    \end{split}
\end{equation}
for $\varsigma \in \{-1, 0, 1\}$. This establishes that $\tilde\beta_{z,\varsigma}$ is an eigenfunction of $\hat{L}_z$ with eigenvalue $-\varsigma\hbar$, confirming that the scattered electron acquires well-defined OAM from the spin transition. The characteristic azimuthal phase $e^{\pm i\phi_r}$ represents the topological signature of this transfer of orbital angular momentum from the spin to the electron.

Further approximation in the paraxial regime can be applied but now considering that the position probability distribution involves a wide range of spatial frequencies. Using Eq.~\eqref{eq:kz_qz_sol_approx}, we find that the ratio between longitudinal and transverse momentum transfer is
\begin{equation}\label{eq:qz_sol_complete_ratio}
    \frac{q_{z,\varsigma}^{{\rm sol},0}}{q_\perp} 
    \approx- \varsigma\frac{\delta k^2}{k_{z,0}q_\perp} 
        +\frac{q_\perp}{2k_{z,0}} - \frac{k_\perp}{k_{z,0}} \cos(\phi) .
\end{equation}
Note that $\vec{k}_\perp=k_{z,0}\vartheta\vu(\phi_k)$ in Eq.~\eqref{eq:L_nvarsigma_deflection_3level} is centered at $\vec{q}_\perp$ due to the Gaussian wave function of the initial electron state $\varphi_{{\rm in},\perp}$ with width $\Delta k_\perp$ for which we consider values that are at most of the order of $10^{-4} k_{z,0}$ in this work. Furthermore, the smearing function in Eq.~\eqref{eq:FT_spindens} sets the scale 
$q_\perp \lesssim k_a \sim 10^{10}~\text{m}^{-1}$ {for the parameters in Table~\ref{tab:simulation_constants}. Over most of the integration domain, this implies the hierarchy $\Delta k_\perp \ll q_\perp \ll k_{z,0}$.} Taking into account that $\delta k\sim 10^8~\text{m}^{-1}$, we find that the first term in Eq. \eqref{eq:qz_sol_complete_ratio} is vanishingly small for most of the 2-dimensional $\vec{q}_\perp$-integration range. For the second and the third term, we find that they are at most of the order of $k_a/k_{z,0}\sim10^{-2}$. {Then, we conclude that $q^{\rm sol,0}_{z, \varsigma}/q_\perp\sim 10^{-2}$ and ${q_{z,\varsigma}^{\rm sol,0}}/{k_{z,\varsigma}^{\rm sol,0}}\sim 10^{-4}$.

Consequently, the approximations used to derive Eq.~\eqref{eq:L_nvarsigma_deflection_4level} remain valid, allowing us to employ this expression for $\cL_{\varsigma}(k_\perp/k_{z,0};-z_d)$. This validity is justified by two relations established below Eq.~\eqref{eq:qz_sol_complete_ratio}: first, the term $k_\perp {q_{z,\varsigma}^{\rm sol,0}}/{k_{z,\varsigma}^{\rm sol,0}}$ is strongly suppressed by the small ratio ${q_{z,\varsigma}^{\rm sol,0}}/{k_{z,\varsigma}^{\rm sol,0}}$; and second, the condition $q^{\rm sol,0}_{z, \varsigma}\ll q_\perp$ is satisfied.
Furthermore, we derive an expression for $\tilde \cL_{\varsigma}(r_\perp;-z_d)$ by substituting $\cL_\varsigma(\vartheta_k;z_p)$ from Eq.~\eqref{eq:L_nvarsigma_deflection_4level} with the angle $\vartheta_k = k_\perp/k_{z,0}$ and replacing the parameter $z_p$ with $-z_d$}, obtaining
\begin{equation}\label{eq:tildeL_nvar}
    \begin{aligned}
\tilde \cL_{\varsigma}(r_\perp;-z_d)
&\approx e^{-i\varsigma \frac{\delta k^2}{k_{z,0}}z_d}\int_0^\infty \frac{ dq_\perp}{2\Delta k_\perp}\,e^{i\frac{q_{\perp}^{2}}{2k_{z,0}}z_d} \exp\!\left[-\frac{q_\perp^2}{4\Delta k_\perp^2}\right] 
\mathcal I_{\rm es}(q_\perp)\int_0^{\infty} dk_{\perp}\;k_{\perp}\;\exp\left[-\frac{k_\perp^{2}}{4\Delta k_\perp^{2}}\right]\;J_1({{k}_{\perp}{r}_\perp})I_1\left(\frac{k_\perp q_\perp}{2\Delta k_\perp^2} \zeta(-z_d)\right)\\  
&=e^{-i\varsigma \frac{\delta k^2}{k_{z,0}}z_d}\int_0^\infty \frac{ dq_\perp}{2\Delta k_\perp}\,\exp\!\left[-\frac{q_\perp^2}{4\Delta k_\perp^2}\zeta(-z_d)\right] 
\mathcal I_{\rm es}(q_\perp)2\Delta k_\perp^2\,
\exp\!\left[\frac{q_\perp^2}{4\Delta k_\perp^2}\zeta(-z_d)^2 - r_\perp^2\Delta k_\perp^2\right]\,
J_1\!\!\left(q_\perp \,r_\perp \zeta(-z_d)\right)\\
&=\, \Delta k_\perp e^{-i\varsigma \frac{\delta k^2}{k_{z,0}}z_d}\exp\!\left[- r_\perp^2\Delta k_\perp^2\right]\,\int_0^\infty { dq_\perp}\,\exp\!\left[-i\frac{q_\perp^2}{2 k_{z,0}}z_d\zeta(-z_d)\right] 
\mathcal I_{\rm es}(q_\perp)J_1\left(q_\perp r_\perp \zeta(-z_d)\right),
    \end{aligned}
\end{equation}
where, in the second step, we have used the integral identity from Ref.~\cite{gradshteyn2007} (Eq.~6.633-4, p.~707):
\begin{equation}\label{eq:identity_GaussIJ}
    \int_0^\infty x\,e^{-c x^2}\,I_\nu(b x)\,J_\nu(a x)\,dx
=\frac{1}{2c}\exp\!\left(\frac{b^2-a^2}{4c}\right)\,
J_\nu\!\left(\frac{ba}{2c}\right),
\end{equation}
valid for $\Re (c)>0$ and $\nu=1$, with the substitutions $c=\tfrac{1}{4\Delta k_\perp^2}$, $a=r_\perp$, and $b=\tfrac{q_\perp}{2\Delta k_\perp^2}\zeta(-z_d)$.
The dependence on the defocus distance $z_d$ in this expression manifests as a phase term integrated over the transverse momentum transfer, encompassing both the interaction effects and the momentum-space broadening. For numerical evaluations, we impose a cutoff determined by the momentum-space density of the spin smearing, such that $q_\perp\in[0, 10/a_0]$.

Finally, it is worth noting that the assumptions justifying the large-probe-size regime are not applicable in this case. In contrast to diffraction-mode imaging, the position-space scattering amplitudes require inclusion of a broad range of spatial frequencies, as the process is not constrained by interference with the zero-deflection amplitude. Therefore, $q_\perp$ cannot be further limited.

\subsection{Single spin position space imaging}

Furthermore, to determine the probability of detecting an electron at a position $\vec{r}_\perp$ on a transverse detection plane located at a defocus distance $z_d$, we account for the fact that the real-space probability distribution is time-dependent. This leads to
\begin{equation}
\begin{aligned}
 P(\vec{r}_\perp,z_d,t) &= \int d^3k_{1} \,d^3k_{2}\, e^{i(\vec{k}_{1} - \vec{k}_{2})\cdot \vec{r}} e^{-\frac{i}{\hbar} (E(\vec{k}_1)-E(\vec{k}_2)) t} \rho_{{\rm sc,e}\,\vec{k}_1;\vec{k}_2}M(\vec{k}_{\perp,1})M(\vec{k}_{\perp,2})\,,
\end{aligned}
\end{equation}
where $M(\vec{k}_\perp)$ encapsulates both the mask function and the contrast transfer function of the imaging system. 
This comprehensive treatment incorporates the point spread function, which accounts for the finite spatial resolution introduced by the microscope's lens system. Additionally, the application of a mask in the far field can be considered, effectively blocking specific regions of momentum space. 
{For example, one may select either the high-angle dark-field region (dominated by large scattering angles, where strongly scattered electrons are collected) or the bright-field region (dominated by small scattering angles, where unscattered or weakly scattered electrons are collected), and subsequently reconstruct the image in real space from the corresponding electron distributions in these angular domains.}

We restrict our consideration to situations where the single electron wave functions are much shorter than the experimental time resolution. Therefore, we average the probability over the whole temporal domain leading to
\begin{equation}\label{eq:prob_bar}
    \begin{aligned}
    P_{\rm img}(\vec{r}_\perp,z_d) &=  \,\mathcal{N}_{\rm img}\int dt\,  P(\vec{r}_\perp,z_d,t)\\
    &=  \,\mathcal{N}_{\rm img} \int d^3k_{1} \,d^3k_{2}\, e^{i(\vec{k}_{1} - \vec{k}_{2})\cdot \vec{r}}  2\pi\hbar \delta(E(\vec{k}_1)-E(\vec{k}_2))   \rho_{{\rm sc,e}\,\vec{k}_1;\vec{k}_2}M(\vec{k}_{\perp, 1})M(\vec{k}_{\perp, 2}),
    \end{aligned}
\end{equation}
where $\mathcal{N}_{\rm img}$ renormalizes the probability distribution.

From the first-order expansion of the density matrix in Eq.~\eqref{eq:scatt_state_density_electron}, {the probability distribution in position can be written as
\begin{equation}\label{eq:prob_pos_paraxial}
    \begin{aligned}
&P_{\rm img} \approx \tilde\Pi_0+\Re(\tilde\Pi_{n, +1}+\tilde\Pi_{n, -1})\expval{\osigma_{m,0}}-\Im(\tilde\Pi_{n, +1}-\tilde\Pi_{n, -1})\expval{\osigma_{l,0}}+2\Re(\tilde\Pi_{n, 0})\expval{\osigma_{n,0}}\,,
    \end{aligned}
\end{equation}
where we have dropped the explicit dependence on $(\vec{r}_\perp, z_d)$ for brevity}. Assuming that the function $M(\vec{k})=M(\vec{k}_\perp)$, we find the following expressions for the functions appearing in Eq.~\eqref{eq:prob_pos_paraxial}:
\begin{subequations}\label{eq:pi_pos_def}
\begin{align}
    \tilde\Pi_{0}(\vec{r}_\perp,z_d):=&\,\mathcal{N}_{\rm img} \int d^3k_{1} \,d^3k_{2}\, e^{i(\vec{k}_{1} - \vec{k}_{2})\cdot \vec{r}} 2\pi\hbar \delta(E(\vec{k}_1)-E(\vec{k}_2)) \beta_{1}(\vec{k}_1) \beta_{1}(\vec{k}_2)^*M(\vec{k}_{\perp,1})M(\vec{k}_{\perp,2})\,,\\
    \tilde\Pi_{n,\zeta}(\vec{r}_\perp,z_d):=&\,\mathcal{N}_{\rm img} \int d^3k_{1} \,d^3k_{2}\, e^{i(\vec{k}_{1} - \vec{k}_{2})\cdot \vec{r}} 2\pi\hbar \delta(E(\vec{k}_1)-E(\vec{k}_2)) \beta_{n,\varsigma}(\vec{k}_1) \beta_{1}(\vec{k}_2)^*M(\vec{k}_{\perp,1})M(\vec{k}_{\perp,2})\,.
\end{align} 
\end{subequations}
Furthermore, we decouple the transversal and longitudinal component of the wavevector using $\vec{k}_{j}=(\vec{k}_{\perp, j}, k_{z,j})$ for $j=\{1,2\}$, and then we rewrite the delta function in energy into one in longitudinal momentum and obtain
\begin{subequations}\label{eq:pi_pos_1level}
    \begin{align}
    \begin{split}
        \tilde\Pi_{0}(\vec{r}_\perp,z_d)& = \,\mathcal{N}_{\rm img}\int d^3k_{1} \,d^3k_{2}\,e^{i(\vec{k}_{\perp,1} - \vec{k}_{\perp,2})\cdot \vec{r}_\perp} e^{i (k_{z,1}-k_{z,2}) z_d}  \frac{2\pi E(\vec{k}_{ 2})}{c^2\hbar k_{z,1}} \delta(k_z^{\rm con}-k_{z,1})  \beta_{1}(\vec{k}_1) \beta_{1}(\vec{k}_2)^*M(\vec{k}_{\perp,1})M(\vec{k}_{\perp,2})   \\
    &= \,\mathcal{N}_{\rm img}\int d^2k_{\perp,1} \,d^3k_{2} ~ e^{i(\vec{k}_{\perp,1} - \vec{k}_{\perp,2})\cdot \vec{r}_\perp} e^{i (k_z^{\rm con}-k_{z,2}) z_d}  \frac{2\pi E(\vec{k}_{2})}{c^2\hbar k_z^{\rm con}} \bar\varphi_{{\rm in}}({k}_{\perp,1},k_z^{\rm con}) \bar\varphi_{{\rm in}}({k}_{\perp,2},k_{z,2})^*M(\vec{k}_{\perp,1})M(\vec{k}_{\perp,2}) \,,
    \end{split}\\
    \begin{split}
        \tilde\Pi_{n,\zeta}(\vec{r}_\perp,z_d)& = \,\mathcal{N}_{\rm img}\int d^2k_{\perp,1} \,d^3k_{2} ~ e^{i(\vec{k}_{\perp,1} - \vec{k}_{\perp,2})\cdot \vec{r}_\perp} e^{i (k_z^{\rm con}-k_{z,2}) z_d}  \frac{2\pi E(\vec{k}_{ 2})}{c^2\hbar k_z^{\rm con}} \\
        & \qquad\qquad\qquad\qquad\qquad\qquad\qquad\qquad \beta_{n,\varsigma}((\vec{k}_{\perp,1},k_z^{\rm con})) \bar\varphi_{{\rm in}}(k_{\perp,2},k_{z,2})^* M(\vec{k}_{\perp,1})M(\vec{k}_{\perp,2})\,,
    \end{split}
    \end{align}
\end{subequations}
where we defined the $z$-component of the electron under energy conservation
\begin{equation}
    \begin{aligned}\label{eq:kzcon}
         k_z^{\rm con}&= \frac{1}{\hbar c}\sqrt{E(\vec{k}_2)^2-\hbar^2k^{2}_{\perp,1} c^2 - (mc^2)^2} = \sqrt{ k_{z,2}^{2} + k^{2}_{\perp,2} - k^{2}_{\perp,1} }\,.
    \end{aligned}
\end{equation}

Based on the expressions in Eq.~\eqref{eq:pi_pos_1level}, we employ the paraxial approximation following a similar procedure as for the wavenumber in Eq.~\eqref{eq:kzsol_paraxial}.
From Eq.~\eqref{eq:elldef}, we conclude that the relevant scale for the angles is set by the Fourier representation $I_{\rm es}(q)$ of the smearing function, which implies $q_\perp \lesssim (2a_0)^{-1} \sim 10^{10}~\text{m}^{-1}\sim 10^{-2} k_{z,0}$ for spins on the atomic scale. Since we consider initial wave functions with small momentum spread $\Delta k_\perp \lesssim 10^{-4}k_{z,0}$, we can then expand the expression in Eq.~\eqref{eq:kzcon}
\begin{equation}
    \begin{aligned}
         k_z^{\rm con}& \approx k_{z,2} + \frac{k^{2}_{\perp,2} - k^{2}_{\perp,1}}{2 k_{z,2}} \,,
    \end{aligned}
\end{equation}
and find the term in the integrand can be approximated as
\begin{equation}
    e^{i (k_z^{\rm con}-k_{z,2}) z_d}  \frac{2\pi E({\vec{k}_2})}{c^2\hbar k_z^{\rm con}}\approx e^{i \frac{k^{2}_{\perp,2} - k^{2}_{\perp,1}}{2 k_{z,2}} z_d}  \frac{2\pi E(\vec{k}_{ 2})}{c^2\hbar k_{z,2}} \,.
\end{equation}
Furthermore, we can employ the small spread of longitudinal momentum $k_{z,2}$ around $k_{z,0}$ imposed by $\varphi_{\rm in,\|}(k_{z,2})$ 
to restrict to the approximate expression
\begin{equation}
    \begin{aligned}\label{eq:kzconapp0}
         k_z^{\rm con}& \approx k_{z,2} + \frac{k^{2}_{\perp,2} - k^{2}_{\perp,1}}{2 k_{z,0}} 
    \end{aligned}
\end{equation} 
in all following steps, and we find
\begin{equation}
      e^{i \frac{k^{2}_{\perp,2} - k^{2}_{\perp,1}}{2 k_{z,2}} z_d}\frac{2\pi E(\vec{k}_{ 2})}{c^2\hbar k_z^{\rm con}} \approx e^{i \frac{k^{2}_{\perp,2} - k^{2}_{\perp,1}}{2 k_{z,0}} z_d}\frac{2\pi E_0}{c^2\hbar k_{z,0}}=e^{i \frac{k^{2}_{\perp,2} - k^{2}_{\perp,1}}{2 k_{z,0}}z_d}\frac{2\pi}{v_0}\,.
\end{equation}
Let us consider the $dk_{z,2}$-integral for $\tilde{\Pi}_0$. With Eq.~\eqref{eq:kzconapp0}, we find 
\begin{equation}\label{eq:overlap_pos_nonscatt}
    \begin{aligned}
\int_{-\infty}^\infty dk_{z,2}\,\bar\varphi_{{\rm in}}({k}_{\perp,1},k_z^{\rm con}) \bar\varphi_{{\rm in}}({k}_{\perp,2},k_{z,2})^*=& \,
\varphi_{{\rm in, \perp}}({k}_{\perp,1}) \varphi_{{\rm in, \perp}}({k}_{\perp,2})^*\int_{-\infty}^\infty dk_{z,2}\,\varphi_{{\rm in, \|}}(k_{z}^{\rm con}) \varphi_{{\rm in, \|}}(k_{z,2})^*\\
=&\varphi_{{\rm in, \perp}}({k}_{\perp,1}) \varphi_{{\rm in, \perp}}({k}_{\perp,2})^*e^{i \frac{k_{\perp,2}^2-k_{\perp,1}^2}{2k_{z,0}}\,
 z_p}\,
\exp\!\left[-\frac{1}{8\Delta k_z^2}\left(\frac{k_{\perp,2}^2-k_{\perp,1}^2}{2k_{z,0}}\right)^2\right],
    \end{aligned}
\end{equation}
where we decouple the wavefunction using Eq.~\eqref{eq:incident_Gaussian_wavefunction}.
{Due to the transverse Gaussian wave functions of $k_{\perp,1}$ and $k_{\perp,2}$ appearing in the above expression, the integrand in the expression for $\tilde\Pi_{0}(\vec{r}_\perp,z_d)$ is only of relevant size for values of $k_{\perp,1}$ and $k_{\perp,2}$ of the order of $\Delta k_\perp$ or smaller for which
\begin{equation}\label{eq:overlap_argument_bound} 
\frac{1}{8\Delta k_z^2}
\left(\frac{k_{\perp,2}^2 - k_{\perp,1}^2}{2k_{z,0}}\right)^2
\lesssim
\frac{\Delta k_\perp^4}{\Delta k_z^2 k_{z,0}^2}\,.
\end{equation}
The right hand side of this inequality is much smaller than 1 for the parameter considered in this paper. Therefore, we can consider the longitudinal overlap in Eq.~\eqref{eq:overlap_pos_nonscatt} as unity.}
{Consequently, defining $\tilde{\mathcal{N}}_{\rm img} = 2\pi\mathcal{N}_{\rm img}/v_0$,} we obtain
\begin{subequations}\label{eq:tildePi0_B1}
   \begin{align}
 \tilde\Pi_0(\vec{r}_\perp, z_d)
 \approx & {\tilde{\mathcal{N}}_{\rm img}}
 \left| \mathcal{B}_1(\vec{r}_\perp;z_d)\right|^2\,,\label{eq:Pi0_pos}\\
 \mathcal{B}_1(\vec{r}_\perp;z_d)=&\int d^2k_{\perp}\;e^{i\vec{k}_{\perp}\cdot\vec{r}_\perp}\, e^{-i\frac{k_{\perp}^2}{2k_{z,0}}(z_d+z_p)}\varphi_{{\rm in},\perp}({k}_{\perp})M(\vec{k}_{\perp}). \label{eq:B1_pos}
   \end{align}
\end{subequations}

{A similar analysis can be performed  for the first order scattered contributions in the probability. Again} considering the $dk_{z,2}$-integral, we obtain
\begin{equation}\label{eq:integraldkz_betavarphi}
    \begin{aligned}
    \quad \int dk_{z,2}~ \tilde\beta_{n,\varsigma}((\vec{k}_{\perp,1},k_z^{\rm con}))& \varphi_{\rm in}((k_{\perp,2},k_{z,2}))^* 
    = -2^{|\varsigma|/2}\frac{r_{\rm e}}{2\pi} 
    \ve_{n, -\varsigma} \cdot \int dk_{z,2}~ {\vec{\ell}}(\vec{k}_{\perp,1},k_z^{\rm con},0)  \\
    &= -2^{|\varsigma|/2}\sqrt{\frac{2}{\pi}}   r_{\rm e} \vu^*(\phi_{k,1})\cdot\ve_{n, -\varsigma}\varphi_{\rm in, \perp}(k_{\perp, 2})^*\int dk_{z,2}~ \ell(k_{\perp,1},k_z^{\rm con};\varsigma)\varphi_{\rm in, \|}(k_{z, 2})^*\,.
    \end{aligned}
\end{equation}
In the following, we introduce the longitudinal wavenumber $k_{z,\varsigma}^{{\rm sol-con}}$ which corresponds to $k_{z,\varsigma}^{{\rm sol}}$ evaluating $k_z=k_z^{\rm con}$ and performing the replacements $k'_\perp\rightarrow k_{\perp,1}$ and $k_\perp^2\rightarrow  k_{\perp,1}^2+q_\perp^2 - 2k_{\perp,1} q_\perp \cos(\phi_q - \phi_{k_1})$ in Eq~\eqref{eq:kz_qz_sol_approx}, where $\phi_{k_1}$ is the azimuthal angle of $\vec{k}_{\perp,1}$ analogously to the azimuthal representation of $\vec{k}_\perp$. Using the expansion in Eq.~\eqref{eq:kzsol_paraxial}, we find in the paraxial approximation
\begin{equation}
k_{z,\varsigma}^{{\rm sol-con}} { \approx}\, k_{z}^{{\rm con}}  {  - \frac{q_\perp^2 - 2k_{\perp,1} q_\perp \cos(\phi_q - \phi_{k_1})}{2k_{z,0}}
            + \varsigma\,\frac{\delta k^2}{k_{z,0}} \approx k_{z,2} - q_{z,\varsigma}^{{\rm sol-con}} }\,,
\end{equation}
where
\begin{equation}
\begin{split}
    q_{z,\varsigma}^{{\rm sol-con}}:=& -\frac{k^{2}_{\perp,2} - k^{2}_{\perp,1}}{2 k_{z,0}} + \frac{q_\perp^2 - 2k_{\perp,1} q_\perp \cos(\phi_q - \phi_{k_1})}{2k_{z,0}}
            - \varsigma\,\frac{\delta k^2}{k_{z,0}}.
\end{split}
\end{equation}
As in Eq.~\eqref{eq:Pi_nvarsig_momentum_2level_pre}, the $k_{z,2}$-integral in Eq.~\eqref{eq:integraldkz_betavarphi} is dominated by the narrow Gaussian distributions in longitudinal momentum. All other factors in the integrand can therefore be evaluated at $k_{z,2} = k_{z,0}$, allowing $k_{z,\varsigma}^{{\rm sol-con}} \approx k_{z,\varsigma}^{{\rm sol, 0}}$ to be factored out of the integral, leading to
\begin{align}\label{eq:integraldkz_ellvarphi}
     \int dk_{z,2}~ \ell(k_{\perp,1},k_z^{\rm con};\varsigma)\varphi_{\rm in, \|}(k_{z, 2})^* \approx& \int_0^\infty dq_\perp\, \frac{ 
     { q_\perp}}{2\sqrt{2\pi}}\int_0^{2\pi}\!d\phi\, { \frac{k_{z,0}}{k_{z, \varsigma}^{\rm sol,0}}}\left(q_\perp\cos{\phi}- {\frac{k_{\perp,1}}{k_{z,0}}} q_{z,\varsigma}^{{\rm sol,0}}\right)\,\varphi_{{\rm in},\perp}(
     { \sqrt{k_{\perp,1}^2+q_\perp^2 - 2k_{\perp,1} q_\perp \cos\phi }})\,
        \nonumber \\
    & \quad   \left.\frac{1}{|\vec{q}|^2}
        \mathcal{I}_{\rm es}(|\vec{q}|) 
        \right|_{q_z = q_{z,\varsigma}^{{\rm sol,{0}}}}  \underbrace{\int dk_{z,2}~ \varphi_{{\rm in},\|}(k_{z,\varsigma}^{{\rm sol-con}})\varphi_{{\rm in},\|}(k_{z,2})^*}_{:=\tilde D_{\varsigma}(k_{\perp,1}, k_{\perp,2})}. 
\end{align}
where we used the substitution $\phi=\phi_q-\phi_{k_1}$.
The overlap integral of the longitudinal parts of the electron quantum state results in the following expressions
{
\begin{equation}\label{eq:overlap_pos_scatt}
    \begin{aligned}
\tilde D_{\varsigma}(k_{\perp,1}, k_{\perp,2})=&\int dk_{z,2}~ \varphi_{{\rm in},\|}\left(k_{z,2}+
            q_{z,\varsigma}^{{\rm sol-con}} \right)\varphi_{{\rm in},\|}(k_{z,2})^*\\
            =&e^{-i q_{z,\varsigma}^{{\rm sol-con}} \,
 z_p}\,
\exp\!\left[-\frac{(q_{z,\varsigma}^{{\rm sol-con}})^2}{8\Delta k_z^2}\right]\,.
		    \end{aligned}
\end{equation}}
Using analogue arguments as around Eq~\eqref{eq:overlap_argument_bound}, can conclude that only values of $k_{\perp,2}$ and $\sqrt{k_{\perp,1}^2+q_\perp^2 - 2k_{\perp,1} q_\perp}$ of the order of $\Delta k_\perp$ or smaller are relevant for the the integral in $\tilde\Pi_{n,\zeta}(\vec{r}_\perp,z_d)$. Since also $\delta k^2/(k_{z,0}\Delta k_z)$ is much smaller than one, we can approximate the Gaussian in the last line of Eq.~\eqref{eq:overlap_pos_scatt} as unity.

Consequently, we find
\begin{equation}
    \begin{aligned}
\int dk_{z,2}~ \ell(k_{\perp,1},k_z^{\rm con};\varsigma)\varphi_{\rm in, \|}(k_{z, 2})^* \approx& \,e^{i \left(\frac{k_{\perp,2}^2-k_{\perp,1}^2}{2k_{z,0}}\right)\,
 z_p} {\bar{\cL}_{\varsigma}}(k_{\perp, 1}/k_{z,0}; z_p)\,,
    \end{aligned}
\end{equation}
{
where
\begin{equation}\label{eq:tildeL_nvarsigma_deflection_3level}
    \begin{aligned}
\bar{\cL}_{\varsigma}(  k_{\perp, 1}/k_{z,0}; z_p)\approx& \int_0^\infty dq_\perp\, \frac{ 
     { q_\perp}}{2\sqrt{2\pi}}\int_0^{2\pi}\!d\phi\, { \frac{k_{z,0}}{k_{z, \varsigma}^{\rm sol,1}}}\left(q_\perp\cos{\phi}- {\frac{k_{\perp,1}}{k_{z,0}}} q_{z,\varsigma}^{{\rm sol,1}}\right)\\
     &\qquad \qquad\qquad\qquad\qquad \varphi_{{\rm in},\perp}(
     { \sqrt{k_{\perp,1}^2+q_\perp^2 - 2k_{\perp,1} q_\perp\cos\phi}})\,
        e^{- i q_{z,\varsigma}^{{\rm sol},1}z_p}\left.\frac{1}{|\vec{q}|^2}
        \mathcal{I}_{\rm es}(|\vec{q}|) 
        \right|_{q_z = q_{z,\varsigma}^{{\rm sol,{1}}}} \,, 
    \end{aligned}
\end{equation}
and
\begin{equation}
	 k_{z,\varsigma}^{{\rm sol},1}
    = k_{z,0} - q_{z,\varsigma}^{{\rm sol},1},
    \qquad q_{z,\varsigma}^{{\rm sol},1}:= \frac{q_\perp^2 - 2k_{\perp,1} q_\perp \cos\phi}{2k_{z,0}}
            - \varsigma\,\frac{\delta k^2}{k_{z,0}} .
\end{equation}
}
This yields the compact factorized form
\begin{equation}\label{eq:Pi_nvar_pos}
    \begin{aligned}
\tilde{\Pi}_{n, \varsigma}(\vec{r}_\perp; z_d)=& { \tilde{\mathcal{N}}_{\rm img} }\mathcal{B}^*_1(\vec{r}_\perp;z_d) \mathcal{B}_{n,\varsigma}(\vec{r}_\perp;z_d),
    \end{aligned}
\end{equation}
where the time-integrated scattered amplitudes are defined as
\begin{equation}\label{eq:Bnvar_pos}
\mathcal{B}_{n,\varsigma}(\vec{r}_\perp;z_d)=-2^{|\varsigma|/2}\;\sqrt{\frac{2}{\pi}}r_{\rm e}\int d^2k_{\perp}\,e^{i\vec{k}_{\perp}\cdot\vec{r}_\perp}\;e^{-i\frac{k_{\perp}^{2}}{2k_{z,0}}(z_d+z_p)}\;
 \vu^*(\phi_{k})\!\cdot\!\mathbf e_{n,-\varsigma}  { \bar{L}}_{\varsigma}(k_{\perp}/k_{z,0}; z_p)M(\vec{k}_{\perp}).
\end{equation}

Since the transverse profile of the incident beam and the spin's spatial probability density are azimuthally symmetric and concentric, we can exploit this symmetry. To preserve it, we {restrict our considerations to masks} and contrast transfer functions that are also azimuthally symmetric, i.e., $M(\vec{k}_\perp)=M(k_\perp)$. Under this condition, we can evaluate the azimuthal-angle integral $\phi_k$ in the time-integrated probability terms.
We obtain
\begin{subequations}
    \begin{align}
    \begin{split}
\mathcal{B}_1(\vec{r}_\perp;z_d)=&\int d^2k_{\perp}\;e^{i{k}_{\perp}{r}_\perp \cos(\phi_{k}-\phi_{r})}\, e^{-i\frac{k_{\perp}^2}{2k_{z,0}}(z_d+z_p)}\varphi_{{\rm in},\perp}({k}_{\perp})M({k}_{\perp})\\
=&8\pi\Delta k^2_\perp\int \frac{dk_{\perp}\;k_\perp}{(2\Delta k_\perp)^2} J_0({{k}_{\perp}{r}_\perp})\, e^{-i\frac{k_{\perp}^2}{2k_{z,0}}(z_d+z_p)}\varphi_{{\rm in},\perp}({k}_{\perp})M({k}_{\perp}),
    \end{split}\\
    \begin{split}
\mathcal{B}_{n,\varsigma}(\vec{r}_\perp;z_d)=&-2^{|\varsigma|/2}\sqrt{\frac{2}{\pi}}r_{\rm e} \int d^2k_{\perp}\;e^{i{k}_{\perp}{r}_\perp\cos(\phi_k-\phi_r)}\;
e^{-i\frac{k_{\perp}^{2}}{2k_{z,0}}(z_d+z_p)}  \vu^*(\phi_{k})\!\cdot\!\mathbf e_{n,-\varsigma}  {\bar{L}}_{\varsigma}(k_{\perp}/k_{z,0}; z_p)M({k}_{\perp})\\
 =&-8i \sqrt{{2}{\pi}}2^{|\varsigma|/2}\;r_{\rm e}\Delta k_\perp^2 \vu^*(\phi_r)\!\cdot\!\mathbf e_{n,-\varsigma}\int \frac{dk_{\perp}\;k_{\perp}}{(2\Delta k_\perp)^2}e^{-i\frac{k_{\perp}^{2}}{2k_{z,0}}(z_d+z_p)}J_1({{k}_{\perp}{r}_\perp})\;
   \bar \cL_{\varsigma}(k_{\perp}/k_{z,0}; z_p)\,M({k}_{\perp}),
    \end{split}
    \end{align}
\end{subequations}
where we employ the same relations as in Eq.~\eqref{eq:beta1_pos_paraxial_approx} and Eq.~\eqref{eq:betanvar_pos_paraxial_approx} for solving the integral over $\phi_k$.

{Note that we can employ the same arguments as above Eq.~\eqref{eq:tildeL_nvar} to use $\bar{\cL}_{\varsigma}(  k_\perp/k_{z,0}; z_p)\approx \cL_{\varsigma}(  k_\perp/k_{z,0}; z_p)$ as given in Eq.~\eqref{eq:L_nvarsigma_deflection_4level}, i.e.
\begin{equation}\label{eq:barL_nvarsigma_deflection_4level}
    \begin{aligned}
\bar{\cL}_{\varsigma}(k_\perp/k_{z,0}; z_p)\approx& \int_0^\infty \, \frac{ dq_\perp}{2\Delta k_\perp}\exp\left[-\frac{k_\perp^2+q_\perp^2}{4\Delta k_\perp^2}\right]e^{i\left(\varsigma \frac{\delta k^2}{k_{z,0}}-\frac{q_\perp^2}{2k_{z,0}}\right)z_p}
        I_1\left[\frac{k_\perp q_\perp }{2\Delta k_\perp^2}\zeta(z_p)\right]\mathcal{I}_{\rm es}(q_\perp).
    \end{aligned}
\end{equation}} 
In the following, we consider two specific cases: first, a model where the spatial envelope of the transfer function is represented by a Heaviside function that vanishes for $k_\perp>k_{\rm max}$ (or equivalently by a mask with dominant spatial frequencies much smaller than the inverse spatial resolution); and second, the ideal imaging case with perfect resolution and no mask, where all spatial frequencies contribute to the image.

In the first case, we model the transfer function as $M(k_\perp)=\Theta(k_{\rm max}-k_\perp)$, where we fix $k_{\rm max}=2\pi/r_{\rm res}$, with $r_{\rm res}$ representing the minimal spatial resolution. For state-of-the-art TEM systems, we take $r_{\rm res}=50~\text{pm}$. The integrals appearing in the functions $\bar{\cL}_{\varsigma}$ and $\mathcal{B}_{n,\varsigma}$ are then evaluated numerically. {For this numerical evaluation, we assume $z_p=0$. The general case relevant for defocused incident beam imaging may be treated elsewhere. We note that 
\begin{equation}
	\exp\left[-\frac{k_\perp^2+q_\perp^2}{4\Delta k_\perp^2} \right] I_1\left[\frac{k_\perp q_\perp }{2\Delta k_\perp^2} \right]= \int_0^{2\pi} d\phi \exp\left[-\frac{k_\perp^2+q_\perp^2 - 2k_\perp q_\perp\cos\phi}{4\Delta k_\perp^2} \right]\,.
\end{equation}
We can immediately see that the integrand becomes vanishingly small if $|q_\perp-k_\perp|/\Delta k_\perp \gg 2$. Therefore, we restrict the integration range to $q_\perp\in[\max(0,\,k_\perp - 10\Delta k_\perp),k_\perp + 10\Delta k_\perp]$.}
{In contrast, for the non-scattered time-integrated probability $\mathcal{B}_1$, the aperture limit $k_{\rm max}$ is effectively infinite compared to the momentum width of the incident beam, and we can approximate
\begin{equation}\label{eq:B1pos_defocus_infinity}
    \begin{aligned}
\mathcal{B}_1(\mathbf r_\perp;z_d)
&\approx\frac{8\pi}{\sqrt{2\pi}} \Delta k_\perp
\int_0^{\infty} \frac{k_\perp\,dk_\perp}{(2\Delta k_\perp)^2}\;J_0(k_\perp r_\perp)\;e^{-\frac{1}{4\Delta k_\perp^2}\zeta(z_d+z_p) k_\perp^2}\\
&=2\sqrt{2\pi}\Delta k_\perp\frac{1}{\zeta(z_d+z_p)}\;\exp\!\left[-\frac{r^2\Delta k_\perp^2}{\zeta(z_d+z_p)}\right],
\end{aligned}
\end{equation}
where we solve the $k_\perp$ integral as in Eq.~\eqref{eq:beta1_pos_paraxial_approx} for $\zeta(z_d+z_p)=1+2i\frac{z_d+z_p}{k_{z,0}}\Delta k_\perp^2$.

In analogy to the diffraction mode analysis, the scattered time-integrated probability is small compared to the non-scattered one, implying that the total image intensity is dominated by the non-scattered term $\mathcal{B}_1$. We therefore determine the normalization constant $\tilde{\mathcal{N}}_{\rm img}$ by requiring the background probability distribution to be normalized to unity. Integrating the squared modulus of the non-scattered amplitude (Eq.~\eqref{eq:B1pos_defocus_infinity}) over the detector plane, we obtain:
\begin{equation}
    \begin{aligned}
\tilde{\mathcal{N}}_{\rm img}^{-1} \approx& \int d^2\mathbf{r}_\perp \, |\mathcal{B}_1(\mathbf{r}_\perp; z_d)|^2 \\
=& \int_0^\infty r_\perp dr_\perp \int_0^{2\pi} d\phi \, \frac{8\pi \Delta k_\perp^2}{|\zeta(z_d+z_p)|^2} \exp\left[-\frac{2\Delta k_\perp^2}{|\zeta(z_d+z_p)|^2}r_\perp^2\right] = 4\pi^2.
    \end{aligned}
\end{equation}
This factor of $4\pi^2$ arises consistently from the Fourier transform convention used in the free-space propagation. Consequently, to ensure probability conservation, we define the normalization constant as $\tilde{\mathcal{N}}_{\rm img} = (2\pi)^{-2}$.}

In the second case, the contribution $\mathcal{B}_1$ is exactly given by Eq.~\eqref{eq:B1pos_defocus_infinity}.
For the scattered time-integrated probabilities, we use Eq.~\eqref{eq:barL_nvarsigma_deflection_4level} and we evaluate the integral
\begin{equation}
    \begin{aligned}
\int_0^{\infty} dk_{\perp}\;k_{\perp}e^{-i\frac{k_{\perp}^{2}}{2k_{z,0}}(z_d+z_p)}J_1({{k}_{\perp}{r}_\perp})\;
   \cL_{\varsigma}(k_{\perp}/k_{z,0}; z_p)
&=e^{i\varsigma \frac{\delta k^2}{k_{z,0}}z_p}\int_0^\infty \frac{ dq_\perp}{2\Delta k_\perp}\,e^{-i\frac{q_{\perp}^{2}}{2k_{z,0}}z_p} \exp\!\left[-\frac{q_\perp^2}{4\Delta k_\perp^2}\right] 
\mathcal I_{\rm es}(q_\perp)\\
&\qquad\int_0^{\infty} dk_{\perp}\;k_{\perp}J_1({{k}_{\perp}{r}_\perp})\;e^{-\frac{1}{4\Delta k_\perp^{2}}\zeta(z_d+z_p) k_\perp^{2}}\;I_1\left(\frac{k_\perp q_\perp}{2\Delta k_\perp^2} \zeta(z_p)\right)\\  
&=e^{i\varsigma \frac{\delta k^2}{k_{z,0}}z_p}\int_0^\infty \frac{ dq_\perp}{2\Delta k_\perp}\,e^{-i\frac{q_{\perp}^{2}}{2k_{z,0}}z_p} \exp\!\left[-\frac{q_\perp^2}{4\Delta k_\perp^2}\right] 
\mathcal I_{\rm es}(q_\perp)\\
&\qquad\frac{2\Delta k_\perp^2}{\zeta(z_d+z_p)}\,
\exp\!\left[\frac{q_\perp^2 \zeta^2(z_p)/\Delta k_\perp^2 - 4r_\perp^2\Delta k_\perp^2}{4\zeta(z_d+z_p)}\right]\,
J_1\!\!\left(\frac{q_\perp \,r_\perp \zeta(z_p)}{\zeta(z_d+z_p)}\right),
    \end{aligned}
\end{equation}
where, in the last step, we have used Eq.~\eqref{eq:identity_GaussIJ}
with the substitutions $c=\tfrac{1}{4\Delta k_\perp^2}\zeta(z_d+z_p)$, $a=r_\perp$, and $b=\tfrac{q_\perp}{2\Delta k_\perp^2}\zeta(z_p)$.
Hence, the effective scattered amplitude takes the form
\begin{equation}\label{eq:Bnpos_defocus_infinity}
    \begin{aligned}
\mathcal{B}_{n,\varsigma}(\vec{r}_\perp;z_d)=-4 i& \sqrt{2\pi}2^{|\varsigma|/2}\;r_{\rm e}\Delta k_\perp^2e^{i\varsigma \frac{\delta k^2}{k_{z,0}}z_p} \vu^*(\phi_r)\!\cdot\!\mathbf e_{n,-\varsigma}\exp\left[-\frac{r_\perp^2\Delta k_\perp^2}{\zeta(z_d+z_p))}\right]\\
&\int_0^\infty \frac{ dq_\perp}{2\Delta k_\perp}\exp\left[-\frac{q_\perp^2}{4\Delta k_\perp^2}\zeta(z_p)\left(1-\frac{\zeta(z_p)}{\zeta(z_d+z_p)}\right)\right] 
\mathcal I_{\rm es}(q_\perp)\frac{1}{\zeta(z_d+z_p)}\,
J_1\!\!\left(\frac{q_\perp \,r_\perp \zeta(z_p)}{\zeta(z_d+z_p)}\right),
    \end{aligned}
\end{equation}
which constitutes the most accurate analytical description of the scattered state amplitudes, including both defocus and longitudinal displacement effects, relevant for the reconstruction of the distribution along the beam axis through a series of focus planes.

\section{Backaction effect and loss of purity of the spin state}\label{ap:backaction}

To investigate the backaction effects that a free electron may induce on a spin system, and the resulting potential loss of purity, we analyze the spin quantum state immediately after the interaction with the beam electron.
In addition, we extend our analysis to second order in the interaction strength, focusing specifically on the case of a longitudinal bias magnetic field oriented along the $z$-axis, which corresponds to the dominant field component generated by the microscope pole pieces. 
Following this, we apply the leading-order Magnus expansion to derive the density matrix up to second order. Owing to the slow dynamics of the spin system and the validity of the paraxial approximation for the electron beam, the commutators of the interaction Hamiltonian at different times can be safely neglected~\cite{metrologySpinTEM}. Under these assumptions, the scattering operator in Eq.~\eqref{eq:scattoperator_first_order} can be extended to
\begin{equation}
    \hat{S} \approx \id + \hat{\tau}^{(1)} + \frac{1}{2} \hat{\tau}^{(1)\,2}.
\end{equation}
The inclusion of the second-order contributions in the Magnus expansion ensures the normalizability of the spin quantum state.

Furthermore, we work initially with the scattering amplitudes in Eq.~\eqref{eq:beta_mom_smear_n}. However, we transform them into a combined transversal momentum and longitudinal position space because the dependence on $k_{z, \varsigma}^{\rm sol}$ appears linearly in a phase in the real $z$ space and not quadratic in the argument of the exponential in Fourier space. As we will show later, this simplifies the expression, allowing us to utilize the approximations presented in Section~\ref{ap:general_diffraction}. This representation of the scattering amplitudes corresponds to 
\begin{equation}
    \eta_j((\vec{k}_\perp,z)):=\frac{1}{\sqrt{2\pi}}\int_{-\infty}^\infty dk_z\, e^{ik_z z}\beta_j(\vec{k})\,,
\end{equation}
for $j\in\{ 1, (n,0), (n, \pm)\}$ and the factor $(2\pi)^{-1/2}$ guarantees the normalization of the unscattered part. In the longitudinal bias field case, we use the expressions in Eq.~\eqref{eq:beta_mom_smear_n} for $n=z$ with Eq.~\eqref{eq:ell_vect_split}, finding that $\eta_{z,0}((\vec{k}_\perp,z))=0$ and the inelastic amplitudes correspond to
\begin{subequations}\label{eq:etapm_momzd_smear_long}
    \begin{align}
\eta_{ z, \pm}((\vec{k}'_\perp, z))=& \pm \frac{i}{\pi}r_{\rm e} e^{\mp i \phi_k} \tilde\ell({k}'_\perp, z; \pm 1) \,,\\
\tilde\ell({k}'_\perp, z; \pm 1):=& \int_{-\infty}^\infty dk'_z\, e^{ik'_z z} \ell({k}'_\perp, k'_z; \pm 1)\,,
    \end{align}
\end{subequations}
where we highlight the azimuthal angle dependence resulting from the transference of angular momentum between the spin and the electron beam.
As a consequence, we use the non-scattered amplitude and the inelastic scattered amplitudes to describe the joint electron-spin state with $z_p=0$.
The former is explicitly given by
\begin{equation}\label{eq:eta1_momzd_smear_long}
    \eta_1((\vec{k}_\perp, z))=\frac{1}{\sqrt{2\pi}}e^{ik_{z,0}(z)}\tilde\varphi_{\rm in,\|}(z+z_p)\varphi_{\rm in, \perp}(k_\perp).
\end{equation}
Meanwhile, the function $\tilde\ell({k}'_\perp, z; \pm 1)$ is obtained from the second line of Eq.~\eqref{eq:betanvar_pos_paraxial_approx} by performing the integral over $k_z$ while omitting the subsequent Fourier transform into transverse coordinates, which results in
\begin{equation}
    \begin{aligned}
    \tilde\ell({k}'_\perp, z; \pm 1) \approx e^{i k_{z,0} z}\,\tilde\varphi_{{\rm in, \|}}(z)\,\cL_{\varsigma}(k'_\perp/k_{z,0}; -z).
    \end{aligned}
\end{equation}
This expression applies within the paraxial framework of Eq.~\eqref{eq:L_nvarsigma_deflection_3level}, where complete longitudinal overlap renders the longitudinal dependence irrelevant. Alternatively, under the conditions established from Eq.~\eqref{eq:qz_sol_complete_ratio}, the function $\cL_{\varsigma}(k'_\perp/k_{z,0}; -z)$ is approximated by Eq.~\eqref{eq:L_nvarsigma_deflection_4level}.

Considering an azimuthally symmetric incident electron wavepacket focused to the sample plane
($z_p=0$), we find the following scattering amplitudes in $(\vec{k}'_\perp, z_d)$ representation to second order
\begin{subequations}
    \begin{align}
{S}^{(2)}_{\vec{k}'_\perp, z, s_1;{\rm in}, s}=&\eta_1((\vec{k}'_\perp, z))\delta_{s_1,s}+{\tau}^{(1)}_{\vec{k}'_\perp, z, s_1;{\rm in}, s}+{\tau}^{(2)}_{\vec{k}'_\perp, z, s_1;{\rm in}, s},\\
{\tau}^{(1)}_{\vec{k}'_\perp, z, s_1;{\rm in}, s}=&\bra{\vec{k}'_\perp, z, s_1}\hat{\tau}^{(1)}|\varphi_{\rm in},s\ra=\la s_1|{\eta_{z,+}((\vec{k}'_\perp, z))\osigma_{z,+}+\eta_{z,-}((\vec{k}'_\perp, z))\osigma_{z,-}}|s\ra,\\
{\tau}^{(2)}_{\vec{k}'_\perp, z, s_1;{\rm in}, s}=&\frac{1}{2}\bra{\vec{k}'_\perp, z, s_1}\hat{\tau}^{(1)~2}|\varphi_{\rm in},s\ra=\frac{1}{2}\int d^2 k_\perp^{\prime\prime}\int dz^{\prime\prime}_d\sum_{s''} {\tau}^{(1)}_{\vec{k}'_\perp, z, s_1;\vec{k}^{\prime\prime}_\perp, z^{\prime\prime}, s^{\prime\prime}}{\tau}^{(1)}_{\vec{k}^{\prime\prime}_\perp, z^{\prime\prime}, s^{\prime\prime};{\rm in}, s},
    \end{align}
\end{subequations}
where ${\tau}^{(1)}_{\vec{k}'_\perp, z, s_1;\vec{k}^{\prime\prime}_\perp, z^{\prime\prime}, s^{\prime\prime}}=\bra{\vec{k}'_\perp, z, s_1}\hat{\tau}^{(1)}|\vec{k}^{\prime\prime}_\perp, z^{\prime\prime}, s^{\prime\prime}\ra$.
Then, the reduced density matrix of the spin state is obtained from the final 
state given in Eq.~\eqref{eq:psi_final_pure} as:
\begin{equation}
    \hat{\rho}_{\rm s, f} := \operatorname{Tr}_{\rm e} \left[ |\tilde{\psi}_{\rm out}\rangle \langle \tilde{\psi}_{\rm out}| \right]
    = \frac{1}{\mathcal{N}} \sum_{s_1, s_2} \rho_{\rm s, f}(s_1, s_2) \, |s_1\rangle \langle s_2|\,,
\end{equation}
where the un-normalized matrix elements are given by
\begin{equation}
    \tilde\rho_{\rm s, f}(s_1, s_2) = \int d^3 \xi \, 
    S_{\vec{\xi}, s_1; {\rm in}, s} \, 
    S^*_{\vec{\xi}, s_2; {\rm in}, s}\,,
\end{equation}
and the normalization factor is
$\mathcal{N} = \operatorname{Tr}_{s'}[\hat{\rho}_{\rm s, f}] =  \tilde\rho_{\rm s, f}(\uparrow, \uparrow) +  \tilde\rho_{\rm s, f}(\downarrow, \downarrow)$. Working with an initially pure spin state and expanding up to second order in the interaction strength, we obtain
\begin{equation}\label{eq:rhotilde_sf_pre}
    \begin{aligned}
        \tilde\rho_{\mathrm{s}, f}(s_1, s_2) 
        \approx& 
        \int d^2k'_\perp\, dz \,
        \Big(
            \eta_1((\vec{k}'_\perp, z))\, \delta_{s_1,s}
            + {\tau}^{(1)}_{\vec{k}'_\perp, z, s_1; {\rm in}, s}
            + {\tau}^{(2)}_{\vec{k}'_\perp, z, s_1; {\rm in}, s}
        \Big)
        \Big(
            \eta_1^*((\vec{k}'_\perp, z))\, \delta_{s_2,s}
            + {\tau}^{(1)*}_{\vec{k}'_\perp, z, s_2; {\rm in}, s}
            + {\tau}^{(2)*}_{\vec{k}'_\perp, z, s_2; {\rm in}, s}
        \Big) \\[4pt]
        \approx& 
        \int d^2k'_\perp\, dz\, 
        |\eta_1((\vec{k}'_\perp, z))|^2 \, \delta_{s_1,s_2}\delta_{s,s_1}
        + \int d^2k'_\perp\, dz\,
        {\tau}^{(1)}_{\vec{k}'_\perp, z, s_1; {\rm in}, s}
        {\tau}^{(1)*}_{\vec{k}'_\perp, z, s_2; {\rm in}, s} \\[4pt]
        &+ \left[
            \int d^2k'_\perp\, dz\, 
            \eta_1((\vec{k}'_\perp, z))\, \delta_{s_1,s}\,
            \left(
                {\tau}^{(1)*}_{\vec{k}'_\perp, z, s_2; {\rm in}, s}
                + {\tau}^{(2)*}_{\vec{k}'_\perp, z, s_2; {\rm in}, s}
            \right)
            + \text{(c.c. and $s_1 \leftrightarrow s_2$)}
        \right],
    \end{aligned}
\end{equation}
where in the second line we retain only terms up to second order in the interaction strength. 
The notation (c.c. and $s_1 \leftrightarrow s_2$) denotes the complex conjugate of the preceding terms with the spin indices exchanged.

Then, each of the contributions appearing in Eq.~\eqref{eq:rhotilde_sf_pre} can be explicitly developed as follows.
We begin with the first two terms:
\begin{subequations}
    \begin{align}
\int d^2k'_\perp\, dz\, 
        |\eta_1((\vec{k}'_\perp, z))|^2  =&\, 1,
        \\[3pt]
\begin{split}    
        \int d^2k'_\perp dz\,{\tau}^{(1)}_{\vec{k}'_\perp, z, s_1;{\rm in}, s}\,
        {\tau}^{(1)~*}_{\vec{k}'_\perp, z, s_2;{\rm in}, s}
        =& \int d^2k'_\perp dz\,
        \langle s_1|\eta_{z,+}((\vec{k}'_\perp, z))\osigma_{z,+}
        +\eta_{z,-}((\vec{k}'_\perp, z))\osigma_{z,-}|s\rangle\\
        &\qquad \qquad \qquad
        \langle s|\eta^*_{z,+}((\vec{k}'_\perp, z))\osigma_{z,-}
        +\eta^*_{z,-}((\vec{k}'_\perp, z))\osigma_{z,+}|s_2\rangle\\
        \approx & \Delta \mathcal{P}
\big(\langle s_1|\osigma_{z,+}|s\rangle\langle s|\osigma_{z,-}|s_2\rangle
+\langle s_1|\osigma_{z,-}|s\rangle\langle s|\osigma_{z,+}|s_2\rangle\big),
\end{split}
    \end{align}
\end{subequations}
where in the second line of the last expression, we have employed the relations
\begin{equation}
    \begin{aligned}
\int d^2k'_\perp dz\, 
\eta_{z,\pm}((\vec{k}'_\perp, z))\eta^*_{z,\mp}((\vec{k}'_\perp, z))=0
\,,\\[4pt]
\Delta \mathcal{P}:= 
\int d^2k'_\perp dz 
|\eta_{z,+}((\vec{k}'_\perp, z))|^2
\approx&\,\int d^2k'_\perp dz 
|\eta_{z,-}((\vec{k}'_\perp, z))|^2 .
    \end{aligned}
\end{equation}
The first relation follows from the fact that the integrand contains phase terms proportional to 
$e^{\pm 2i \phi_k}$, which vanish upon integration over the azimuthal angle $\phi_k$. 
This cancellation arises because the incident electron beam is azimuthally symmetric, 
so only contributions sharing the same $\phi'_k$-dependence survive in the trace operation. 
The second relation results from Eq.~\eqref{eq:etapm_momzd_smear_long} with $\cL_{\varsigma}(k'_\perp/k_{z,0}; -z)$ from Eq.~\eqref{eq:L_nvarsigma_deflection_4level}, which yield 
$|\eta_{z,+}((\vec{k}'_\perp, z))| \approx |\eta_{z,-}((\vec{k}'_\perp, z))|$.
Next, we consider the mixed terms that appear in the last line of Eq.~\eqref{eq:rhotilde_sf_pre}:
\begin{subequations}
    \begin{align}
    \begin{split}
\int d^2k'_\perp dz\,\eta_1((\vec{k}'_\perp, z))\, \delta_{s_1,s}\,
{\tau}^{(1)*}_{\vec{k}'_\perp, z, s_2; {\rm in}, s}
=&\, \int d^2k'_\perp dz\,\eta_1((\vec{k}'_\perp, z))
\langle s_1|
\eta^*_{z,+}((\vec{k}'_\perp, z))\osigma_{z,-}\\        &\qquad\qquad\qquad\qquad\qquad\qquad\qquad+\eta^*_{z,-}((\vec{k}'_\perp, z))\osigma_{z,+}
|s_2\rangle
=0,
    \end{split}\\
\begin{split}
\int d^2k'_\perp dz\,\eta_1((\vec{k}'_\perp, z))\,
\delta_{s_1,s}\,{\tau}^{(2)~*}_{\vec{k}'_\perp, z, s_2;{\rm in}, s}
=&\,\frac{1}{2} \int d^2 k_\perp^{\prime\prime}\!\int dz^{\prime\prime}_d\sum_{s''}
\int d^2k'_\perp dz\, 
\eta_1((\vec{k}'_\perp, z))\,\delta_{s_1,s}\,
{\tau}^{(1)~*}_{\vec{k}'_\perp, z, s_2;\vec{k}^{\prime\prime}_\perp, z^{\prime\prime}, s^{\prime\prime}}
{\tau}^{(1)~*}_{\vec{k}^{\prime\prime}_\perp, z^{\prime\prime}, s^{\prime\prime};{\rm in}, s}\\
=&\,\frac{1}{2} \int d^2 k_\perp^{\prime\prime}\!\int dz^{\prime\prime}_d\sum_{s''}
\delta_{s_1,s} \,
\langle{\vec{k}^{\prime\prime}_\perp, z^{\prime\prime}, s^{\prime\prime}}|
\tau^{(1)~\dagger}
\!\int d^2k'_\perp dz\, 
|\vec{k}'_\perp, z\rangle 
\langle\vec{k}'_\perp, z|\varphi_{\rm in}, s_2\rangle\\
&\qquad\qquad\qquad\qquad\qquad\qquad
\langle \varphi_{\rm in}, s|
\tau^{(1)~\dagger}
|{\vec{k}^{\prime\prime}_\perp, z^{\prime\prime}, s^{\prime\prime}}\rangle\\
=&\,-\frac{1}{2} \int d^2 k_\perp^{\prime\prime}\!\int dz^{\prime\prime}_d
\sum_{s''}\delta_{s_1,s}\,
\langle{\vec{k}^{\prime\prime}_\perp, z^{\prime\prime}, s^{\prime\prime}}|
\tau^{(1)}|\varphi_{\rm in}, s_2\rangle\,
\big(\langle\vec{k}^{\prime\prime}_\perp, z^{\prime\prime}, s^{\prime\prime} |
\tau^{(1)}|\varphi_{\rm in}, s_1\rangle\big)^*\\
=&\,-\frac{1}{2}
|\eta_{z,\pm}((\vec{k}^{\prime\prime}_\perp, z^{\prime\prime}))|^2
{\Delta\mathcal{P}}
\,\delta_{s_1, s_2}\,\delta_{s_1,s},
\end{split}
    \end{align}
\end{subequations}
where we have used the relation $\tau^{(1)~\dagger} = -\,\tau^{(1)}$ and
\begin{equation}
\int d^2k'_\perp dz\, 
\eta_1((\vec{k}'_\perp, z))\eta^*_{z,\mp}((\vec{k}'_\perp, z))
=  0 
\end{equation}
which follows because the integrand contains phase terms proportional to $e^{\pm i \phi_k}$.

By inserting all the previously derived results into Eq.~\eqref{eq:rhotilde_sf_pre}, 
the matrix elements of the reduced spin density operator can be compactly expressed as
\begin{equation}\label{eq:rhotilde_sf}
    \begin{aligned}
        \tilde\rho_{\mathrm{s}, f}(s_1, s_2) 
        \approx& (1-\Delta\mathcal{P})\delta_{s_1,s_2}\delta_{s_1,s}+ \Delta\mathcal{P}
        \left(\langle s_1|\osigma_{z,+}|s\rangle\langle s|\osigma_{z,-}|s_2\rangle+\langle s_1|\osigma_{z,-}|s\rangle\langle s|\osigma_{z,+}|s_2\rangle\right).
    \end{aligned}
\end{equation}
In this Appendix, we focus on the on-resonance configuration illustrated in Fig.~\ref{fig:spectroscopy_imaging_TEM}, 
where the spin state after the microwave driving is given by $\ket{s} = \ket{\uparrow_y}$. 
Using the operator identity
\[
\osigma_{z,\pm}
= -\frac{i}{2}(\osigma_{y, +}-\osigma_{y,-})
\pm \frac{i}{2}\osigma_y,
\quad
\text{with} \quad
\osigma_{y, \pm}=\frac{1}{2}(\osigma_z\pm i\osigma_x),
\]
we obtain the relations
\begin{equation}
    \begin{aligned}
\langle s_1|\osigma_{z,\pm}|\uparrow_y\rangle
= \langle s_1|
-\tfrac{i}{2}(\osigma_{y, +}-\osigma_{y,-})
\pm \tfrac{i}{2}\osigma_y
|\uparrow_y\rangle
= \frac{i}{2}\left(\delta_{s_1, \downarrow_y}\pm\delta_{s_1, \uparrow_y}\right).
    \end{aligned}
\end{equation}
Substituting this result into Eq.~\eqref{eq:rhotilde_sf}, we find that 
$\tilde\rho_{\mathrm{s}, f}(\uparrow_y, \downarrow_y) \approx 0$, 
indicating that off-diagonal spin coherences vanish in this configuration. 
Consequently, the reduced spin density matrix takes the diagonal form
\begin{equation}
    \tilde{\rho}_{{\rm s}, f}\approx
    \begin{pmatrix}
        1 -\tfrac{1}{2}\Delta \mathcal{P} & 0\\[4pt]
        0 & \tfrac{1}{2}\Delta \mathcal{P}
    \end{pmatrix}.
\end{equation}
Since both the diagonal elements contain terms of second-order in the interaction strength, the purity
\begin{equation}
    \begin{aligned}
        \mathcal{P}\approx\left( 
    | \rho_{\rm s, f}(\uparrow_y, \uparrow_y)|^2 
    + | \rho_{\rm s, f}(\downarrow_y, \downarrow_y)|^2 \right) 
        &\approx 1-\Delta \mathcal{P}.
    \end{aligned}
\end{equation}
The loss of purity can be written as
\begin{equation}
    \begin{aligned}
        \Delta \mathcal{P} &   \approx \int d^2k'_\perp \, d{z} \, {\frac{1}{\pi^2}} r^2_{\rm e} 
        \left|\tilde{\varphi}_{{\rm in}, \|}(z)\right|^2 
        {| \cL_1(k'_\perp/k_{z,0}; -z)|^2}
        \approx \frac{2}{\pi} r_{\rm e}^2 \int  dk'_\perp dz   k'_\perp\, 
        \left|\tilde{\varphi}_{{\rm in}, \|}(z)\right|^2 
         {| \cL_1(k'_\perp/k_{z,0}; -z)|^2}\\
        &= \frac{2}{\pi} r_{\rm e}^2 (2 \Delta k_\perp)^2 \int_{-\infty}^\infty d{z} \, 
        \left|\tilde{\varphi}_{{\rm in}, \|}(z)\right|^2 
        \int_0^\infty d\tilde{q}_\perp \int_0^\infty d\tilde{q}'_\perp 
        e^{-(\tilde{q}_\perp^2+\tilde{q}_\perp^{'2})} 
        e^{-i\frac{\tilde{z}}{2\tilde{k}_{z,0}}(\tilde{q}_\perp^2-\tilde{q}_\perp^{'2})} 
        \mathcal{I}_{\rm es}(2\Delta k_\perp\tilde{q}_\perp)
        \mathcal{I}_{\rm es}(2\Delta k_\perp\tilde{q}'_\perp) \\
        &\qquad\qquad\qquad\qquad \int_0^\infty  d\tilde{k}'_\perp  \tilde{k}'_\perp 
        e^{-2\tilde{k}_\perp^{'2}}
        I_1\left(\tilde{k}'_\perp \tilde{q}_\perp \left(2 + \frac{i\tilde{z}}{\tilde{k}_{z,0}}\right)\right)
        I_1\left(\tilde{k}'_\perp \tilde{q}'_\perp \left(2 - \frac{i\tilde{z}}{\tilde{k}_{z,0}}\right)\right) \\
        &= 4r_{\rm e}^2 (2 \Delta k_\perp)^2 \int_0^\infty d\tilde{q}_\perp \int_0^\infty d\tilde{q}'_\perp 
        \mathcal{I}_{\rm es}(2\Delta k_\perp\tilde{q}_\perp) 
        \mathcal{I}_{\rm es}(2\Delta k_\perp\tilde{q}'_\perp) \\
        &\qquad\qquad\qquad\qquad \frac{\tilde{\Delta k_z}}{\sqrt{2\pi}} 
        \int_{-\infty}^\infty d\tilde{z} \, 
        e^{-\frac{1}{2} \tilde{z}^2 \tilde{\Delta k}_z^2} 
        \exp\left[-\left(\frac{1}{2} + \frac{\tilde{z}^2}{8\tilde{k}_{z,0}^2}\right) (\tilde{q}_\perp^2+\tilde{q}_\perp^{'2})\right]
        I_1\left(\tilde{q}_\perp \tilde{q}'_\perp \left(1 + \frac{\tilde{z}^2}{4 \tilde{k}_{z,0}^2}\right)\right),
    \end{aligned}
\end{equation}
where we introduce the scaled variables $\tilde{k}'_\perp= k'_\perp/(2\Delta k_\perp)$,
$\tilde{q}_\perp= q_\perp/(2\Delta k_\perp)$, $\tilde{q}'_\perp = q_\perp/(2 \Delta k_\perp)$, $\tilde{z} = 2 z \Delta k_\perp$, $\tilde{k}_{z,0} = k_{z,0}/(2 \Delta k_\perp)$, and $\tilde{\Delta k}_z = \Delta k_z / \Delta k_\perp$. 

Assuming a hydrogen-like atomic density with Bohr radius $a_0 = 52.9~\mathrm{pm}$ and an incident beam with $E_{\rm kin} = 200$~keV and longitudinal $\Delta k_\perp=4.22\cdot 10^{-7} k_{z,0}$ (FWHM~$=1.1~\text{\textmu}$m), numerical estimates for the loss of purity are:
\begin{itemize}
    \item $\Delta k_\perp=4.22\times 10^{-6} k_{z,0}$ (FWHM=$110$~nm): $ \Delta \mathcal P \approx 3.81 \times 10^{-14}$,
    \item $\Delta k_\perp=4.22\cdot 10^{-5} k_{z,0}$ (FWHM=$11$~nm): $ \Delta \mathcal P \approx 3.79 \times 10^{-12}$,
    \item $\Delta k_\perp=4.22\cdot 10^{-4} k_{z,0}$ (FWHM=$1.1$~nm): $ \Delta \mathcal P \approx 2.71 \times 10^{-10}$.
\end{itemize}
We conclude that even in the strongly focused case, continuous detection of $\sim 10^{7}$ electrons would reduce the spin purity by only about 
$10^{-3}$. This demonstrates that the loss of spin purity depends on the beam focusing, but is highly suppressed due to the $r_{\rm e}^2$ scaling. Hence, spin coherence remains essentially intact under realistic beam parameters, and decoherence induced by scattering is negligible in this regime.

\subsection{Coherent pure final state}\label{app:coherent_wavefunction}

The effective first-order probability amplitudes for the joint electron-spin state when the bias field is applied longitudinally and the spin state is initialized at $\ket{s}=\ket{\uparrow_y}$ are in position representation
\begin{equation}\label{eq:scatt_amp_coherent_incoherent}
    \begin{split}
S_{\vec{r}_\perp, z,s';{\rm in},s}\approx
    \la s'| \tilde\beta_1((\vec{r}_\perp, z))&+\tilde\beta_{z,+}((\vec{r}_\perp, z))\osigma_{z,+}+\tilde\beta_{z,-}((\vec{r}_\perp, z))\osigma_{z,-}|s\ra\\
    = 
    \langle s' | 
    \tilde\beta_1((\vec{r}_\perp, z))
    &+ \frac{i}{2} \left( \tilde\beta_{z,+}((\vec{r}_\perp, z)) - \tilde\beta_{z,-}((\vec{r}_\perp, z)) \right) \hat{\sigma}_y \\
    &    - \frac{i}{2} \left( \tilde\beta_{z,+}((\vec{r}_\perp, z)) + \tilde\beta_{z,-}((\vec{r}_\perp, z)) \right) \left( \hat{\sigma}_{y,+} - \hat{\sigma}_{y,-} \right) 
    | s \rangle,
    \end{split}
\end{equation}
where clearly the final state corresponds to
\begin{equation}
    \ket{\psi_{\rm out}(z)} = \int d^2r_\perp \sum_{s'} S_{\vec{r}_\perp, z,s';{\rm in},s}(z, t)\ket{\vec{r}_\perp, z, s'}.    
\end{equation}
If we only consider the coherent interaction, where no spin flip is produced due to the interaction with the electron beam, we define a coherent electron state given by:
\begin{equation}\label{eq:psi_ecoh}
    \ket{\psi_{\rm e, coh}(z)}=\frac{1}{\sqrt{\mathcal{N}_{\rm coh}}} \int d^2r_\perp \left[\tilde\beta_1((\vec{r}_\perp, z))+\frac{i}{2}(\tilde\beta_{z,+}((\vec{r}_\perp, z))-\tilde\beta_{z,-}((\vec{r}_\perp, z))) \right] \ket{\vec{r}_\perp}\,,
\end{equation}
where {the norm of this state is given by:
\begin{equation}\label{eq:normalization_Ncoh}
\begin{aligned}
\mathcal{N}_{\rm coh}&\approx\int d^2r_\perp \left|\tilde\beta_1((\vec{r}_\perp, z))+\frac{i}{2}(\tilde\beta_{z,+}((\vec{r}_\perp, z))-\tilde\beta_{z,-}((\vec{r}_\perp, z)))\right|^2\approx \int d^2r_\perp \left|\tilde\beta_1((\vec{r}_\perp, z))\right|^2\\
&\approx \int d^2r_\perp \left| 2\sqrt{2\pi}\Delta k_\perp \tilde{\varphi}_{{\rm in},\|}(z) e^{-r_\perp^2\Delta k_\perp^2} \right|^2 \\
        &= 4\pi^2 |\tilde{\varphi}_{{\rm in},\|}(z)|^2 = 4\pi^2 \sqrt{8\pi}\Delta k_z e^{-2z^2 \Delta k_z^2},
    \end{aligned}
\end{equation}
where we use, in the first line, the fact that the un-scattered amplitude is much larger than the scattered ones.
The scattering amplitudes and phases derived in Eq.~\eqref{eq:psi_ecoh} are visualized in Figure~\ref{fig:coherent_wavefunction}. Graphs (a) and (b) display the scattered amplitudes $\tilde\beta_{z,\pm}$ normalized to the range $[0,1]$. The remaining graphs characterize the total coherent amplitude, defined as $\Psi_{\rm coh} = \tilde\beta_1 + \frac{i}{2}(\tilde\beta_{z,+} - \tilde\beta_{z,-})$. Specifically, Figure~\ref{fig:coherent_wavefunction}~c plots the normalized magnitude $|\Psi_{\rm coh}|/\sqrt{\mathcal{N}_{\rm coh}}$, while Figure~\ref{fig:coherent_wavefunction}~d displays the corresponding phase argument $\text{arg}[\Psi_{\rm coh}]$.}

\section{Classical Fisher Information and Optimization}\label{ap:CFI_optimal}

The observed signal from a single spin is inherently small due to the weak spin-orbit interaction. The probability density of the images presented in the main text ($\vec{\xi}_\perp=\vartheta\vu(\phi_k)$ for diffraction mode and $\vec{\xi}_\perp=\vec{r}_\perp$ for image mode) can be expressed as $P(\vec{\xi}_\perp) \approx P_0(\vec{\xi}_\perp) + P_1(\vec{\xi}_\perp)$, where $P_0(\vec{\xi}_\perp)$ represents the zero-deflection peak distribution, and $P_1(\vec{\xi}_\perp)$ corresponds to the first-order contribution.

To evaluate the sensitivity of the differential measurement procedure to the value of the magnetic moment of the electron spin (whose value is known to be $\mu_{\rm B}$), we evaluate the classical Fisher information (CFI) in the subset $X$ of the sample space
\begin{equation}\label{eq:cfi_first_order}
    \begin{aligned}
        \mu_{\rm B}^2 \, \mathrm{CFI}
        &:= \int_X d^2\xi_\perp \,
        \frac{\mu_{\rm B}^2}{P(\vec{\xi}_\perp)}
        \left( \frac{\partial P(\vec{\xi}_\perp)}{\partial \mu_{\rm B}} \right)^2 
        \approx
        \int_X d^2\xi_\perp \, \frac{\big(P_1(\vec{\xi}_\perp)\big)^2}{P_0(\vec{\xi}_\perp)}.
    \end{aligned}
\end{equation}
In the approximation, we used the facts that $P_1 \propto \mu_{\rm B}$ and $P_1 \ll P_0$. Note that the CFI above is directly related to the conditional CFI that can be calculated from the renormalized probability distribution $\tilde{P}(\vec{\xi}_\perp) = P(\vec{\xi}_\perp)/\mathcal{N}_X$ (where $\mathcal{N}_X \approx N_{\rm e,det}/N_{\rm e}$) restricted to $X$ by multiplication with the detection efficiency factor $N_{\rm e,det}/N_{\rm e}$, where the number of detected electrons $N_{\rm e,det}$ can be approximated as
\begin{equation}
    N_{\rm e,det} \approx N_{\rm e} \int_X d^2\xi_\perp \, P(\vec{\xi}_\perp)  \approx N_{\rm e} \int_X d^2\xi_\perp \, P_0(\vec{\xi}_\perp).
\end{equation}

The total SNR of the image can be bounded using the Cram{\'e}r-Rao inequality~\cite{cramer1946,rao1945}, which sets a limit on the variance of the estimator
\begin{equation}\label{eq:SNR_bound_ap}
    \begin{aligned}
        \text{SNR}=\frac{\mu_{\rm B}}{\Delta \mu_{\rm B}}\leq \mu_{\rm B}\sqrt{N_{\rm e}\text{CFI}}\,.
    \end{aligned}
\end{equation}
For our simulations, we fix the total electron count at $N_{\rm e}=10^{10}$, which corresponds to an acquisition time of $t_{\rm acq}=1~\mathrm{s}$ given a beam current of $1.6~\mathrm{nA}$. 
{The SNR upper bound is evaluated on resonance, where the spin state is defined by $\langle \hat{\sigma}_x \rangle = \langle \hat{\sigma}_z \rangle = 0$ and $\langle \hat{\sigma}_y \rangle = 1$. Under these conditions, we calculate the CFI via Eq.~\eqref{eq:cfi_first_order}. For diffraction mode imaging, we model the probability density using a background term $P_0=\Pi_0$ and a signal term $P_1=-\Im(\Pi_{z,1}-\Pi_{z,-1})$, with the normalization constant $\mathcal{N}_{\rm diff}=1$. Conversely, for the defocused image mode, we utilize the corresponding position-space components $\tilde{\Pi}_j$ (where $j\in\{0, z\pm 1\}$) with the normalization $\tilde{\mathcal{N}}_{\rm img}=(2\pi)^{-2}$. These calculations follow the simulation procedures detailed in the main text and Appendices~\ref{ap:general_diffraction} and \ref{ap:general_position_mode}, respectively.}

In diffraction mode, nearly all transmitted electrons contribute to the signal since features are analyzed near the zero-deflection peak (typically up to $8\Delta k_\perp/k_{z,0}$). In contrast, image mode is spatially restricted, recording only electrons within a region $X=[-x_{\rm max}, x_{\rm max}]\times [-x_{\rm max}, x_{\rm max}]$, where $x_{\rm max}$ represents the half-width of the square detection region. For a Gaussian beam illuminating the sample, the number of detected electrons is given by $N_{\rm e,det}= N_{\rm e} \, \left[\erf\left({\sqrt{2}}{x_{\rm max} \, \Delta k_\perp}\right)\right]^2$, where $\Delta k_\perp$ is the transverse momentum width of the incident beam. Specifically, using $x_{\rm max}=10~\mathrm{\AA}$ and $\Delta k_\perp\sim 10^7-10^8~\mathrm{m}^{-1}$ yields $N_{\rm e,det}\sim 10^5-10^7$.

\subsection{Optimized data analysis}\label{ap:optimal_single_spins}

We propose a simple data analysis scheme based on the pixelation of the electron sensor. We consider the number of electrons detected in a specific pixel $j$ of a detection region $X$, denoted as $N_{\rm px}[j]$ and the reference value $N_{0,{\rm px}}[j]$ measured at far off-resonant MW driving conditions. These quantities are related to the continuous probability densities by integrating over the pixel area and scaling by the total number of interacting electrons, $N_{\rm e}$:
\begin{equation}
    N_{{\rm px}}[j] = N_{\rm e} \int_{\text{pixel } j} d^2\xi_\perp \, P(\vec{\xi}_\perp), \quad \text{and}\quad N_{0,{\rm px}}[j] = N_{\rm e} \int_{\text{pixel } j} d^2\xi_\perp \, P_0(\vec{\xi}_\perp)\,.
\end{equation}
The signal is then defined by the deviations from the off-resonant reference signal, i.e., $N_{{\rm px}}[j] - N_{0,{\rm px}}[j] $, and can be approximated as
\begin{equation}
    N_{1,{\rm px}}[j] := N_{\rm e} \int_{\text{pixel } j} d^2\xi_\perp \, P_1(\vec{\xi}_\perp)\,. 
\end{equation}
Note that $N_{1,{\rm px}}[j]$ can have negative values. Therefore, we define the signal magnitude for pixel $j$ as $|N_{1,{\rm px}}[j]|$ and we find that the pixel-resolved SNR, $\text{SNR}_\text{px}$, which relates the signal to the shot noise inherent to the total counts in that pixel, can be predicted by the expression
\begin{equation}
    \text{SNR}_\text{px}[j] \approx \frac{|N_{\rm 1, px}[j]|}{\sqrt{2N_{\rm 0, px}[j]+N_{\rm 1, px}[j]}}.
\end{equation}

To optimize data analysis and achieve the theoretical bound of information retrieval from the images, we  
use the following approach: 
\begin{enumerate}
    \item Pixelation: The raw images undergo a pixelation process, as shown in Fig.~\ref{fig:optimal_mask_image}. In image mode, this step reduces spatial resolution to enhance signal strength, while in diffraction mode, it accounts for the resolution limitations of the detectors.
    \item Masking Strategy: A selective mask is applied to the image such that only the uncovered pixels contribute to the signal analysis. The uncovered region $D_{\rm max}$ is determined by defining a threshold { $\text{SNR}_{\rm px, min }$ for $\text{SNR}_\text{px}$, where $D_{\rm max}=\{j\in X_{\rm px}|| \text{SNR}_{\rm px}[j]>\text{SNR}_{\rm px, min }\}$, where $X_{\rm px}$ is the pixelation of the detected region.}  
    \item Threshold Optimization: The optimal threshold
    {$\text{SNR}_{\rm px, min }>0$} is chosen to maximize the overall signal-to-noise ratio, using the following optimization function:
\begin{equation}
       \text{SNR} = \max_{\text{SNR}_{\rm px, min }}\left( \frac{\sum_{j\in D_{\rm max} } |N_{\rm 1, px}[j]|}{\sqrt{\sum_{j\in D_{\rm max} }(2N_{\rm 0, px}[j]+N_{\rm 1, px}[j]})} \right).
   \end{equation}
 
\end{enumerate}

\subsection{Position space imaging}\label{app:position_space_imaging}

\begin{figure}[h]
    \centering
    \includegraphics[width=0.9\linewidth]{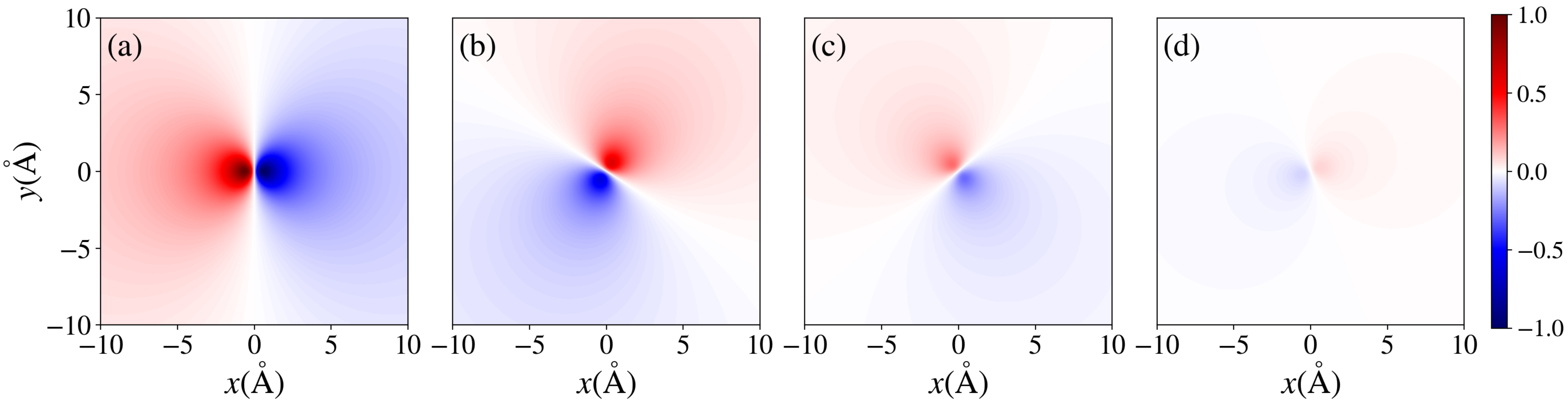}
    \caption{Zernike image mode: differential probability density from the interaction between a free electron and a single spin driven by a $\pi/2$ pulse, for varying detunings. Electron beam FWHM: $0.11~\text{\textmu m}~( \Delta k_\perp=4.22\times 10^{-6} k_{z,0})$; defocus: $z_d = 0~\text{\AA}$. (a)~$\delta = 0.000~\omega_0$, (b)~$\delta = 0.025~\omega_0$, (c)~$\delta = 0.050~\omega_0$, (d)~$\delta = 0.075~\omega_0$.} 
    \label{fig:spectroscopy_zernike_image_mode}
\end{figure}

In position-space imaging, the spatial distribution of the electron beam can be probed at arbitrary image planes by applying a suitable defocus. Additional insight into the spin-dependent scattering is obtained by sweeping the driving frequency across resonance, as illustrated in Fig.~\ref{fig:spectroscopy_imaging_TEM}~III for defocused imaging and Fig.~\ref{fig:spectroscopy_zernike_image_mode} for Zernike phase contrast. The Zernike-mode images were derived by imparting a $\pi/2$ phase shift to the unscattered amplitude; this corresponds to the substitution $\mathcal{B}_1(\vec{r}_\perp; z_d) \to i\mathcal{B}_1(\vec{r}_\perp; z_d)$ in Eq.~\eqref{eq:tildePi0_B1} and Eq.~\eqref{eq:Pi_nvar_pos}. Although Zernike phase imaging is experimentally challenging at the small scattering angles relevant for spin detection, our simulations indicate that it can reveal complementary features. As shown in Fig.~\ref{fig:spectroscopy_zernike_image_mode}, the Zernike-mode images display broader contrast regions concentrated near the optical axis. While the overall spatial structure resembles that of defocused imaging, the Zernike method lacks the interference fringe patterns since no propagation displacement was considered $z_d=0$.
In both imaging modes, off-resonant microwave driving reduces signal intensity and causes a rotation of the image pattern, consistent with the spin orientation set by the detuning. This rotation reflects the coherent spin dynamics and provides an additional handle for frequency-resolved spin-state discrimination.

A natural question arises: how does the informational content of the image vary with defocus, and how does this compare to alternative techniques such as Zernike phase imaging?

\begin{figure}[h]
    \centering
    \includegraphics[width=0.85\linewidth]{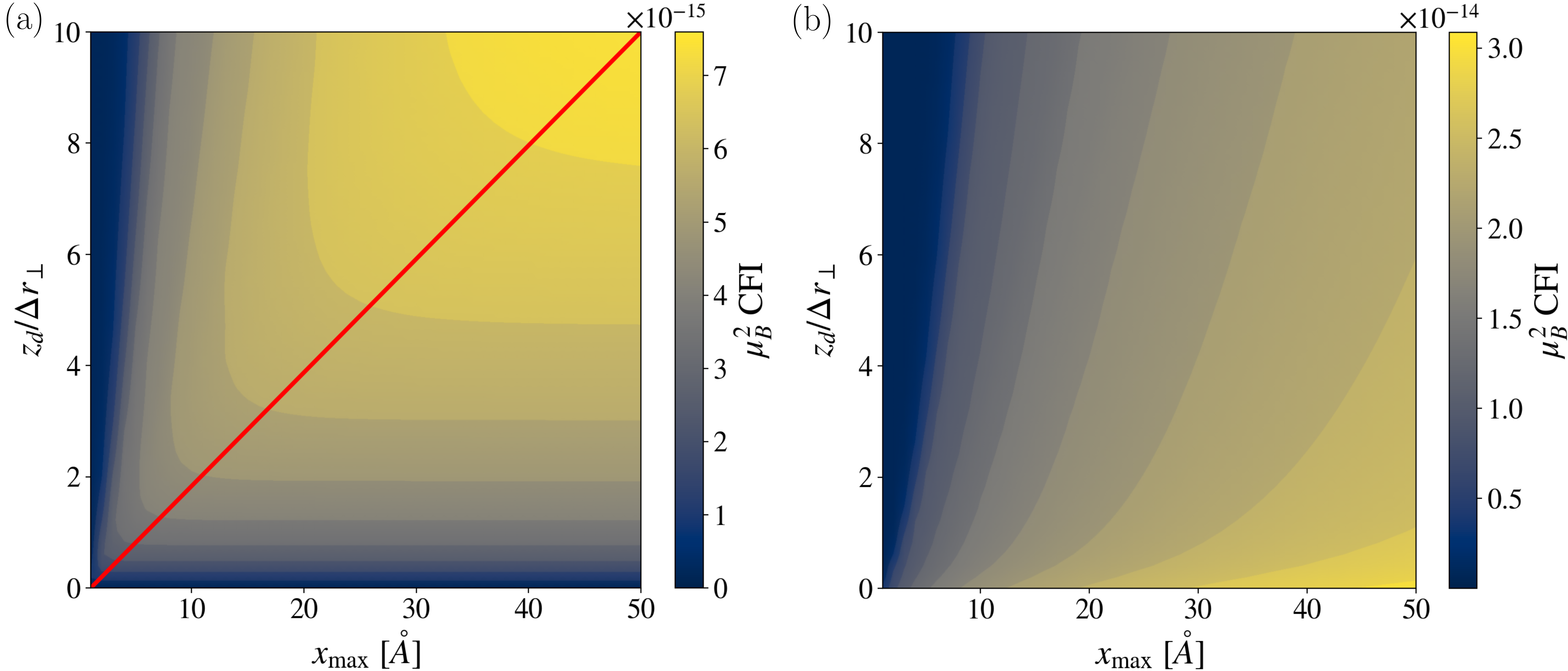}
    \caption{CFI as a function of the defocus distance $z_d$ and the half-side length $x_{\rm max}$ of the square detection region $X = [-x_{\rm max}, x_{\rm max}]\times [-x_{\rm max}, x_{\rm max}]$. The defocus is normalized to the transverse spread $\Delta r_\perp = 470$~\AA~($\Delta k_\perp=4.22\times 10^{-6} k_{z,0}$). The trends are shown for (a) standard defocused imaging and (b) Zernike phase imaging. In (a), the CFI increases with both defocus and detection size, though the values remain generally lower than the Zernike maximum CFI. The red curve marks the saturation threshold; to the right of this curve, expanding the detection region ($x_{\rm max}$) for a fixed defocus yields no additional information. In contrast, the Zernike configuration (b) exhibits a maximum at zero defocus ($z_d=0$) and increases monotonically with the detection region size.}\label{fig:CFI_defocus_series}
\end{figure}

\begin{figure}[h]
    \centering
\includegraphics[width=0.85\linewidth]{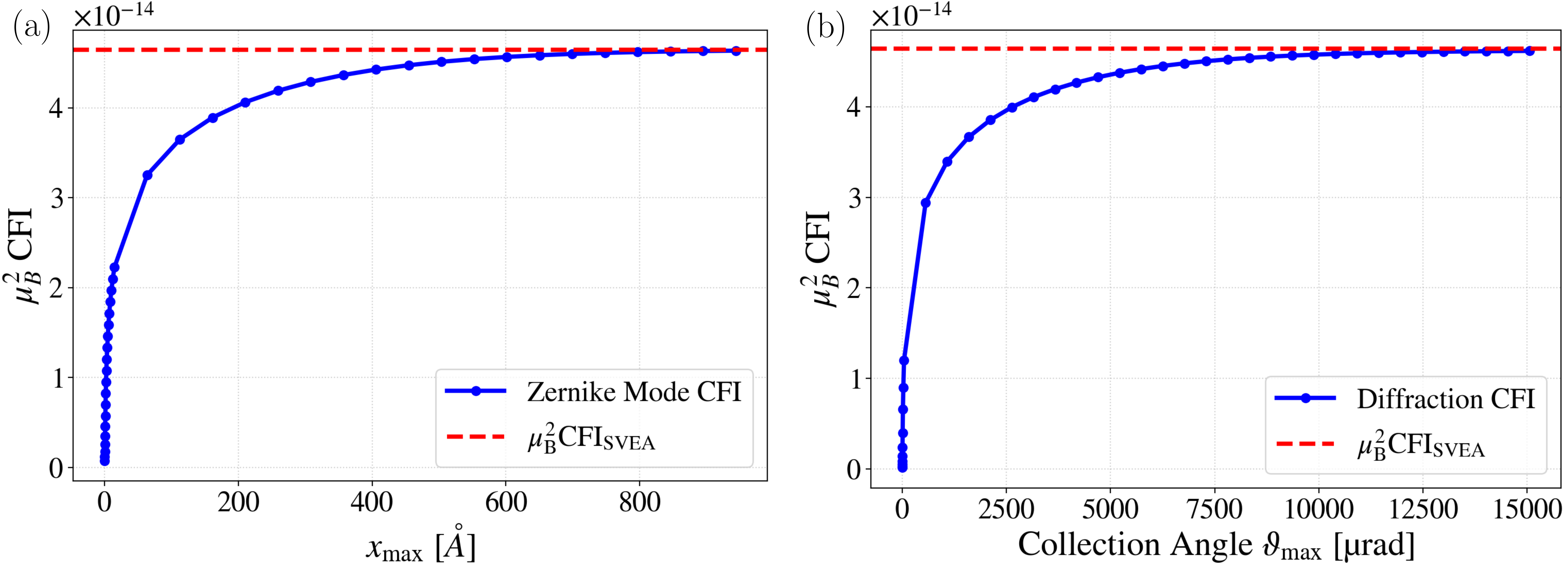}
    \caption{CFI dependence on detection region size for an incident beam with FWHM $= 110$~nm ($\Delta k_\perp=4.22\times 10^{-6} k_{z,0}$). The trends are displayed for (a) Zernike phase imaging at $z_d=0$ (as a function of half-side length $x_{\rm max}$) and (b) diffraction mode imaging (as a function of collection angle $\vartheta_{\rm max}$). In both configurations, convergence towards the classical limit~\cite{metrologySpinTEM} is achieved once the detection window resolves the governing physical features: the beam spread in position space ($x_{\rm max}>\Delta r_\perp=470~\text{\AA}$) and the spin size in reciprocal space ($\vartheta_{\rm max}>(a_0 k_{z,0})^{-1}=7.5$~mrad).}
    \label{fig:asympthotic_cfi}
\end{figure}

\begin{figure}[h]
    \centering
\includegraphics[width=0.66\linewidth]{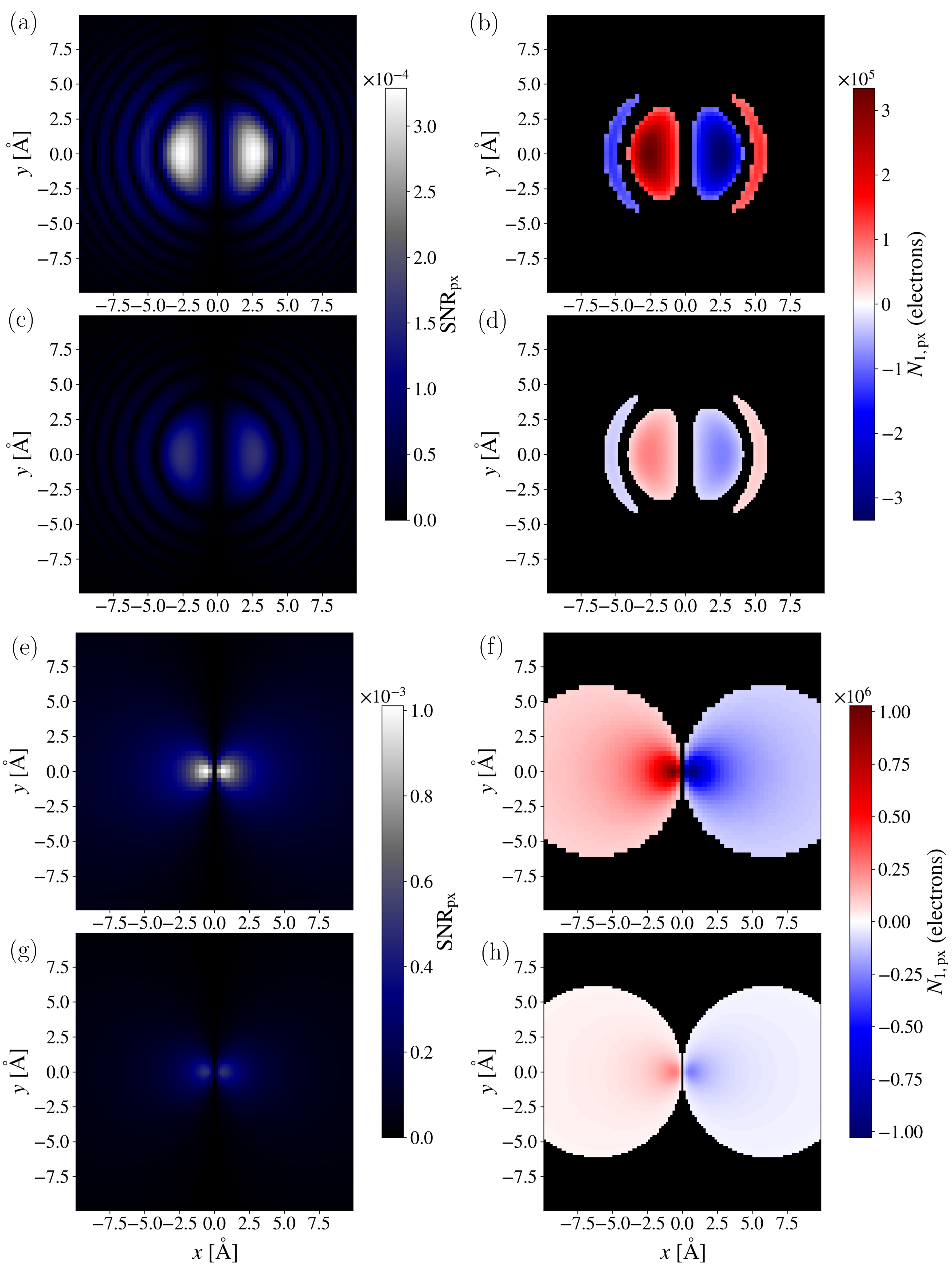}
    \caption{{The left column displays the SNR per pixel calculated from the images in the right column, which present the pixelated differential images. The color scale in the latter indicates how the average number of detected electrons per detector (or virtual) pixel changes when comparing two scenarios: a spin driven by an MW pulse before interacting with the electron beam, and a spin in the ground state aligned with the bias field. Black regions in the right column plots denote the optimized mask applied to maximize the total SNR. (a)--(d) Image mode with defocus $z_d = 1.7 \Delta r_\perp =800~\text{\AA}$. Pixel sizes are $0.32~\text{\AA}$ for (a, b) and $0.16~\text{\AA}$ for (c, d). (e)--(h) Zernike phase contrast imaging at $z_d=0~\text{\AA}$. Pixel sizes are $0.32~\text{\AA}$ for (e, f) and $0.16~\text{\AA}$ for (g, h). The calculation was done for an incident Gaussian electron beam wavepacket with $\text{FWHM} = 0.11~\text{\textmu m}$ ($\Delta k_\perp=(2\Delta r_\perp)^{-1}=4.22\times 10^{-6} k_{z,0}$). In addition, we assume $10^{10}$ electrons (corresponding to a $1$~s acquisition time with $1.6$~nA current) passing through the interaction plane, resulting in an electron dose of $7.2\times 10^{3}~\mathrm{e}^-/\text{\AA}^2$.}  }
    \label{fig:optimal_mask_image}
\end{figure}

To quantify these dependencies, we analyze the classical bound from Eq.~\eqref{eq:SNR_bound_ap} under various imaging configurations, varying both the defocus and the detection region size. We consider both conventional defocused imaging and Zernike phase imaging, specifically restricting the analysis to the on-resonance case.
As shown in Fig.~\ref{fig:CFI_defocus_series}~a, the CFI increases with both defocus and the detection size $x_{\rm max}$ in conventional image mode. This behavior arises from the spatial spreading of the signal away from the optical axis. As defocus increases, the localized information originally encoded near the origin disperses over a wider area, facilitating a trade-off between spatial resolution and signal integration. However, the decay of interference fringes at large spatial coordinates establishes a saturation threshold, marked by the red curve in Fig.~\ref{fig:CFI_defocus_series}~a. To the right of this curve, the information content saturates, meaning no additional information is extracted by further enlarging the detection region. This broadening enhances detectability at the expense of spatial sharpness, a trade-off that proves advantageous for capturing low-amplitude signals. Crucially, in this limit of large defocus and detection area, the CFI converges to the metrological maximum established by diffraction mode imaging, consistent with findings in~\cite{metrologySpinTEM}.

In contrast, Zernike phase imaging exhibits a different behavior, as illustrated in Fig.~\ref{fig:CFI_defocus_series}~b. In this case, the highest CFI is observed at zero defocus and when the detection region is larger. Introducing defocus degrades the contrast and reduces the intensity of critical features, thereby diminishing the CFI and SNR. This suggests that Zernike imaging is most effective when phase contrast is maximized without propagation.
In the limit of a large detection region, as depicted in Fig.~\ref{fig:asympthotic_cfi}, the CFI converges to the maximum value associated with diffraction mode~\cite{metrologySpinTEM}. This convergence occurs provided the imaging setup possesses sufficient resolution to resolve features of size $a_0$ and a field of view exceeding the beam's characteristic transverse length.
We compute this theoretical maximum ($\mu_{\rm B}^2\mathrm{CFI}_{\text{SVEA}}\approx 4.64\times 10^{-14}$ for $\chi=892$) under the slowly varying envelope approximation (SVEA) using Eq.~(28) from~\cite{metrologySpinTEM}, incorporating the expression for a spin in the Hydrogen 1s orbital (Eq.~(G.8)).

For on-resonance defocused imaging ($z_d=800~\text{\AA}$ and FWHM~$=110$~nm; see Fig.~\ref{fig:spectroscopy_imaging_TEM}~III~a), we find that a detection window of $x_{\rm max}=10~\text{\AA}$ captures the dominant spatial features, ensuring the effective convergence of the CFI. Under these conditions, the CFI converges to $\mu_{\rm B}^2\,\mathrm{CFI} \approx 4.6\times10^{-15}$. At a beam current of $1.6~\text{nA}$, this yields an SNR upper bound of $\sim6.8\times 10^{-3}\,\text{s}^{-1/2}\sqrt{t_{\rm acq}}$; consequently, achieving unity SNR would require an acquisition time of approximately $6$ hours. 
Maintaining the same detuning and defocus but narrowing the beam to FWHM~$=1.1$~nm increases the CFI to $\mu_{\rm B}^2\text{CFI}=4.6\times 10^{-11}$. This corresponds to an improved SNR upper bound of $\sim 0.67\,\text{s}^{-1/2}\sqrt{t_{\rm acq}}$, which reduces the acquisition time for unity SNR to 3 seconds. Remarkably, this improvement requires only one-third of the electron dose, lowering it to $5.1\times 10^{7}\rm{e}^-/\text{\AA}^2$.

For Zernike phase imaging, we operate in-focus ($z_d=0$) with a fixed detection window of $x_{\rm max}=10~\text{\AA}$ to maintain spatial resolution comparable to the defocused case. This configuration yields a sensitivity of $\mu_{\rm B}^2\,\mathrm{CFI} \sim 1.9 \times 10^{-14}$ and an SNR upper bound of $\sim1.4 \times 10^{-2}\,\text{s}^{-1/2}\sqrt{t_{\rm acq}}$ at a beam current of $1.6~\text{nA}$. Consequently, achieving unity SNR requires an integration time of approximately $1.5$ hours.Although these results confirm that Zernike phase contrast is more efficient than standard defocused imaging, since it offers roughly $4.2$ times the Fisher information and a twofold SNR improvement~\cite{Koppell2022TransmissionEM, Dwyer2023}, the sensitivity is currently limited by the small detection range. As shown in Fig.~\ref{fig:asympthotic_cfi}~a, expanding the window could substantially increase sensitivity, as the current $10~\text{\AA}$ window captures only a small fraction ($7.2\times10^{-3}$) of the interacting electrons. Approaching the diffraction-mode CFI limit is possible by narrowing the focused beam to FWHM~$=1.1$~nm (with $x_{\rm max}=10$~\AA and $\Delta r_\perp=8.7~a_0$). This configuration reaches $\mu_{\rm B}^2\text{CFI} \approx 1.37\times 10^{-10}$, corresponding to an SNR upper bound of $\sim 1.2~\text{s}^{-1/2}\sqrt{t_{\rm acq}}$. This improvement dramatically reduces the acquisition time for unity SNR to just 1 second at $1.6$~nA, while simultaneously halving the required electron dose to $1.7\times 10^{7}\rm{e}^-/\text{\AA}^2$.

\begin{table}[h]
    \centering
    \begin{tabular}{l|c c}\hline \hline
     & \multicolumn{2}{c}{SNR$/\sqrt{t_{\rm acq}}$ ($10^{-2}\text{s}^{-1/2}$)} \\
    \multicolumn{1}{c|}{Imaging mode} & FWHM $= 110$~nm & FWHM $= 1.1$~nm \\ \hline
        Image mode ($z_d=800$~\AA, pixel $0.32$~\AA) & $0.403$ & $9.55$ \\
        Image mode ($z_d=800$~\AA, pixel $0.16$~\AA) & $0.403$ & $9.52$ \\
        Zernike phase imaging (pixel $0.32$~\AA)     & $0.781$ & $15.8$ \\
        Zernike phase imaging (pixel $0.16$~\AA)     & $0.781$ & $15.9$ \\\hline\hline
    \end{tabular}
    \caption{Comparison of the SNR per squared root of acquisition time (at $1.6$~nA) for the broad beam (FWHM~$=110$~nm) versus the narrow focused beam (FWHM~$=1.1$~nm). In the narrow beam regime, the detection converges towards the diffraction-limited upper bound ($\sim 0.67\,\text{s}^{-1/2}$), effectively equalizing the performance across the different imaging configurations.}
    \label{tab:SNR_position_imaging}
\end{table}

Despite its theoretical advantages, the practical implementation of Zernike phase imaging remains challenging. The method requires isolating the unscattered beam from the spin-induced scattered component, despite their angular separation being on the order of nanoradians. Moreover, applying the necessary phase shift exclusively to the unscattered beam without perturbing the scattered amplitudes demands exceptional precision in phase plate fabrication and alignment, posing significant experimental hurdles.

\begin{figure}[h]
    \centering
\includegraphics[width=0.66\linewidth]{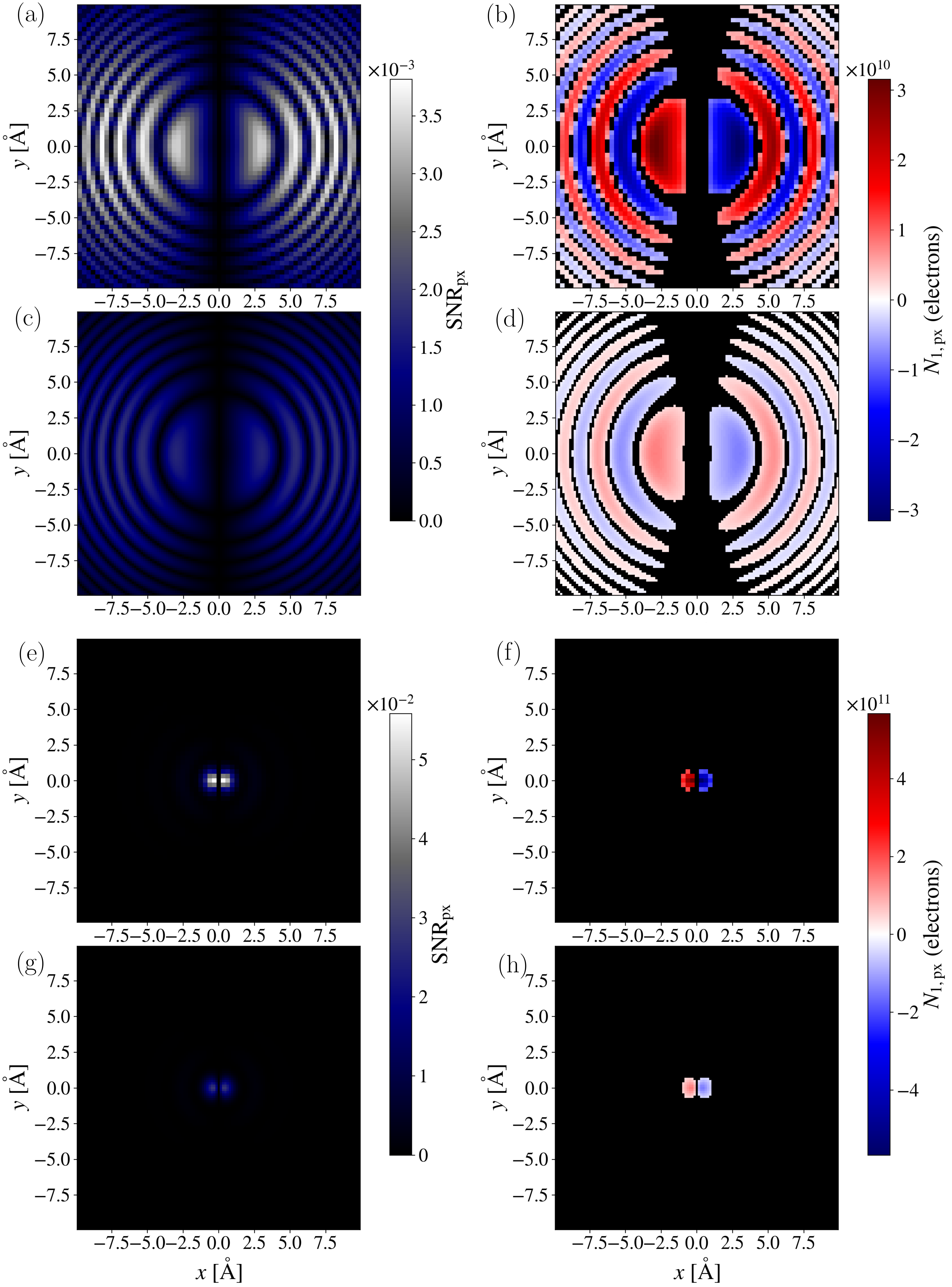}
    \caption{SNR analysis for a narrow electron beam ($\text{FWHM} = 1.1~\text{nm}$). The right column displays the pixelated differential signal, showing the change in electron counts between an MW-excited spin and a ground-state spin. The left column maps the corresponding SNR per pixel derived from these signals. Black overlays in the right column indicate the pixels selected by the optimization mask to maximize the total SNR.
(a)--(d) Defocused imaging ($z_d =800~\text{\AA}$) with pixel sizes of $0.32~\text{\AA}$ (a, b) and $0.16~\text{\AA}$ (c, d).
(e)--(h) Zernike phase contrast ($z_d=0~\text{\AA}$) with pixel sizes of $0.32~\text{\AA}$ (e, f) and $0.16~\text{\AA}$ (g, h).
Calculations assume an incident flux of $10^{10}$ electrons ($1$~s acquisition at $1.6$~nA), corresponding to a dose of $5.1\times 10^{7}~\mathrm{e}^-/\text{\AA}^2$.  }
    \label{fig:optimal_mask_image_nmprobe}
\end{figure}

 The behavior of the signal features is further elucidated in panels (a) and (b) from Fig.~\ref{fig:optimal_mask_image} and Fig.~\ref{fig:optimal_mask_image_nmprobe} for the defocused imaging case, where the pixel-resolved SNR ($\text{SNR}_\text{px}$) is plotted for two different pixel sizes. These plots demonstrate that the majority of the distinguishable signal originates from regions near the expected spin location, while the surrounding interference fringes provide relatively little contribution to the total information content. This spatial localization of the signal implies that only a limited region of the image plane is critical for optimizing detection sensitivity in defocused imaging.
In contrast, the Zernike phase imaging results shown in Figs.~\ref{fig:optimal_mask_image}~e and f reveal a different distribution of $\text{SNR}_\text{px}$. Here, the maximum $\text{SNR}_\text{px}$ remains centered near the spin location, but its peak value is approximately twice as high as that observed in the defocused case. Moreover, the signal extends significantly across a broader region of the image plane, with non-negligible $\text{SNR}_\text{px}$ values persisting far from the origin. 
The broader spatial support of $\text{SNR}_\text{px}$ in Zernike imaging is advantageous for signal integration over extended detector regions, potentially improving robustness against spatial misalignment and enabling more flexible detection schemes. This behavior further demonstrates the effectiveness of Zernike phase contrast imaging for extracting weak phase signals.

To evaluate practical achievability, we applied the masked analysis method described in Appendix~\ref{ap:optimal_single_spins}. As summarized in Table~\ref{tab:SNR_position_imaging}, our simulations indicate no significant dependence on image pixelation. For the broad beam case (FWHM~$=110$~nm), applying this masked method to position-space imaging results in an achievable SNR that falls below the theoretical upper bound by approximately $1.6$ for defocused imaging and $1.8$ for the Zernike configuration. In the narrow beam regime ($\text{FWHM}=1.1$~nm), however, the highly localized signal presents a greater challenge for pixelated detection. Consequently, the masked analysis yields an SNR that falls short of the theoretical limit by factors of $7.0$ and $7.6$ for defocused and Zernike imaging, respectively. Despite this discrepancy, the masked method remains superior to global detection, and as previously noted, increasing the image resolution allows the achievable SNR to effectively converge toward the diffraction-limited theoretical values.

To evaluate practical achievability, we applied the masked analysis method described in Appendix~\ref{ap:optimal_single_spins}. As summarized in Table~\ref{tab:SNR_position_imaging}, our simulations indicate no significant dependence on image pixelation for the broad beam case ($\text{FWHM}=110$~nm); here, the achievable SNR with masking falls below the theoretical upper bound by a factor of approximately $1.6$ for standard imaging and $1.8$ for the Zernike configuration.In the narrow beam regime ($\text{FWHM}=1.1$~nm), however, the highly localized signal presents a greater challenge for pixelated detection. Consequently, the masked analysis yields an SNR that falls short of the theoretical limit by factors of $7.0$ and $7.6$ for standard and Zernike imaging, respectively. These results confirm that optimized information extraction remains feasible and highly effective through the proper selection of informative data regions.

\subsection{Diffraction mode imaging}\label{ap:optimal_single_spin_diffraction}

\begin{figure}[h]
    \centering
    \includegraphics[width=0.6\linewidth]{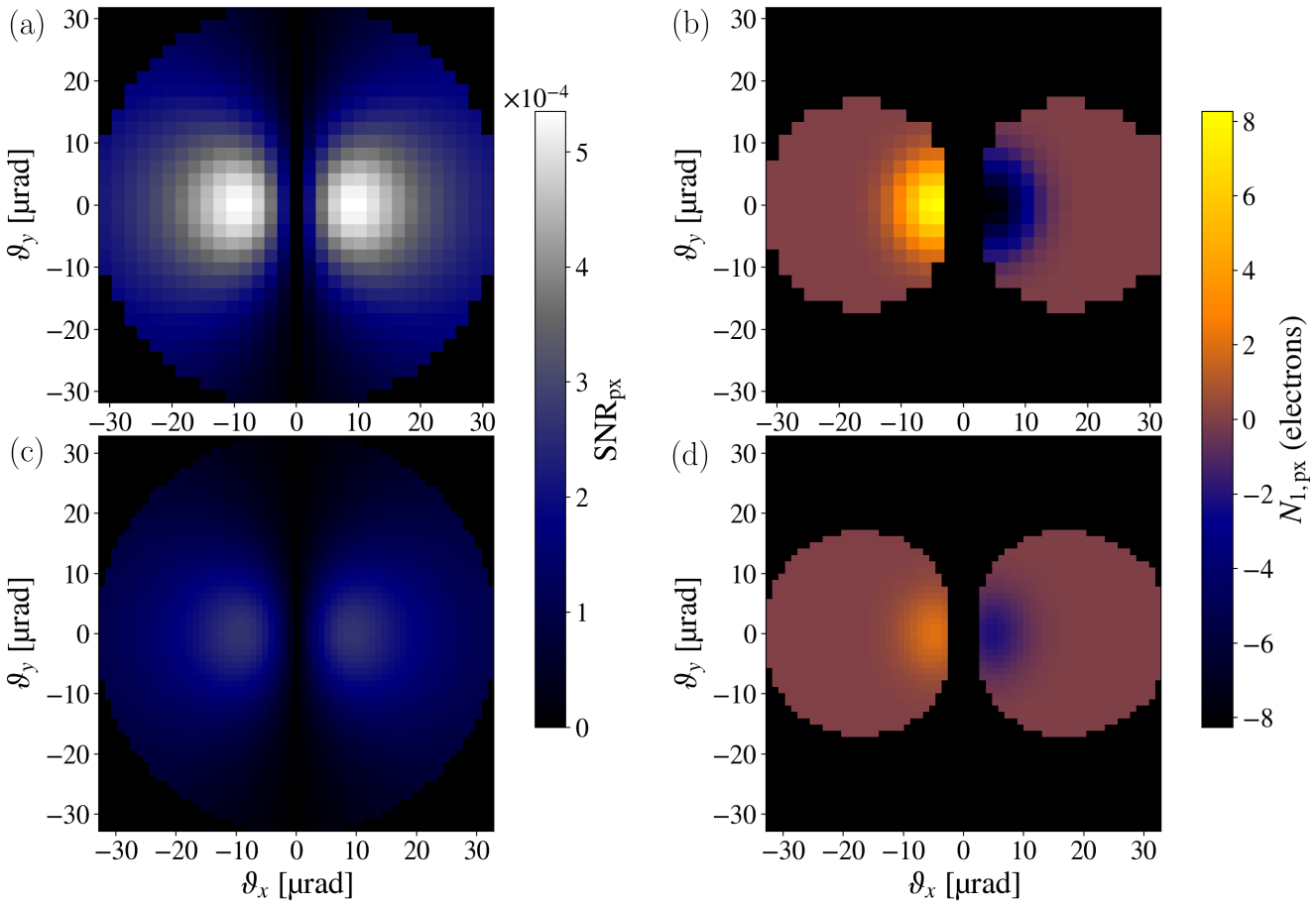}
    \caption{
    Differential diffraction mode imaging of a single spin on-resonance, initially oriented along the $y$-axis at $t_0 = 0$. The color scale represents the difference in the average number of detected electrons per pixel between a spin driven by a MW pulse and a spin in the ground state. While the diffraction pattern rotates over time reflecting spin precession, these panels focus on the optimization of the on-resonance case. (a) and (c) display the SNR per pixel for angular pixel sizes of $2.12~\text{\textmu rad}$ and $1.06~\text{\textmu rad}$, respectively. (b) and (d) show the pixelated images with an optimized mask (black) applied to maximize the total SNR for pixel sizes of $2.12~\text{\textmu rad}$ and $1.06~\text{\textmu rad}$, respectively. The simulation employs an incident Gaussian electron beam wavepacket with $\text{FWHM} = 0.11~\text{\textmu m}$ ($\Delta k_\perp=4.22\times 10^{-6} k_{z,0}$), and a total of $10^{10}$ electrons is assumed to traverse the sample.}
    \label{fig:diffraction_single_spin_optimal}
\end{figure}

For single-spin diffraction mode imaging, Fig.~\ref{fig:spectroscopy_imaging_TEM}~II displays the differential probability distributions across various detunings. Focusing on the on-resonance case, Fig.~\ref{fig:asympthotic_cfi}~b demonstrates that as the maximal collection angle increases, the CFI converges to the metrological maximum derived for diffraction imaging ($\mu_{\rm B}^2\mathrm{CFI}_{\text{SVEA}}$)~\cite{metrologySpinTEM}. Achieving this limit requires a detector with sufficient resolution to resolve the beam spread ($\Delta k_\perp/k_{z,0}$) and a range wide enough to capture the high-momentum components associated with the spin delocalization ($k_a/k_{z,0}$). However, at large angles, our simulation yields unphysical negative probabilities; thus, we restrict the analysis to $\vartheta \leq 8\Delta k_\perp/k_{z,0}$. Since this range encompasses essentially all interacting electrons, the excluded outer pixels correspond primarily to the distribution tails.

Under this restricted detection range, using a broad incident beam (FWHM $= 110$~nm) in the on-resonance configuration, the CFI yield in diffraction mode is approximately double that of standard defocused imaging, yet roughly half that of Zernike phase imaging. However, as illustrated in Fig.~\ref{fig:asympthotic_cfi}, both Zernike and diffraction modes ultimately converge to the same SVEA limit once the full detection range and necessary resolution become accessible.Specifically, for the broad beam case (FWHM $= 110$~nm), we calculate $\mu_{\rm B}^2\mathrm{CFI} \sim 9.4\times 10^{-15}$. This yields an SNR upper bound of $\sim 9.7\times 10^{-3}~\text{s}^{-1/2}\sqrt{t_{\rm acq}}$ at $1.6$~nA, requiring $\sim 3$ hours of acquisition to reach unity SNR. Conversely, tightening the beam focus (FWHM $= 1.1$~nm) increases the sensitivity to $\mu_{\rm B}^2\mathrm{CFI} \sim 8.8\times 10^{-11}$. This value closely approaches the theoretical diffraction-mode limit ($\mu_{\rm B}^2\mathrm{CFI} \sim 1.37\times 10^{-10}$ for $\Delta r_\perp=8.92a_0$) due to the relatively large detection region involved. Consequently, the SNR upper bound rises to $\sim 0.94~\text{s}^{-1/2}\sqrt{t_{\rm acq}}$ at $1.6$~nA. This dramatically reduces the acquisition time for unity SNR to $\sim 1.2$ seconds, with a required electron dose of $2.8\times 10^{7}\rm{e}^-/\text{\AA}^2$, which is roughly three times lower than that of the broad beam case.

\begin{table}[h!]
    \centering
    \begin{tabular}{c|c|c}
    \hline \hline
    FWHM incident beam & Pixelation & SNR/$\sqrt{t_{\rm acq}}$ ($10^{-2}$~s$^{-1/2}$) \\
    \hline
    $110$~nm & 1 px = $2.12$~\textmu rad & 0.338 \\  
    $110$~nm & 1 px = $1.06$~\textmu rad & 0.329\\ 
    $1.1$~nm & 1 px = $212$~\textmu rad & 33.4 \\  
    $1.1$~nm & 1 px = $106$~\textmu rad & 32.5 \\
    \hline \hline
    \end{tabular}
    \caption{SNR per square root of acquisition time at 1.6~nA current for optimized masking in diffraction mode, showing results for two beam sizes and pixelations.}
    \label{tab:snr_diffraction}
\end{table}

\begin{figure}[h]
    \centering
    \includegraphics[width=0.6\linewidth]{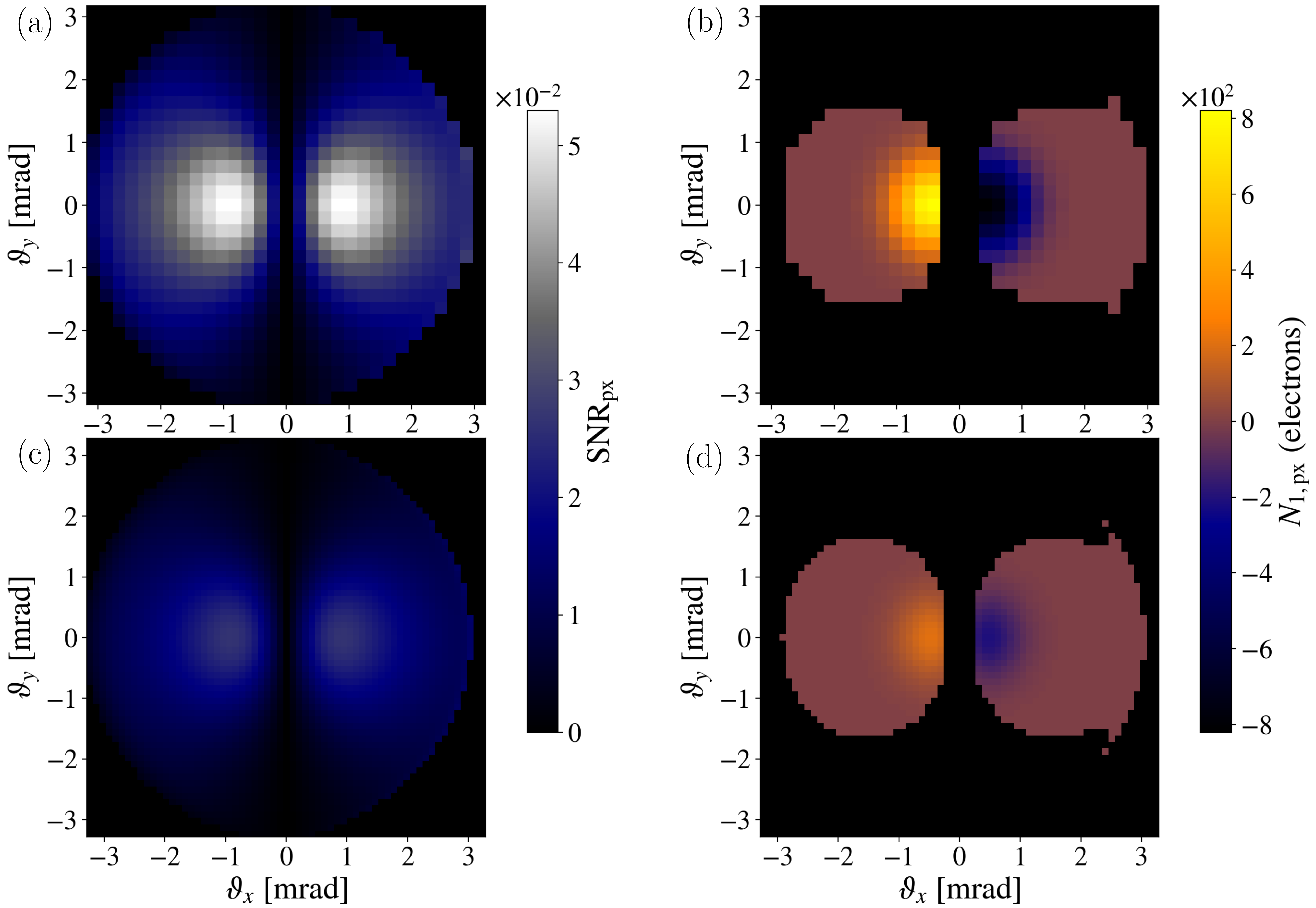}
    \caption{Differential diffraction mode imaging of a single spin. on-resonance condition, with the spin initially oriented along the $y$-axis at $t_0 = 0$. 
    The color scale indicates how the average number of detected electrons per detector (or virtual) pixel changes when comparing two scenarios: a spin driven by an MW pulse before interacting with the electron beam, and a spin in the ground state aligned with the bias field. 
As time progresses, the diffraction pattern undergoes rotation, reflecting the spin precession within the transverse plane. Then, we analyze only the on-resonance case for SNR and optimization. The SNR per pixel is analyzed for angles with pixel sizes of (a) $212 $~\textmu rad and (c) $106 $~\textmu rad. An optimized mask (black) is applied to the pixelated images, enhancing the total SNR for pixel sizes of (b) $212 $~\textmu rad and (d) $106 $~\textmu rad. The simulation employs an incident electron beam as a Gaussian wavepacket with FWHM = $1.1$~nm ($\Delta k_\perp=4.22\cdot 10^{-4} k_{z,0}$). Furthermore, a total of $10^{10}$ electrons is assumed to traverse the sample.}
    \label{fig:diffraction_single_spin_optimal104}
\end{figure}

In addition, we depicted the $\text{SNR}_\text{px}$ analysis in Figs.~\ref{fig:diffraction_single_spin_optimal}~a,~\ref{fig:diffraction_single_spin_optimal}~c,~\ref{fig:diffraction_single_spin_optimal104}~a and~\ref{fig:diffraction_single_spin_optimal104}~c reveals that the upper bound SNR is not necessarily located where the signal is highest, due to the dominance of noise in this region. Instead, the optimal detection region is approximately $2 \Delta k_\perp/k_{z,0}$ from the beam center, where the signal sufficiently exceeds the noise originating from the Gaussian zero-deflection peak.
Applying the optimized masking procedure to diffraction images (Figs.~\ref{fig:diffraction_single_spin_optimal}~b,d and Figs.~\ref{fig:diffraction_single_spin_optimal104}~b,d) results in unavoidable signal loss due to significant noise contributions from the zero-deflection peak. The analysis masks exhibit slight asymmetry, arising from reduced counts in first-order terms that simultaneously decrease noise (visible in Fig.~\ref{fig:diffraction_single_spin_optimal104}~c). At larger angles, the probability distribution develops negative entries, which we exclude by restricting analysis to $\vartheta \leq 8\Delta k_\perp/k_{z,0}$, detecting essentially all  electrons that interact with the sample. Consequently, a portion of the covered pixels corresponds primarily to distribution tails.

The resulting SNR values for masked images are summarized in Table~\ref{tab:snr_diffraction}. We observe negligible variation in SNR with respect to pixel size. Notably, the analysis falls short by a factor $3$ with respect to the theoretical upper bounds.
These results indicate that the pixel classification strategy based on $\text{SNR}_{\text{px}}$ and the subsequent segmentation approach, while suboptimal, approaches the classical bound. The simulations provide a foundation for refining data analysis methods to more closely approach theoretical sensitivity bounds and guide future signal extraction improvements.

\end{document}